%% file: arxiv.tex
\documentclass[journal]{IEEEtran}
\input{eucapconfig}
\begin{document}
\title{Wireless Propagation Parameter Estimation with Convolutional Neural Networks}

\author{Steffen~Schieler\orcidlink{0000-0003-4480-234X},~\IEEEmembership{Student Member,~IEEE},
        Sebastian~Semper\orcidlink{0000-0002-2610-7389},
        and~Reiner~Thom\"a\orcidlink{0000-0002-9254-814X},~\IEEEmembership{Life Fellow,~IEEE}
\thanks{Corresponding author: Steffen Schieler,\href{mailto:steffen.schieler@tu-ilmenau.de}{steffen.schieler@tu-ilmenau.de}}
\thanks{All authors are with the Technische Universit\"at Ilmenau, FG EMS, Ilmenau, Germany}%
\thanks{This is the accepted version of the article published in the International Journal of Microwave and Wireless Technologies with the DOI 10.1017/S1759078725000431.}}

\vskip0.5\baselineskip

\maketitle

\begin{abstract}
    Wireless channel propagation parameter estimation forms the foundation of channel sounding, estimation, modeling, and sensing.
    This paper introduces a Deep Learning approach for joint delay- and Doppler estimation from frequency and time samples of a radio channel transfer function.
    
    Our work estimates the two-dimensional path parameters from a channel impulse response containing an unknown number of paths.
    Compared to existing deep learning-based methods, the parameters are not estimated via classification but in a quasi-grid-free manner.
    We employ a deterministic preprocessing scheme that incorporates a multi-channel windowing to increase the estimator's robustness and enables the use of a \gls{cnn} architecture.
    The proposed architecture then jointly estimates the number of paths along with the respective delay and Doppler-shift parameters of the paths.
    Hence, it jointly solves the model order selection and parameter estimation task.
    We also integrate the \gls{cnn} into an existing maximum-likelihood estimator framework for efficient initialization of a gradient-based iteration, to provide more accurate estimates.

    In the analysis, we compare our approach to other methods in terms of estimate accuracy and model order error on synthetic data.
    Finally, we demonstrate its applicability to real-world measurement data from a anechoic bi-static RADAR emulation measurement.
\end{abstract}


\section{Introduction}\label{sec:introduction}
\input{tex/introduction}
\section{Signal Model}\label{sec:signalmodel}
\input{tex/signalmodel}

\section{Neural Network}\label{sec:neurealnetwork}
\input{tex/neuralnetwork}

\section{Analysis}\label{sec:analysis}
\input{tex/analysis}

\section{Conclusion}\label{sec:conclusion}
\input{tex/conclusion}

\section*{Funding}
The authors acknowledge the financial support by the Federal Ministry of Education and Research of Germany in the project “Open6GHub” (grant number: 16KISK015), “KOMSENS-6G” (grant number: 16KISK125), and DFG project HoPaDyn with Grant-No. TH 494/30-1.

\section*{Declaration of Interests}
The authors report no conflicts of interest.

\newpage
\section*{Author Biographies}
\begin{IEEEbiography}[{\includegraphics[width=1in,height=1.25in,clip,keepaspectratio]{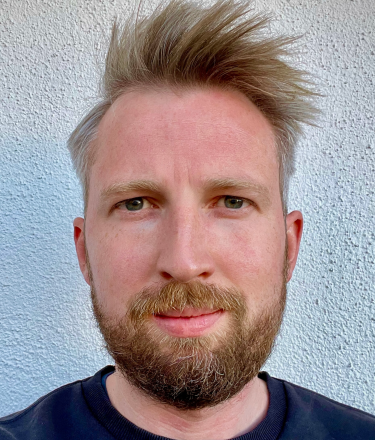}}]{Steffen Schieler}
is a researcher and Ph.D. student with the Electronic Measurement and Signal Processing Group at Technische Universität Ilmenau, Germany.
His research focus is multidimensional wireless channel parameter estimation, and the application of signal processing algorithms with Machine- and Deep Learning methods.
\end{IEEEbiography}
\begin{IEEEbiography}[{\includegraphics[width=1in,height=1.25in,clip,keepaspectratio]{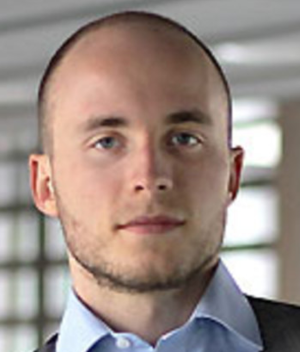}}]{Sebastian Semper}
studied mathematics at Technische Universit\"at Ilmenau, (TU Ilmenau), Ilmenau, Germany. He received the Master of Science degree in 2015. Since 2015, he has been a Research Assistant with the Electronic Measurements and Signal Processing Group, which is a joint research activity between the Fraunhofer Institute for Integrated Circuits IIS and TU Ilmenau, Germany,
In 2022 he finished his doctoral studies and received the doctoral degree with honors in electrical engineering. Since then, he has been a post doctoral student in the Electronic Measurements and Signal Processing Group.
His research interests consist of compressive sensing, parameter estimation, optimization, numerical methods, and algorithm design.
\end{IEEEbiography}
\begin{IEEEbiography}[{\includegraphics[width=1in,height=1.25in,clip,keepaspectratio]{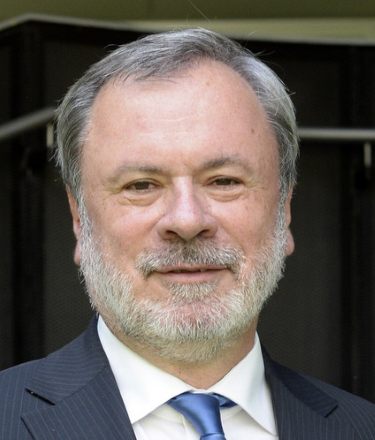}}]{Reiner Thom\"a}
 received his degrees in electrical engineering and information technology from TU Ilmenau, Germany. 
Since 1992, he has been a Professor at the same university and retired in 2018. 
In 2007, he received the Thuringian State Research Award for Applied Research and in 2014 the Vodafone Innovation Award, both for his contributions to high-resolution multidimensional channel sounding, and in 2020, the EurAAP Propagation Award. 
His research interests include multidimensional channel sounding, propagation measurement and model-based parameter estimation, MIMO system over-the-air testing in virtual electromagnetic environments, MIMO radar, passive coherent location, and integrated sensing and communication.
\end{IEEEbiography}
\vfill
\printbibliography

\end{document}

%% file: eucapconfig.tex
\usepackage{amsmath,amssymb,amsfonts}
\usepackage[hidelinks]{hyperref}
\usepackage{afterpage}
\usepackage{orcidlink}
\usepackage{mathtools}
\usepackage{bm}
\usepackage{algorithmic}
\usepackage{graphicx}
\usepackage{textcomp}
\usepackage{xcolor}
\usepackage[acronym,toc]{glossaries}
\usepackage{cleveref}
\usepackage{siunitx}
\usepackage{tabularx}
\usepackage{booktabs}
\usepackage{pgfplots, tikz}
\pgfplotsset{compat=1.17}
\usetikzlibrary{arrows.meta}
\usepackage{tikzit}
\usepackage{environ}
\usepackage{placeins}
\ifCLASSOPTIONcompsoc
    \usepackage[caption=false, font=normalsize, labelfont=sf, textfont=sf]{subfig}
\else
    \usepackage[caption=false, font=footnotesize]{subfig}
\fi
\usepackage[activate={true,nocompatibility},
    final,
    tracking=true,
    kerning=true,
    spacing=true,
    factor=1100,
    stretch=25,
    shrink=25,
    letterspace=-30
]{microtype}

\makeatletter
\newsavebox{\measure@tikzpicture}
\NewEnviron{scaletikzpicturetowidth}[1]{%
    \def\tikz@width{#1}%
    \begin{lrbox}{\measure@tikzpicture}%
        \BODY
    \end{lrbox}%
    \pgfmathparse{#1/\wd\measure@tikzpicture}%
    \BODY
}
\makeatother

\loadglsentries{acronyms.tex}
\def\BibTeX{{\rm B\kern-.05em{\sc i\kern-.025em b}\kern-.08em
T\kern-.1667em\lower.7ex\hbox{E}\kern-.125emX}}

\newcommand{\N}{\mathbb{N}}
\newcommand{\R}{\mathbb{R}}
\newcommand{\C}{\mathbb{C}}

\newcommand{\Norm}[1]{{\left\Vert #1\right\Vert}}

\newcommand{\reddot}{\tikz{\fill[red, draw=black] (0,0) circle (2pt);}}
\newcommand{\whitecircle}{\tikz{\fill[white, draw=black] (0,0) circle (3pt);\fill[white, draw=black] (0,0) circle (1.6pt);}}
\newcommand{\blackcircle}{\tikz{\fill[black, draw=black] (0,0) circle (3pt);\fill[white, draw=black] (0,0) circle (2.4pt);}}

\definecolor{coolor1}{HTML}{ff9900}
\definecolor{coolor2}{HTML}{29789b}
\definecolor{coolor3}{HTML}{f1f7ed}
\definecolor{coolor4}{HTML}{1a936f}
\definecolor{coolor5}{HTML}{cf3155}

\definecolor{QC1}{HTML}{0077BB}
\definecolor{QC2}{HTML}{33BBEE}
\definecolor{QC3}{HTML}{009988}
\definecolor{QC4}{HTML}{EE7733}
\definecolor{QC5}{HTML}{CC3311}
\definecolor{QC6}{HTML}{EE3377}
\definecolor{QC7}{HTML}{BBBBBB}

\ifx\qty\undefined
    \let\oldnum\num
    \renewcommand{\num}[2][ignored]{\oldnum{#2}}

    \ifx\qty\undefined
        \newcommand{\qty}[3][ignored]{\SI{#2}{#3}}
    \fi

    \ifx\qtyrange\undefined
        \newcommand{\qtyrange}[4][ignored]{\SIrange{#2}{#3}{#4}}
    \fi
\fi

\usepackage[
    sorting=none,
    maxbibnames=99,
    defernumbers=true,
    style=ieee,
    giveninits=true,
]{biblatex}
\AtEveryBibitem{
    \clearlist{address}
    \clearfield{url}
    \clearfield{month}
    \clearfield{editor}
    \clearfield{eprint}
    \clearfield{volume}
    \clearfield{isbn}
    \clearfield{issn}
    \clearfield{pages}
    \clearfield{addendum}
    \clearlist{location}
    \clearlist{pages}
    \clearlist{editor}
}
\makeatletter
\let\blx@rerun@biber\relax
\makeatother

%% file: acronyms.tex
\newacronym{admm}{ADMM}{Alternating Directions of Multipliers Method}
\newacronym{anm}{ANM}{Atomic Norm Minimization}
\newacronym{adc}{ADC}{Analog-to-Digital Converter}
\newacronym{awgn}{AWGN}{Additive White Gaussian Noise}
\newacronym{asic}{ASIC}{Application Specific Integrated Circuit}
\newacronym{arpack}{ARPACK}{ARnoldi PACKage}
\newacronym{api}{API}{Application Programmable Interface}
\newacronym{aut}{AUT}{Antenna Under Test}
\newacronym[plural=AOI, firstplural=Areas of Interest (AOI)]{aoi}{AOI}{Area of Interest}
\newacronym{ai}{AI}{Artifical Intelligence}
\newacronym{aic}{AIC}{Akaike Information Criterion}

\newacronym{bp}{BP}{Basis Pursuit}
\newacronym{bpdn}{BPDN}{Basis Pursuit Denoising}
\newacronym{blue}{BLUE}{Best Linear Unbiased Estimator}

\newacronym{cnn}{CNN}{Convolutional Neural Network}
\newacronym{crb}{CRB}{Cramér-Rao Bound}
\newacronym{cs}{CS}{Compressed Sensing}
\newacronym{cf}{CF}{Crest factor}
\newacronym{cr}{CR}{Compression Ratio}
\newacronym{cmos}{CMOS}{Complementary Metal Oxide Semiconductor}
\newacronym{cots}{COTS}{Commercial-Off-The-Shelf}
\newacronym{cpu}{CPU}{Central Processing Unit}
\newacronym{cfar}{CFAR}{Constant False Alarm Rate}
\newacronym{cvx}{CVX}{Convex Optimization toolboX}

\newacronym{doa}{DoA}{Direction of Arrival}
\newacronym{dod}{DoD}{Direction of Departure}
\newacronym[plural=DMC,firstplural=Dense Multipath Components (DMC)]{dmc}{DMC}{Dense Multipath Components}
\newacronym{das}{DAS}{Delay-and-Sum}
\newacronym{dft}{DFT}{Discrete Fourier Transform}
\newacronym{idft}{IDFT}{Inverse Discrete Fourier Transform}
\newacronym{dct}{DCT}{Discrete Cosine Transform}
\newacronym{dsp}{DSP}{Digital Signal Processor}
\newacronym{dnn}{DNN}{Deep Neural Network}

\newacronym{eadf}{EADF}{Effective Aperture Distribution Function}
\newacronym{etadf}{ETADF}{Effective Time-Aperture Distribution Function}
\newacronym{esprit}{ESPRIT}{Estimation of Signal Parameters via Rotational Invariance Techniques}
\newacronym{ett}{ETT}{Eigenvalue Threshold Test}
\newacronym{edc}{EDC}{Efficient Detection Criterion}
\newacronym{eft}{EFT}{Exponential Fitting Test}

\newacronym{fc}{FC}{fully-connected}
\newacronym{fht}{FHT}{Fast Hadamard Transform}
\newacronym[longplural={Fast Fourier Transforms}]{fft}{FFT}{Fast Fourier Transform}
\newacronym{fmcw}{FMCW}{Frequency-Modulated Continuous-Wave}
\newacronym{fpga}{FPGA}{Field Programmable Gate Array}
\newacronym{fri}{FRI}{Finite Rate of Innovation}
\newacronym{fim}{FIM}{Fisher Information Matrix}
\newacronym{fmc}{FMC}{Full Matrix Capture}
\newacronym{fista}{FISTA}{Fast Iterative Shrinkage-Thresholding Algorithm}
\newacronym{frvm}{FRVM}{Fast Relevance Vector Machine}

\newacronym{gfcs}{grid-free CS}{grid-free compressive sensing}
\newacronym{gpu}{GPU}{Graphical Processing Unit}
\newacronym{gtd}{GTD}{Geometrical Theory of Diffraction}
\newacronym{gan}{GAN}{Generative Adversarial Network}

\newacronym{hrpe}{HRPE}{High Resolution Parameter Estimator}
\newacronym{hfss}{Ansys HFSS}{Ansys High Frequency Electromagnetic Simulation Software}

\newacronym[longplural={Inverse Fast Fourier Transforms}]{ifft}{IFFT}{Inverse Fast Fourier Transform}
\newacronym{ir}{IR}{Impulse Response}
\newacronym{iid}{iid}{independent and identically distributed}
\newacronym{iir}{IIR}{Infinite Impulse Response}
\newacronym{irf}{IRF}{Impulse Response Function}
\newacronym{ista}{ISTA}{Iterative Shrinkage-Thresholding Algorithm}
\newacronym{isac}{ISAC}{Integrated Sensing and Communication}

\newacronym{kld}{KLD}{Kullback-Leibler Divergence}

\newacronym{ls}{LS}{Least Squares}
\newacronym{lasso}{LASSO}{Least Absolute Shrinkage and Selection Operator}
\newacronym{lse}{LSE}{Line Spectral Estimation}
\newacronym{lfsr}{LFSR}{Linear Feedback Shift Register}
\newacronym{lo}{LO}{Local Oscillator}
\newacronym{los}{LOS}{Line of Sight}
\newacronym{lti}{LTI}{Linear Time-Invariant}
\newacronym{ltv}{LTV}{Linear Time-Variant}
\newacronym{lam}{LAM}{Large Area Monitoring}

\newacronym{mimo}{MIMO}{multiple input multiple output}
\newacronym{mmv}{MMV}{multiple measurement vectors}
\newacronym{ml}{ML}{maximum likelihood}
\newacronym{mmse}{MMSE}{misspecified mean squared error}
\newacronym{mse}{MSE}{mean squared error}
\newacronym{bce}{BCE}{Binary Crossentropy}
\newacronym{mlbs}{MLBS}{Maximum Length Binary Sequence}
\newacronym{mpc}{MPC}{Multipath Component}
\newacronym{msm}{MSM}{M-Sequence Method}
\newacronym{mwc}{MWC}{Modulated Wideband Converter}
\newacronym{mpm}{MPM}{Matrix Pencil Method}
\newacronym{mpu}{MPU}{Microprocessor Unit}
\newacronym{mri}{MRI}{Magnetic resonance imaging}
\newacronym{music}{MUSIC}{Multiple Signal Classification}
\newacronym{mkl}{MKL}{Math Kernel Library}
\newacronym{mcrb}{MCRB}{Misspecified Cramér-Rao Bound}
\newacronym{mmle}{MMLE}{Misspecified Maximum-Likelihood Estimator}
\newacronym{ndt}{NDT}{Nondestructive Testing}
\newacronym{nde}{NDE}{Nondestructive Evaluation}
\newacronym{nn}{NN}{Neural Net}
\newacronym{nist}{NIST}{National Institute of Standards and Technology}

\newacronym{omp}{OMP}{Orthogonal Matching Pursuit}
\newacronym{oop}{OOP}{Object Oriented Programming}
\newacronym{ofdm}{OFDM}{Orthogonal Frequency Division Multiplexing}

\newacronym{pdp}{PDP}{Power Delay Profile}
\newacronym{pn}{PN}{Pseudo-Noise}
\newacronym{pwc}{PWC}{Plane Wave Compounding}
\newacronym{pwi}{PWI}{Plane Wave Imaging}
\newacronym{pura}{PURA}{Patch Uniform Rectangular Array}
\newacronym{pymax}{PyMAX}{Python Maximization Approach}
\newacronym{pts}{PTS}{Pseudo-True Solution}
\newacronym{pdf}{pdf}{probability density function}
\newacronym{pinn}{PINN}{Physics-Informed Neural Network}

\newacronym{relu}{ReLU}{Rectified Linear Unit}
\newacronym{resnet}{ResNet}{Residual Neural Network}
\newacronym{ram}{RAM}{Random Access Memory}
\newacronym{rcs}{RCS}{Radar Cross Section}
\newacronym{rd}{RD}{Random Demodulator}
\newacronym{rx}{Rx}{receiver}
\newacronym{rem}{REM}{Reconstruction Error Metric}
\newacronym{rmse}{RMSE}{Root Mean Squared Error}
\newacronym{rms}{RMS}{root mean squared}
\newacronym{ric}{RIC}{Restricted Isometry Constant}
\newacronym{rip}{RIP}{Restricted Isometry Property}
\newacronym{rc}{RC}{Raised Cosine}
\newacronym{roi}{ROI}{Region of Interest}
\newacronym{rimax}{RIMAX}{Richter Maximization Approach}
\glsunset{rimax}
\newacronym{rvm}{RVM}{Relevance Vector Machine}

\newacronym{scf}{SCF}{spatial correlation function}
\newacronym{samurai}{SAMURAI}{Synthetic Aperture Measurements of Uncertainty in Angle of Incidence}
\newacronym[plural=SC,firstplural=Specular Components (SC)]{sc}{SC}{Specular Components}
\newacronym{sdp}{SDP}{semi-definite program}
\newacronym{sdr}{SDR}{Signal to Diffuse Ratio}
\newacronym{simd}{SIMD}{Single Instruction Multiple Data}
\newacronym{svd}{SVD}{singular value decomposition}
\newacronym{svm}{SVM}{Support Vector Machine}
\newacronym{soe}{SOE}{Sparsity Order Estimation}
\newacronym{sgd}{SGD}{Stochastic Gradient Descent}
\newacronym{stuca}{StUCA}{Stacked Uniform Circular Array}
\newacronym{spucpa}{SPUCPA}{Stacked Polarimetric Uniform Circular Patch Array}
\newacronym{suca}{SUCA}{Stacked Uniform Circular Array}
\newacronym{saft}{SAFT}{Synthetic Aperture Focusing Technique}
\newacronym{sota}{SOTA}{State of the Art}
\newacronym{ssd}{SSD}{Solid State Device}
\newacronym{ssr}{SSR}{Sparse Signal Recovery}
\newacronym{sa}{SA}{Synthetic Aperture}
\newacronym{spw}{SPW}{Single Plane Wave}
\newacronym{shm}{SHM}{Structural Health Monitoring}
\newacronym{snr}{SNR}{Signal-to-Noise Ratio}
\newacronym{stela}{STELA}{Soft-Thresholding with Exact Line Search Algorithm}
\newacronym{siso}{SISO}{Single Input Single Output}
\newacronym{simo}{SIMO}{Single Input Multiple Output}
\newacronym{swe}{SWE}{Spherical Wave Expansion}
\newacronym{sme}{SME}{Spherical Mode Expansion}
\newacronym{sage}{SAGE}{Space-Alternating Generalized Expectation-Maximization}
\newacronym{spp}{SPP}{Spatial Pyramid Pooling}

\newacronym{th}{T\&H}{Track and Hold}
\newacronym{tf}{TF}{Transfer Function}
\newacronym{tx}{TX}{transmitter}
\newacronym{twista}{TWISTA}{Two-step Iterative Shrinkage-Thresholding Algorithm}
\newacronym{tof}{ToF}{Time-of-Flight}

\newacronym{uca}{UCA}{uniform circular array}
\newacronym{ura}{URA}{Uniform Rectangular Array}
\newacronym{ula}{ULA}{Uniform Linear Array}
\newacronym{uwb}{UWB}{Ultra-Wideband}
\newacronym{usndt}{US-NDT}{Ultrasonic Non-destructive Testing}

\newacronym{vna}{VNA}{Vector Network Analyser}
\newacronym{vsh}{VSH}{Vector Spherical Harmonics}
\hyphenation{op-tical}
\hyphenation{net-works}
\hyphenation{semi-conduc-tor}

%% file: tex/introduction.tex
Wireless channel parameter estimation is a problem encountered in many signal processing tasks, e.g., channel estimation~\cite{thomae2005multidim_cs}, radar localization, and direction finding.
Available solutions for the task include subspace algorithms, like \gls{music}~\cite{schmidt1986MUSIC} or \gls{esprit}~\cite{roy1989esprit}, iterative \gls{ml}~\cite{richter_estimation_2005}, and \gls{ssr}~\cite{malioutov2005ssr_source_loc} and \gls{dnn}-based algorithms~\cite{izacard_data-driven_2019, papageorgiou_deep_2020, chen_robust_2022, naseri_machine_2022, liu_super_2020, barthelme_machine_2021}.
We focus on the latter category, presenting a \gls{cnn} approach that is also compatible with existing model-based optimization techniques.
Our review of similar works in this category, namely those discussed below, includes spectral- and \gls{doa}-estimation due to their similar algebraic structure.
\subsection{State of the Art}\label{sec:introduction:sota}
In~\cite{izacard_data-driven_2019} a \gls{cnn} is used to estimate frequencies by predicting a super-resolution pseudo-spectrum from a superposition of up to ten components.
A separate, subsequent network performs the model order estimation task.
The results show a performance improvement when compared to \gls{music}, especially in the low-\gls{snr} domain.
In~\cite{papageorgiou_deep_2020} a \gls{cnn} is trained to perform \gls{doa} estimation of up to three unknown paths by determining their location on a grid, i.e., solving a classification problem.
Similarly, the authors of~\cite{chen_robust_2022} address the problem of \gls{doa} estimation by combining a denoising autoencoder with another \gls{dnn} for the estimation.
As in~\cite{izacard_data-driven_2019}, both approaches show performance improvements in the low-\gls{snr} domain compared to \gls{music}.\par
However, classification methods suffer from inherent estimation bias due to grid mismatch and poor scaling of the desired resolution, as higher grid granularity increases classification complexity exponentially.
Naturally, approaches not based on grids are highly interesting, especially if we wish to achieve super-resolution.
Comparably,~\cite{barthelme_machine_2021} performs \gls{doa} estimation by combining a regression task, solved by a \gls{dnn}, with gradient steps on the data's likelihood function.
The grid-free estimates from the \gls{dnn} are used as initial guesses for a second-order Newton method.
Hence, it combines a \gls{dnn} and maximum-likelihood methods to improve the fast, but approximate initial estimates from the \gls{dnn} with an iterative high-resolution method.
The results highlight the decrease in computational complexity and improvements in performance from the combination.
However, the presented method considers only up to $3$ stochastic paths estimated from multiple snapshots, which is insufficient in many practical scenarios, especially if paths are modeled as deterministic.
More significantly, the used antenna array geometries have a very small aperture, rendering the initialization of the Newton method very well conditioned since it tolerates high deviations of the initialization from the true solution.
\subsection{Our Contributions}\label{sec:introduction:contributions}
Comparably, we estimate an unknown number of deterministic paths (also known as model order selection) from a single snapshot of frequency-time data recorded from a \gls{siso} system.
Additionally, we jointly estimate each path's propagation parameters, namely delay and Doppler-shift.%
\footnote{An earlier version of this paper was presented at the \num{18}th Conference of Antennas and Propagation (EuCAP \num{2024}) and was published in its Proceedings~\cite{schieler2024_eucap}.}.
We use supervised learning, combined with a signal model to generate synthetic data, where a full groundtruth for the estimates is available compared to measurement data.
For the architecture and training, we take inspiration from computer vision, particularly~\cite{redmon_you_2016} (and subsequent iterations), where a \gls{cnn} estimates an unknown number of objects and their positions in an image.
To alleviate the aforementioned grid dependencies, we exploit sparsity in the parameter space by dividing it up into a significantly smaller number of auxiliary grid cells.
Then we employ the \gls{cnn} to estimate the number of paths in each cell alongside the parameters, encoded as real-valued differences from the respective cells' centers.

To ease the learning task and render the training more robust, we apply multiple windowing functions and the \gls{dft} in a preprocessing scheme to the original frequency and time domain input data.
For performance analysis, we consider the \gls{mse} of the raw \gls{cnn} estimates and compare them to existing methods, i.e., an iterative maximum-likelihood approach (similar to \gls{rimax}~\cite{richter_estimation_2005}), and the \gls{crb} as the theoretical lower bound.
We also conduct an isolated analysis of the model order selection performance and compare our approach to the \gls{edc}.
Then, we use the raw estimates to initialize a second-order Newton scheme to achieve higher accuracy, such that the final estimates follow the \gls{crb}.
The result is a robust $2$D parameter estimator with only moderate computational complexity and a tunable accuracy-runtime trade-off, suitable for real-time applications, such as \gls{isac}.
Finally, we demonstrate that the \gls{cnn} trained with synthetic data works on measurement data from a bi-static RADAR emulation with two moving objects~\cite{Schwind2019}.
\begin{figure*}[t]
    \centering
    \begin{scaletikzpicturetowidth}{\textwidth}
        \input{figures/architecture.tikz}
    \end{scaletikzpicturetowidth}
    \caption{The signal processing chain with preprocessing, trainable blocks, and postprocessing, where the lower parts detail the steps in the upper part. The postprocessing features an optional \textit{optimization}-step, which adds a second-order Gauss Newton method to refine the \gls{cnn} estimates.}
    \label{fig:architecture}
\end{figure*}
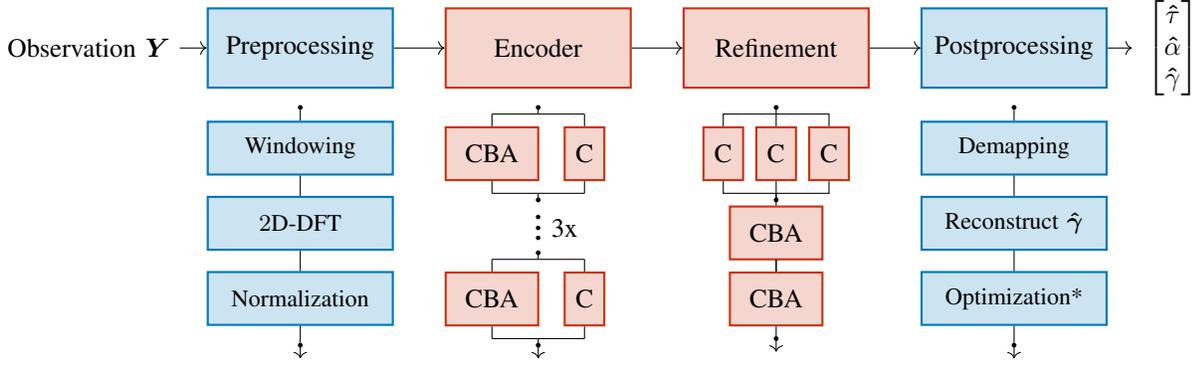

%% file: figures/architecture.tikz

\newcommand{\titleface}[1]{\normalsize{\textbf{#1}}}
\def\blockdist{9em}
\def\boxwidth{7em}
\def\boxheight{3em}

\tikzstyle{dp upper}=[minimum width=\boxwidth, minimum height=\boxheight, align=center, draw=QC1, fill=QC1!20, thick, anchor=center]
\tikzstyle{dp lower}=[minimum width=\boxwidth, minimum height=0.75*\boxheight, align=center, draw=QC1, fill=QC1!20, thick, anchor=center]
\tikzstyle{dp fc}=[minimum width=0.5*\boxwidth, minimum height=0.75*\boxheight, align=center, draw=QC5, fill=QC5!20, thick, anchor=center]
\tikzstyle{dp cba}=[minimum width=0.5*\boxwidth, minimum height=0.75*\boxheight, align=center, draw=QC5, fill=QC5!20, thick, anchor=center]
\tikzstyle{dp sum}=[fill=black, draw=black, shape=circle, inner sep=0pt, minimum size=0.1em, align=center, thick, anchor=center]

\tikzset{
    pics/cba/.style n args={3}{%
        code={%
            \node[inner sep=0pt] (#1) at (0em,1.5em) {};
            \node[minimum width=3.5em, minimum height=2em, align=center, draw=QC5, fill=QC5!20, thick, anchor=center] (CBA) at (0em,0em) {#3};
            \node[inner sep=0pt] (#2) at (0em,-1.5em) {};

            \draw[-] (#1.center) to (CBA.north);
            \draw[-] (CBA.south) to (#2.center);
        }
    },
    pics/cba/.default={In}{Out}{CBA}
}

\tikzset{
  pics/cbaskip/.style n args={3}{
        code={
            \pic at (-1.75em, 0) {cba={In/CBA}{Out/CBA}{CBA}};
            \node[dp sum, inner sep=0pt] (#1) at (0em, 1.75em) {};
            \node[inner sep=0pt] (P1) at (0em, 1.5em) {};
            \node[inner sep=0pt] (R1) at (1.75em, 1.5em) {};
            \node[inner sep=0pt] (L1) at (-1.75em, 1.5em) {};
            \node[inner sep=0pt] (R2) at (1.75em, -1.5em) {};
            \node[inner sep=0pt] (L2) at (-1.75em,-1.5em) {};
            \node[inner sep=0pt] (P2) at (0em, -1.5em) {};
            \node[dp sum, inner sep=0pt] (#2) at (0em, -1.75em) {};

            \draw[-] (#1.center) -- (P1.center);
            \draw[-] (P1.center) -- (R1.center);
            \draw[-] (P1.center) -- (L1.center);
            \draw[-] (P2.center) -- (R2.center);
            \draw[-] (P2.center) -- (L2.center);
            \draw[-] (P2.center) -- (#2.center);
            
            \ifthenelse{#3=1}{
                \draw[-] (R1.center) to (R2.center);
            }{
                \node[minimum width=1.5em, minimum height=2em, align=center, draw=QC5, fill=QC5!20, thick, anchor=center](C) at (1.75em,0em) {C};
                \draw[-] (R1.center) to (C.north);
                \draw[-] (C.south) to (R2.center);
            }
        }
    },
    pics/cbaskip/.default={In}{Out}{1}
}

\tikzset{
  pics/spp/.style n args={2}{
        code={
            \node[dp sum, inner sep=0pt] (#1) at (0em,1.75em) {};
            \node[inner sep=0pt] (P1) at (0em,1.5em) {};
            \node[minimum width=1.5em, minimum height=2em, align=center, draw=QC5, fill=QC5!20, thick, anchor=center] (SPP/1) at (-2em,0) {C};
            \node[minimum width=1.5em, minimum height=2em, align=center, draw=QC5, fill=QC5!20, thick, anchor=center] (SPP/2) at (0,0) {C};
            \node[minimum width=1.5em, minimum height=2em, align=center, draw=QC5, fill=QC5!20, thick, anchor=center] (SPP/3) at (+2em,0) {C};
            \node[inner sep=0pt] (P2) at (0em,-1.5em) {};
            \node[dp sum, inner sep=0pt] (#2) at (0em,-1.75em) {};        
    
            \draw[-] (#1.center) -- (P1.center);
            \foreach \x in {1,...,3}{
                \draw[-] (P1.center) -| (SPP/\x.north);
                \draw[-] (SPP/\x.south) |- (P2.center);
            }
            \draw[-] (P2.center) -- (#2.center);
        }
    },
    pics/spp/.default={In}{Out}
}

\tikzset{
  pics/blockset/.style n args={4}{
        code={
            \foreach \x [count=\xi from 1] in {#3}{
                \node [style=#4, anchor=north, minimum height=2em] (Pre/\xi) at (0,0)[shift={(0em,-(\xi-1)*\boxheight)}] {\small{\x}};
                \xdef\rememberxi{\xi}  
            }
            
            \edef\xi{\rememberxi}  
            \pgfmathsetmacro{\xi}{int(\xi-1)}

            \node[dp sum, inner sep=0pt] (#1) at (0em,0.5em) {};
            \node[dp sum, inner sep=0pt] (#2) at (0em,0em)[xshift=0em,yshift=-(\xi+1)*\boxheight+0.5em] {};
            \draw[-] (#1) to (Pre/1.north);
            \pgfmathsetmacro{\xo}{int(\xi+1)}
            \draw[-] (Pre/\xo.south) to (#2);
            \foreach \x in {1,...,\xi}{
                \pgfmathsetmacro{\y}{int(\x+1)}
                \draw[-] (Pre/\x.south) to (Pre/\y.north);
            }
        }
    },
    pics/spp/.default={In}{Out}{{A, B, C, D}}{dp lower}
}

\tikzset{
    pics/vdots/.style n args={1}{
        code={
            \node[] at (-0.0em, 0em) {$\boldsymbol{\vdots}$};
            \node[] at (+1em, -0.3em) {#1};
        }
    },
    pics/vdots/.default={N}
}

\begin{tikzpicture}
	\begin{pgfonlayer}{nodelayer}
        \node [style=dp upper, minimum width=0em, align=center, draw=none, fill=none, thick] (D) at (0*\blockdist,0)[xshift=1em,yshift=0em] {Observation $\bm Y$};
        \node [style=dp upper, draw=QC1, fill=QC1!20, thick] (Pre) at (1*\blockdist,0)[xshift=0em,yshift=0em] {Preprocessing};
        \node [style=dp upper, draw=QC5, fill=QC5!20, thick] (Enc) at (2*\blockdist,0)[xshift=0em,yshift=0em] {Encoder};
        \node [style=dp upper, draw=QC5, fill=QC5!20, thick] (Eta) at (3*\blockdist,0)[xshift=0em,yshift=0em] {Refinement};
        \node [style=dp upper, draw=QC1, fill=QC1!20, thick] (Pos) at (4*\blockdist,0)[xshift=0em,yshift=0em] {Postprocessing};
        \node [style=dp upper, draw=none, fill=none, thick] (Est) at (5*\blockdist,0)[xshift=-3em,yshift=0em] {%
            $\begin{bmatrix}
                \bm\hat{\tau} \\
                \bm\hat{\alpha} \\
                \bm\hat{\gamma}
            \end{bmatrix}$%
        };

        \pic at (1*\blockdist, -2.75em) {blockset={{In1}{Out1}{Windowing, \num{2}D-DFT, Normalization}{dp lower}}};
        \draw[->] (Out1.center) to ([shift={(0em,-0.5em)}]Out1);

        \pic at (2*\blockdist, -3em)[shift={(0em, -1.0em)}] {cbaskip={In21}{Out21}{2}};
        \pic at (2*\blockdist, -5.5em)[shift={(0em, -1.0em)}] {vdots={3x}};
        \pic at (2*\blockdist, -8.5em)[shift={(0em, -1.0em)}] {cbaskip={In23}{Out23}{2}};
        \draw[->] (Out23.center) to ([shift={(0em,-0.5em)}]Out23);

        \pic at (3*\blockdist, -4.5em)[shift={(0em, 0.5em)}] {spp={In31}{Out31}};
        \pic at (3*\blockdist, -7.5em)[shift={(0em, 0.5em)}] {cba};
        \pic at (3*\blockdist, -10em)[shift={(0em, 0.5em)}] {cba={In33}{Out33}{CBA}};
        \node[dp sum, inner sep=0pt] at (Out33.center) {};
        \draw[->] (Out33.center) to ([shift={(0em,-0.5em)}]Out33);

        \pic at (4*\blockdist, -2.75em) {blockset={{In4}{Out4}{Demapping, Reconstruct $\bm{\hat{\gamma}}$, Optimization*}{dp lower}}};
        \draw[->] (Out4.center) to ([shift={(0em,-0.5em)}]Out4);


	\end{pgfonlayer}
	\begin{pgfonlayer}{edgelayer}
        \draw [->] (D.east) to (Pre.west);
        \draw [->] (Pre.east) to (Enc.west);
        \draw [->] (Enc.east) to (Eta.west);
        \draw [->] (Eta.east) to (Pos.west);
        \draw [->] (Pos.east) to ([xshift=2em]Est.west);
	\end{pgfonlayer}
\end{tikzpicture}

%% file: tex/signalmodel.tex
Our task is to characterize a \gls{siso} channel by estimation of its paths' propagation parameters. 
Therefore, we first define the channel and signal models used to describe the measured observation recorded by the \gls{rx}.

We assume an underspread channel that can be modeled as a \gls{ltv} system composed of $1..P \in \N$ specular paths, each represented by a planar wave impinging on the \gls{rx}.
Each path is parametrized by its magnitude and phase-shift (jointly expressed as $\gamma_p \in \C$), delay $\tau_p \in \R$  and Doppler-shift $\alpha_p \in \R$.
At the \gls{rx} only an electric field proportional to the superposition of all the planar waves is observed, due to the channel linearity.
Our task is to identify each of the $P$ specular propagation paths characterizing the frequency and time impulse response of the \gls{ltv} channel.

At the \gls{rx}, we model the sampled wireless channel transfer-function of bandwidth $B$ with $N_f \in \mathbb{N}$ frequency samples and $N_t \in \mathbb{N}$ snapshots under the narrowband assumption ($B \ll f_c$).
The sampling process is characterized by the sampling intervals in frequency $\Delta f > 0$ and time $\Delta t > 0$ with $\bm S$ sampled $N_f, N_t \in \N$ times at
\begin{equation}
    f_k = f_0 + k \cdot \Delta f,
    t_l = t_0 + l \cdot \Delta t
\end{equation}
where $k = 0,...,N_f-1$, $l = 0,...,N_t-1$, $f_0 = - B/2$, and $t_0 = 0$.
We denote the sampled observation in complex baseband by $\bm S$.
Therefore, the discrete signal model $\bm S \in \C^{N_f \times N_t}$ is formulated as
\begin{equation}\label{eq:signal_discrete}
    S_{k,l}(\bm \gamma, \bm \tau, \bm \alpha) = \sum_{p=1}^{P} \gamma_p \exp{(-2j\pi f_k \tau_p)} \exp{(2j\pi t_l \alpha_p)},
\end{equation}
where the index $p = 1,...,P$ denotes the path index and $\bm\gamma \in \C^{P}$, $\bm\tau \in \R^{P}$, $\bm\alpha \in \R^{P}$ contain the corresponding complex weights, delays, and Doppler-shifts, respectively.
The noisy observation $\bm Y \in \C^{N_f \times N_t}$ is then denoted as
\begin{equation}\label{eq:observation_discrete}
    \bm Y = \bm S(\bm\gamma, \bm\tau, \bm\alpha) + \bm N ,
\end{equation}
where $\bm N \in \C^{N_f \times N_t}$ is a complex, zero-mean, and element-wise uncorrelated Gaussian noise process, where each entry has variance $\sigma^2$.

Our goal is to jointly estimate $P$, $\bm \tau$ and $\bm \alpha$ from the noisy observation $\bm Y$ defined by \eqref{eq:observation_discrete} using a \gls{cnn}, introduced in the next section.
The linear parameters (the complex $\hat{\bm\gamma}$) are reconstructed using the \gls{blue}, i.e., \gls{ls}.

%% file: tex/neuralnetwork.tex
This section introduces the preprocessing scheme applied to the data $\bm Y$, the off-grid parameter encoding used for the labels, and the \gls{cnn} architecture.
\subsection{Preprocessing}\label{sec:neuralnetwork:preprocessing}
Our preprocessing stage aims to provide the neural network with informative and diverse input values via two steps, multi-windowing and subsequent \num{2}D-\gls{dft}.\par
In the first step, multiple views of the data $\bm Y$ are created by filtering with $N_W$ windows and stacking the results into $\bm Y_W \in \C^{N_W \times N_f \times N_t}$.
The motivation for the multi-window approach is to obtain different information from the same observation.
Window functions are typically an application-specific choice, as different windows create different benefits and drawbacks.
For example, rectangular windows achieve the maximum \gls{snr} and provide a narrow pulse shape but also result in high sidelobes after the \gls{dft}, potentially introducing ghost paths.
Other filters, such as Hann windows, reduce the sidelobes, i.e., the probability of ghost paths, but increase the mainlobe width and usually result in higher estimation variances.
Hence, it is recommended to select a Hann window for situations where paths are sufficiently separated from neighboring paths (that would otherwise be masked by the wider mainlobe) and have sufficient \gls{snr}.
Vice versa, a rectangular window is suitable for situations in which a narrow mainlobe is required to separate close paths or the maximum \gls{snr} is required, while strong sidelobes are not a concern.
These two cases demonstrate the choice of the optimal window is always dependent on the situation at hand, and is only obvious if the result is already known.
To alleviate the dependency on one specific window, we apply multiple windows in parallel by processing them as different channels in the \gls{cnn}, similar to color channels of images.
Consequently, the \gls{cnn} can jointly utilize the different views.
We used $N_W=4$ different windowing functions, i.e., a Tukey, Cosine, Hann, and the Rectangular window.

The second preprocessing step is a \num{2}D-\gls{dft} over the two data dimensions, transforming it to the target parameter domain in delay- and Doppler, the Fourier domains corresponding to frequency- and time, respectively.
If desired, the \gls{dft} output can be cropped to those parts of the scattering function which are of interest instead of the full unambiguous range in delay- and/or Doppler-shift.
This enables the incorporation of a priori knowledge from the observed propagation scenario, e.g., as shown later in \Cref{sec:neuralnetwork:training}.
We denote the output of this step as $\bm Y_1 \in \C^{N_W \times H \times W}$, where $H, W \in \mathbb{N}$ underlines the similarity with the height and width of an RGB image.
To obtain real-valued numbers required for the training, we employ two mappings
\begin{equation}
    f_1(\bm Y_1) = \log_{10}(\vert \bm Y_1 \vert),
    f_2(\bm Y_1) = \angle(\bm Y_1)
\end{equation}
to map the complex values in $\bm Y_1$ to real-valued \gls{cnn} input data $\bm Y_2 \in \R^{2 \cdot N_W \times H \times W}$, where $\vert \cdot \vert$ and $\angle$ denote the absolute value and phase of a complex number, respectively.
\subsection{Off-Grid Parameter Encoding}\label{sec:neuralnetwork:offgrid}
The labels for the supervised learning task are provided with a grid-relative parameter encoding, which enables the off-grid estimates.
To this end, we divide the delay-Doppler domain of interest into a few grid-cells.
For each of these, we aim to estimate the number of paths in that cell and their respective displacement from the center.
Our encoding is inspired by~\cite{redmon_you_2016}, and adapted to provide a learnable encoding of the parameters.
It consists of three steps: parameter normalization, cell assignment, and relative parameter encoding.

The normalization maps the parameters into a range between 0 and 1, such that $\bm\tau$ and $\bm\alpha$ are in the interval of $\tau_p \subset \left[0, 1\right)$ and $\alpha_p \subset \left[0, 1\right)$.
Note that this step must adhere to the same range defined by $Y_1$, as it defines the possible parameter range for the estimates.
Next, we define a set of $I \cdot J$ cell centers $\bm x \in (0,1)^{I \times J}$, which each define a non-overlapping rectangular covering of $[0,1) \times [0,1)$.
Paths are mapped to the cells based on the shortest $\ell_\infty$-distance to the cell centroids $x_{i,j}$.
Let $C$ be the maximum number of paths in a single cell; then, to encode the paths' positions and model order in each cell, we define vectors $\bm \eta_{i,j} \in \R^{3 \cdot L_{\rm max}}$ as
\begin{equation}
    \bm\eta_{i,j} = \left[
    \mu_{1}^{[i,j]}, \Delta\tau_{1}^{[i,j]}, \Delta\alpha_{1}^{[i,j]}, \hdots, \mu_{C}^{[i,j]}, \Delta\tau_{C}^{[i,j]}, \Delta\alpha_{C}^{[i,j]}
    \right]^T \label{eq:encoding}
\end{equation}
$\Delta\tau_c^{[i,j]} = \tau_{c} - x_{i,j}$ and $\Delta\alpha_c = \alpha_{c} - x_{i,j}$ denote the distance between the cell centroid $\eta_{i,j}$ (in the respective dimension) and the normalized parameter.
Then, $\Delta\tau_c$ and $\Delta\alpha_c$ are normalized by the cell width and shifted based on each of the cells' centers, such that $\Delta\tau_c, \Delta\alpha_c \subset \left[0, 1\right)$.
Note the model order $P$ can be expressed as $P = \sum_{i,j,c}^{I,J,C} \mu_{c}^{[i,j]}$ and the maximum number of representable paths is $P_{\text{max}} = C\cdot I \cdot J$.

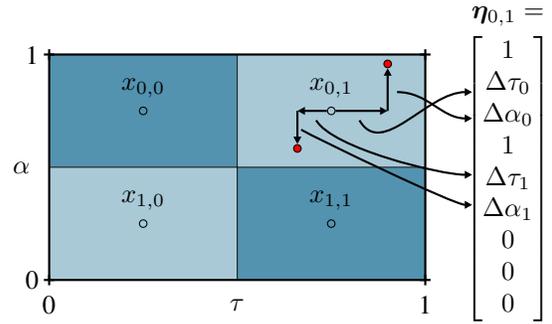
\begin{figure}[t]
    \centering
    \input{figures/label.tikz}
    \caption{Example for the label encoding with $C=3$.The path parameters (\protect\reddot) are encoded relative the closest cell-centroid  $\bm\eta_{i,j}$.}
    \label{fig:labels}
\end{figure}

As the number of paths in each cell can vary, $\mu_{c}^{[i,j]} = \{0, 1\}$ indicates if the path with encoding $\Delta\tau_{c}^{[i,j]}$ and $\Delta\alpha_{c}^{[i,j]}$ is an estimate ($\mu_{c}^{[i,j]} = 1$) or empty ($\mu_{c}^{[i,j]} = 0$).
This enables computing a masked loss during training for different model orders, as detailed in \Cref{sec:neuralnetwork:lossfunctions}.
To enforce a learnable ordering of paths in each cell, the paths are sorted from $c=1...C$ with descending magnitude of the complex path weights $\vert\gamma_p\vert$.
If a cell contains fewer than $C$ paths, the corresponding unassigned parameters are denoted as \num{0} (see \Cref{fig:labels}).
Hence, $C$ denotes the maximum number of paths which can be contained in a single cell.
The result of the encoding is a 3D array $\bm\eta \in \R^{I \times J \times 3 \cdot C}$, which encodes the desired prediction results of our \gls{cnn}.
\subsection{Network Architecture}\label{sec:neuralnetwork:architecture}
Our network architecture is split into two trainable stages, we loosely term them \textit{encoder} and \textit{refinement}, illustrated in \Cref{fig:architecture}.
We embed the convolutional layers into blocks, which each consist of a \num{2}D convolutional layer, followed by batch normalization and a \gls{relu} activation function, denoted as CBA-block in the following.
The \textit{encoder} reduces the dimensionality of the data through downsampling.
For the downsampling, the \num{2}D convolutional layers are parameterized with a stride of 2 and a kernel size of \num{3}\footnote{Kernel sizes refer to all dimensions simultaneously.}.
Hence, the CBA blocks halve the data shape but double the number of channels in each block.
At its output, this stage yields an array whose dimensions already match $\bm\hat{\eta}$ in the first two dimensions $I$ and $J$.

The \textit{refinement}-stage utilizes the results of the \textit{encoder} to produce the grid-relative estimate $\bm\hat{\eta}$.
It utilizes \gls{spp} with four different kernel sizes (\num{3}, \num{5}, \num{7}, and \num{9}) to aggregate the results from the \textit{encoder}.
The result is processed further by a series of CBA blocks to reduce the number of channels until it matches the third dimension of $\bm\hat{\eta}$.
The output of the last CBA block in the \textit{refinement}-stage is of size $I \times J \times 3 \cdot C$ and corresponds to the estimates of the encoded parameters.

To obtain the estimate of the model order, a threshold $\delta$ is applied to all values $\bm\mu^{[i,j]}$ in $\bm\eta$, such that their corresponding parameter estimates $\left[\Delta\tau^{[i,j]} \Delta\alpha^{[i,j]}\right]$ are only evaluated if $\hat{\mu}_c^{[i,j]} \geq \delta$, and disregarded otherwise.
Finding a suitable threshold can be achieved by statistical analysis of the values in $\bm\mu^{[i,j]}$ in the specific application.
Using a threshold reduces the number of trainable parameters and enables flexible tuning of the number of estimated paths in the inference phase without retraining, compared to our previous work presented in~\cite{schieler2024_eucap}.
This makes it particularly suitable when the approach is applied on data, whose distribution does not exactly match those of the training dataset.
For example, when the approach is applied to measurement data, which contain additional disturbances from the measurement equipment~\cite{Semper2023}, or the underlying assumptions of \Cref{eq:observation_discrete} are violated.
\begin{figure*}[t]
    \input{figures/inference/inference.tex}
\end{figure*}
\subsection{Loss Functions}\label{sec:neuralnetwork:lossfunctions}
Our approach combines two different loss functions, denoted $L_{1}$ and $L_{2}$ in a sum.
The first summand $L_{1}$ is the loss for the parameter estimates $\bm\hat{\eta}$ and utilizes a masked \gls{mse} loss function
\begin{align}
    L_{1} = \sum_{i,j=1}^{I,J}\sum_{c=1}^{C}
    \mu_c^{[i,j]}
    \cdot
    \Norm{
        \begin{bmatrix}
            \Delta\hat{\tau}_{c}^{[i,j]} \\
            \Delta\hat{\alpha}_{c}^{[i,j]}
        \end{bmatrix}
        -
        \begin{bmatrix}
            \Delta\tau_{c}^{[i,j]} \\
            \Delta\alpha_{c}^{[i,j]}
        \end{bmatrix}
    }_2^2
    \label{eq:parameterloss},
\end{align}
where $\Hat{\cdot}$ marks predictions.
The second summand $L_{2}$ refers to the estimates for $\bm\hat{\mu}$, which are trained using a \gls{bce} loss defined by
\begin{align}
    L_{2} = \hat{\mu} \cdot \log(\mu) + (1-\hat{\mu}) \cdot \log(1-\mu) \label{eq:bce}.
\end{align}%
The sum of both losses is the metric for the optimization the trainable \gls{cnn} parameters.
\begin{table}[b!]
    \centering
    \small
    \caption{Dataset summary and training hyperparameters.}
    \label{tab:settings}
    \begin{tabularx}{\linewidth}{Xp{4.5cm}}
        \toprule
        Name                                 & Value                                                                                  \\ \midrule
        \textbf{Datasets} & \mbox{}         \\
        Distribution $\tau_p$, $\alpha_p$    & $\mathfrak{U}_{[0,0.025]}$ and $\mathfrak{U}_{[-0.05,0.05]}$                           \\
        Number of Samples                    & $N_f=\num{1024}$, $N_t=\num{100}$                                                      \\
        Input Data $\bm Y_2$                 & $8 \times 256 \times 256$ ($2N_w \times H \times W$)                                   \\
        Magnitudes                           & $\mathfrak{U}_{[0.001, 1]}$                                                            \\
        Phases                               & $\mathfrak{U}_{[0,2\pi]}$                                                              \\
        SNR                                  & \qtyrange{0}{50}{\decibel}                                                             \\
        Number of Paths                      & $\mathfrak{U}_{[1,10]}$                                                                \\
        Trainingset Size                     & \num[exponent-mode=engineering, drop-zero-decimal=true]{500000}                        \\
        Validationset Size                   & \num{1000}                                                                             \\
        Testset Size                         & \num{10000}                                                                            \\ \midrule
        \textbf{Training}                    &                                                                                        \\
        Optimizer                            & Adam~\cite{kingma_adam_2014}, $\gamma=0.0003$, \newline $\beta_1=0.9$, $\beta_2=0.999$ \\
        Mini-Batchsize                       & \num{512}                                                                              \\
        Epochs                               & 100                                                                                    \\
        Trainable Parameters                 & \num[exponent-mode=engineering, drop-zero-decimal=true]{1.3e6} \\\bottomrule
    \end{tabularx}
\end{table}

\subsection{Training Parametrization}\label{sec:neuralnetwork:training}
The synthetic dataset is split into three subsets, a training, validation, and test set.
Each sample in a dataset contains a random number of $P=\numrange{1}{10}$ specular paths.
For each path, the complex path amplitudes $\gamma$ are randomly sampled from a uniform distribution; magnitudes from a range of \qtyrange{0}{-30}{\decibel} and phases from $\numrange[parse-numbers=false]{0}{2\pi}$.
Similarly, the corresponding delay-Doppler parameters are also sampled randomly from uniform distributions.
The additional noise $\bm N$ is generated randomly with different random noise variance $\sigma$ for every snapshot, such that the \gls{snr} is in the range of \qtyrange{-30}{50}{\decibel}.
This way the noise for each snapshot is different in every training epoch, which effectively countered overfitting in our experiments.
\Cref{tab:settings} contains a summary of all the settings for the dataset and other training hyperparameters.\par
Compared to~\cite{schieler2024_eucap}, we limited the range of the non-linear path parameters to reflect a more dense multipath scenario, where $\tau_p \in [0, 0.025)$ and $\alpha_p \in [-0.05, 0.05)$.
This is reflected in the preprocessing, such that the \gls{dft} output (and by that, the \gls{cnn} input) covers this range with $\bm Y_2 \in 8 \times 256 \times 256$ evenly-spaced supports.
Note the original sampling process and resolution defined by $N_f$ and $N_t$ remain unaffected.
Those are chosen in correspondence with the measurement data analysis later in \Cref{sec:analysis:meas}.

\subsection{Postprocessing}\label{sec:neuralnetwork:postprocessing}
The postprocessing recovers the estimated parameters from their encoded form and reconstructs the corresponding complex path gains $\bm{\hat{\gamma}}$.
This is done by first sorting the $\hat{\bm\mu}$ and keeping only those exceeding the threshold $\delta$.
Applying the inverse of the grid encoding \Cref{eq:encoding} returns the non-linear parameters $\hat{\bm\tau}$ and $\hat{\bm\alpha}$.
Illustrating the results at this stage, \Cref{fig:example} shows a single, hand-picked sample from the validation set passed through the network at different \glspl{snr}.
To reconstruct the linear parameters (the complex $\hat{\bm\gamma}$, see \eqref{eq:observation_discrete}), we use the \gls{blue}, i.e., \gls{ls} to obtain $\left[\bm\hat{\gamma}, \bm\hat{\tau}, \bm{\hat{\alpha}}\right]$.
The separate reconstruction of the linear model parameters reduces the number of trainable parameters, avoids coupling with the delay-Doppler parameters during training, and exploits the low computational complexity of \gls{ls}.\par

An additional, though optional step of the postprocessing is the use of further model-based, global optimization using a second-order Gauss-Newton scheme, similar to~\cite{barthelme_machine_2021}.
In it, we utilize the likelihood function of our signal model along with its corresponding Jacobian- and Fisher-Information matrices.
The raw \gls{cnn} estimates initialize a gradient iteration
\begin{equation}\label{eq:gradient_steps}
    (\bm \gamma^{k+1}, \bm \tau^{k+1}, \bm \alpha^{k+1}) =
    (\bm \gamma^{k}, \bm \tau^{k}, \bm \alpha^{k}) - \varepsilon^k \bm z^k
\end{equation}
with descent direction
\[
    \bm z^k = \left[\left(\bm F^k\right)^{-1} \cdot \bm J^k\right](\bm \gamma^{k}, \bm \tau^{k}, \bm \alpha^{k}),
\]
which defines the second-order Gauss-Newton scheme, with the Fisher-Information matrix $\bm F$ and the Jacobian matrix $\bm J$ of the negative $\log$-likelihood function $\lambda$.
Based on the assumption that $\bm Y$ is a Gaussian random variable, it reads as
\begin{equation}\label{eq:llf}
    \lambda(\bm \gamma, \bm \tau, \bm \alpha) = \frac{1}{\sigma^2}\Norm{\bm Y - \bm S(\bm\gamma, \bm\tau, \bm\alpha)}_F^2,
\end{equation}
where $\Norm{\cdot}_F$ denotes the Frobenius norm.

%% file: figures/label.tikz

\tikzstyle{grid}=[fill=none, draw=black, shape=rectangle, minimum width=5cm, minimum height=3cm, thick]
\tikzstyle{xtick}=[fill=white, draw=black, shape=rectangle, minimum height=4pt, minimum width=0pt, inner sep=0pt, thick]
\tikzstyle{ytick}=[fill=white, draw=black, shape=rectangle, minimum height=0pt, minimum width=4pt, inner sep=0pt, thick]
\tikzstyle{cell}=[fill=none, draw=black, shape=rectangle, minimum width=2.5cm, minimum height=1.5cm]
\tikzstyle{param}=[fill=red, draw=black, shape=circle, minimum width=0.1cm, minimum height=0.1cm, inner sep=0pt]
\tikzstyle{center}=[fill=none, draw=black, shape=circle, minimum width=0.1cm, minimum height=0.1cm, inner sep=0pt]
\tikzstyle{arrow}=[->, >={Triangle[length=1mm, width=1mm]}, thick]

\begin{tikzpicture}
	\begin{pgfonlayer}{nodelayer}
		\node [style=cell, fill=coolor2, opacity=0.8] (11) at (-1.25, 0.75) {};
		\node [style=cell, fill=coolor2, opacity=0.4] (12) at (1.25, 0.75) {};
		\node [style=cell, fill=coolor2, opacity=0.8] (13) at (1.25, -0.75) {};
		\node [style=cell, fill=coolor2, opacity=0.4] (14) at (-1.25, -0.75) {};	
		\node [style=grid] (0) at (0, 0) {};
		\node [style=xtick, label={left:1}] (2) at (-2.5, 1.5) {};
		\node [style=xtick, label={below:1}] (3) at (2.5, -1.5) {};
		\node [style=xtick, label={below:0}, label={left:0}] (4) at (-2.5, -1.5) {};
		\node [style=xtick] (5) at (2.5, 1.5) {};
		\node [style=ytick] (7) at (2.5, 1.5) {};
		\node [style=ytick] (8) at (-2.5, 1.5) {};
		\node [style=ytick] (9) at (-2.5, -1.5) {};
		\node [style=ytick] (10) at (2.5, -1.5) {};
		\node [style=param] (15) at (0.8, 0.25) {};
		\node [style=none, below] (16) at (0, -1.75) {$\tau$};
		\node [style=none, left] (17) at (-2.75, 0) {$\alpha$};
		\node [style=param] (18) at (2, 1.375) {};
		\node [style=center, label={above:$x_{0,1}$}] (20) at (1.25, 0.75) {};
		\node [style=center, label={above:$x_{0,0}$}] (21) at (-1.25, 0.75) {};
		\node [style=center, label={above:$x_{1,0}$}] (22) at (-1.25, -0.75) {};
		\node [style=center, label={above:$x_{1,1}$}] (23) at (1.25, -0.75) {};
		\node [style=none] (24) at (2, 0.75) {};
		\node [style=none] (26) at (0.8, 0.75) {};
		\node [style=none] (30) at (1.25, 1.5) {};
		\node [style=none] (31) at (0.75, 1) {};
		\node [style=none, label={[]$\bm \eta_{0,1}=$}] (32) at (3.6, -0.15) {$\begin{bmatrix} 1 \\ \Delta\tau_0 \\ \Delta\alpha_0 \\ 1 \\ \Delta\tau_1 \\ \Delta\alpha_1 \\ 0 \\ 0 \\ 0\end{bmatrix}$};
		\node [style=none] (28) at (2.125, 1) {}; %
		\node [style=none] (29) at (1.625, 0.625) {};
		\node [style=none] (34) at (3.125, 0.65) {};
		\node [style=none] (36) at (3.125, 1) {};
		\node [style=none] (37) at (4.75, 0.75) {}; 
		\node [style=none] (38) at (0.85, 0.5) {};
		\node [style=none] (39) at (3.125, -0.5) {};
		\node [style=none] (40) at (1.05, 0.625) {};
		\node [style=none] (41) at (3.125, -0.1) {};
		
		\draw [style=arrow] (20) to (24.center);
		\draw [style=arrow] (24.center) to (18);
		\draw [style=arrow] (20) to (26.center);
		\draw [style=arrow] (26.center) to (15);
		\draw [style=arrow, in=180, out=0] (28.center) to (34.center);
		\draw [style=arrow, in=-180, out=-60] (29.center) to (36.center);
		\draw [style=arrow, in=180, out=-30, looseness=0.25] (38.center) to (39.center);
		\draw [style=arrow, in=180, out=-60, looseness=0.50] (40.center) to (41.center);		
	\end{pgfonlayer}
\end{tikzpicture}

%% file: figures/inference/inference.tex
\def\scalefig{0.45}
\centering
\subfloat[t][\qty{-40}{\decibel}, \label{fig:example:a}]{%
    \scalebox{\scalefig}{\input{figures/inference/snr=-40.pgf}}
    }%
\subfloat[t][\qty{-20}{\decibel}, \label{fig:example:b}]{%
    \scalebox{\scalefig}{\input{figures/inference/snr=-20.pgf}}
    }%
\subfloat[t][\qty{0}{\decibel}, \label{fig:example:c}]{%
    \scalebox{\scalefig}{\input{figures/inference/snr=0.pgf}}
        }%
\subfloat[t][\qty{20}{\decibel}, \label{fig:example:d}]{%
    \scalebox{\scalefig}{\input{figures/inference/snr=20.pgf}}
        }%
\caption{Inference example at different \glspl{snr}. A validation snapshot with $P=8$ paths at different SNRs (a-d), showing the groundtruth (\protect\whitecircle) and estimates (\protect\blackcircle) alognside the data (rectangular window). The displacement of circles indicates the accuracy of the point estimates (center) and visibly improves with increasing SNR.}
\label{fig:example}

%% file: figures/inference/snr=-40.pgf
\begingroup%
\makeatletter%
\begin{pgfpicture}%
\pgfpathrectangle{\pgfpointorigin}{\pgfqpoint{3.649183in}{3.470650in}}%
\pgfusepath{use as bounding box, clip}%
\begin{pgfscope}%
\pgfsetbuttcap%
\pgfsetmiterjoin%
\definecolor{currentfill}{rgb}{1.000000,1.000000,1.000000}%
\pgfsetfillcolor{currentfill}%
\pgfsetlinewidth{0.000000pt}%
\definecolor{currentstroke}{rgb}{1.000000,1.000000,1.000000}%
\pgfsetstrokecolor{currentstroke}%
\pgfsetdash{}{0pt}%
\pgfpathmoveto{\pgfqpoint{0.000000in}{0.000000in}}%
\pgfpathlineto{\pgfqpoint{3.649183in}{0.000000in}}%
\pgfpathlineto{\pgfqpoint{3.649183in}{3.470650in}}%
\pgfpathlineto{\pgfqpoint{0.000000in}{3.470650in}}%
\pgfpathlineto{\pgfqpoint{0.000000in}{0.000000in}}%
\pgfpathclose%
\pgfusepath{fill}%
\end{pgfscope}%
\begin{pgfscope}%
\pgfsetbuttcap%
\pgfsetmiterjoin%
\definecolor{currentfill}{rgb}{1.000000,1.000000,1.000000}%
\pgfsetfillcolor{currentfill}%
\pgfsetlinewidth{0.000000pt}%
\definecolor{currentstroke}{rgb}{0.000000,0.000000,0.000000}%
\pgfsetstrokecolor{currentstroke}%
\pgfsetstrokeopacity{0.000000}%
\pgfsetdash{}{0pt}%
\pgfpathmoveto{\pgfqpoint{0.782823in}{0.175231in}}%
\pgfpathlineto{\pgfqpoint{3.549183in}{0.175231in}}%
\pgfpathlineto{\pgfqpoint{3.549183in}{3.030372in}}%
\pgfpathlineto{\pgfqpoint{0.782823in}{3.030372in}}%
\pgfpathlineto{\pgfqpoint{0.782823in}{0.175231in}}%
\pgfpathclose%
\pgfusepath{fill}%
\end{pgfscope}%
\begin{pgfscope}%
\pgfpathrectangle{\pgfqpoint{0.782823in}{0.175231in}}{\pgfqpoint{2.766359in}{2.855141in}}%
\pgfusepath{clip}%
\pgfsys@transformshift{0.782823in}{0.175231in}%
\pgftext[left,bottom]{\includegraphics[interpolate=true,width=2.770000in,height=2.860000in]{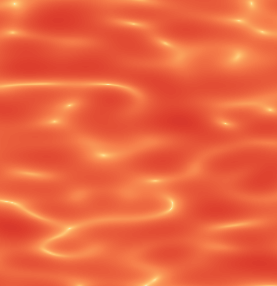}}%
\end{pgfscope}%
\begin{pgfscope}%
\pgfpathrectangle{\pgfqpoint{0.782823in}{0.175231in}}{\pgfqpoint{2.766359in}{2.855141in}}%
\pgfusepath{clip}%
\pgfsetbuttcap%
\pgfsetroundjoin%
\pgfsetlinewidth{2.007500pt}%
\definecolor{currentstroke}{rgb}{1.000000,1.000000,1.000000}%
\pgfsetstrokecolor{currentstroke}%
\pgfsetdash{}{0pt}%
\pgfpathmoveto{\pgfqpoint{2.093753in}{2.630829in}}%
\pgfpathcurveto{\pgfqpoint{2.113927in}{2.630829in}}{\pgfqpoint{2.133279in}{2.638845in}}{\pgfqpoint{2.147544in}{2.653110in}}%
\pgfpathcurveto{\pgfqpoint{2.161810in}{2.667376in}}{\pgfqpoint{2.169825in}{2.686727in}}{\pgfqpoint{2.169825in}{2.706902in}}%
\pgfpathcurveto{\pgfqpoint{2.169825in}{2.727076in}}{\pgfqpoint{2.161810in}{2.746428in}}{\pgfqpoint{2.147544in}{2.760693in}}%
\pgfpathcurveto{\pgfqpoint{2.133279in}{2.774959in}}{\pgfqpoint{2.113927in}{2.782974in}}{\pgfqpoint{2.093753in}{2.782974in}}%
\pgfpathcurveto{\pgfqpoint{2.073578in}{2.782974in}}{\pgfqpoint{2.054227in}{2.774959in}}{\pgfqpoint{2.039961in}{2.760693in}}%
\pgfpathcurveto{\pgfqpoint{2.025696in}{2.746428in}}{\pgfqpoint{2.017680in}{2.727076in}}{\pgfqpoint{2.017680in}{2.706902in}}%
\pgfpathcurveto{\pgfqpoint{2.017680in}{2.686727in}}{\pgfqpoint{2.025696in}{2.667376in}}{\pgfqpoint{2.039961in}{2.653110in}}%
\pgfpathcurveto{\pgfqpoint{2.054227in}{2.638845in}}{\pgfqpoint{2.073578in}{2.630829in}}{\pgfqpoint{2.093753in}{2.630829in}}%
\pgfpathlineto{\pgfqpoint{2.093753in}{2.630829in}}%
\pgfpathclose%
\pgfusepath{stroke}%
\end{pgfscope}%
\begin{pgfscope}%
\pgfpathrectangle{\pgfqpoint{0.782823in}{0.175231in}}{\pgfqpoint{2.766359in}{2.855141in}}%
\pgfusepath{clip}%
\pgfsetbuttcap%
\pgfsetroundjoin%
\pgfsetlinewidth{2.007500pt}%
\definecolor{currentstroke}{rgb}{1.000000,1.000000,1.000000}%
\pgfsetstrokecolor{currentstroke}%
\pgfsetdash{}{0pt}%
\pgfpathmoveto{\pgfqpoint{2.251304in}{0.752429in}}%
\pgfpathcurveto{\pgfqpoint{2.271479in}{0.752429in}}{\pgfqpoint{2.290830in}{0.760445in}}{\pgfqpoint{2.305095in}{0.774710in}}%
\pgfpathcurveto{\pgfqpoint{2.319361in}{0.788976in}}{\pgfqpoint{2.327376in}{0.808327in}}{\pgfqpoint{2.327376in}{0.828502in}}%
\pgfpathcurveto{\pgfqpoint{2.327376in}{0.848676in}}{\pgfqpoint{2.319361in}{0.868027in}}{\pgfqpoint{2.305095in}{0.882293in}}%
\pgfpathcurveto{\pgfqpoint{2.290830in}{0.896559in}}{\pgfqpoint{2.271479in}{0.904574in}}{\pgfqpoint{2.251304in}{0.904574in}}%
\pgfpathcurveto{\pgfqpoint{2.231129in}{0.904574in}}{\pgfqpoint{2.211778in}{0.896559in}}{\pgfqpoint{2.197512in}{0.882293in}}%
\pgfpathcurveto{\pgfqpoint{2.183247in}{0.868027in}}{\pgfqpoint{2.175231in}{0.848676in}}{\pgfqpoint{2.175231in}{0.828502in}}%
\pgfpathcurveto{\pgfqpoint{2.175231in}{0.808327in}}{\pgfqpoint{2.183247in}{0.788976in}}{\pgfqpoint{2.197512in}{0.774710in}}%
\pgfpathcurveto{\pgfqpoint{2.211778in}{0.760445in}}{\pgfqpoint{2.231129in}{0.752429in}}{\pgfqpoint{2.251304in}{0.752429in}}%
\pgfpathlineto{\pgfqpoint{2.251304in}{0.752429in}}%
\pgfpathclose%
\pgfusepath{stroke}%
\end{pgfscope}%
\begin{pgfscope}%
\pgfpathrectangle{\pgfqpoint{0.782823in}{0.175231in}}{\pgfqpoint{2.766359in}{2.855141in}}%
\pgfusepath{clip}%
\pgfsetbuttcap%
\pgfsetroundjoin%
\pgfsetlinewidth{2.007500pt}%
\definecolor{currentstroke}{rgb}{1.000000,1.000000,1.000000}%
\pgfsetstrokecolor{currentstroke}%
\pgfsetdash{}{0pt}%
\pgfpathmoveto{\pgfqpoint{3.144437in}{1.161008in}}%
\pgfpathcurveto{\pgfqpoint{3.164612in}{1.161008in}}{\pgfqpoint{3.183963in}{1.169023in}}{\pgfqpoint{3.198229in}{1.183289in}}%
\pgfpathcurveto{\pgfqpoint{3.212494in}{1.197554in}}{\pgfqpoint{3.220510in}{1.216906in}}{\pgfqpoint{3.220510in}{1.237080in}}%
\pgfpathcurveto{\pgfqpoint{3.220510in}{1.257255in}}{\pgfqpoint{3.212494in}{1.276606in}}{\pgfqpoint{3.198229in}{1.290872in}}%
\pgfpathcurveto{\pgfqpoint{3.183963in}{1.305137in}}{\pgfqpoint{3.164612in}{1.313153in}}{\pgfqpoint{3.144437in}{1.313153in}}%
\pgfpathcurveto{\pgfqpoint{3.124263in}{1.313153in}}{\pgfqpoint{3.104912in}{1.305137in}}{\pgfqpoint{3.090646in}{1.290872in}}%
\pgfpathcurveto{\pgfqpoint{3.076380in}{1.276606in}}{\pgfqpoint{3.068365in}{1.257255in}}{\pgfqpoint{3.068365in}{1.237080in}}%
\pgfpathcurveto{\pgfqpoint{3.068365in}{1.216906in}}{\pgfqpoint{3.076380in}{1.197554in}}{\pgfqpoint{3.090646in}{1.183289in}}%
\pgfpathcurveto{\pgfqpoint{3.104912in}{1.169023in}}{\pgfqpoint{3.124263in}{1.161008in}}{\pgfqpoint{3.144437in}{1.161008in}}%
\pgfpathlineto{\pgfqpoint{3.144437in}{1.161008in}}%
\pgfpathclose%
\pgfusepath{stroke}%
\end{pgfscope}%
\begin{pgfscope}%
\pgfpathrectangle{\pgfqpoint{0.782823in}{0.175231in}}{\pgfqpoint{2.766359in}{2.855141in}}%
\pgfusepath{clip}%
\pgfsetbuttcap%
\pgfsetroundjoin%
\pgfsetlinewidth{2.007500pt}%
\definecolor{currentstroke}{rgb}{1.000000,1.000000,1.000000}%
\pgfsetstrokecolor{currentstroke}%
\pgfsetdash{}{0pt}%
\pgfpathmoveto{\pgfqpoint{2.515704in}{2.044105in}}%
\pgfpathcurveto{\pgfqpoint{2.535879in}{2.044105in}}{\pgfqpoint{2.555230in}{2.052121in}}{\pgfqpoint{2.569495in}{2.066386in}}%
\pgfpathcurveto{\pgfqpoint{2.583761in}{2.080652in}}{\pgfqpoint{2.591776in}{2.100003in}}{\pgfqpoint{2.591776in}{2.120178in}}%
\pgfpathcurveto{\pgfqpoint{2.591776in}{2.140352in}}{\pgfqpoint{2.583761in}{2.159703in}}{\pgfqpoint{2.569495in}{2.173969in}}%
\pgfpathcurveto{\pgfqpoint{2.555230in}{2.188235in}}{\pgfqpoint{2.535879in}{2.196250in}}{\pgfqpoint{2.515704in}{2.196250in}}%
\pgfpathcurveto{\pgfqpoint{2.495529in}{2.196250in}}{\pgfqpoint{2.476178in}{2.188235in}}{\pgfqpoint{2.461912in}{2.173969in}}%
\pgfpathcurveto{\pgfqpoint{2.447647in}{2.159703in}}{\pgfqpoint{2.439631in}{2.140352in}}{\pgfqpoint{2.439631in}{2.120178in}}%
\pgfpathcurveto{\pgfqpoint{2.439631in}{2.100003in}}{\pgfqpoint{2.447647in}{2.080652in}}{\pgfqpoint{2.461912in}{2.066386in}}%
\pgfpathcurveto{\pgfqpoint{2.476178in}{2.052121in}}{\pgfqpoint{2.495529in}{2.044105in}}{\pgfqpoint{2.515704in}{2.044105in}}%
\pgfpathlineto{\pgfqpoint{2.515704in}{2.044105in}}%
\pgfpathclose%
\pgfusepath{stroke}%
\end{pgfscope}%
\begin{pgfscope}%
\pgfpathrectangle{\pgfqpoint{0.782823in}{0.175231in}}{\pgfqpoint{2.766359in}{2.855141in}}%
\pgfusepath{clip}%
\pgfsetbuttcap%
\pgfsetroundjoin%
\pgfsetlinewidth{2.007500pt}%
\definecolor{currentstroke}{rgb}{1.000000,1.000000,1.000000}%
\pgfsetstrokecolor{currentstroke}%
\pgfsetdash{}{0pt}%
\pgfpathmoveto{\pgfqpoint{1.178542in}{1.409253in}}%
\pgfpathcurveto{\pgfqpoint{1.198716in}{1.409253in}}{\pgfqpoint{1.218068in}{1.417269in}}{\pgfqpoint{1.232333in}{1.431534in}}%
\pgfpathcurveto{\pgfqpoint{1.246599in}{1.445800in}}{\pgfqpoint{1.254614in}{1.465151in}}{\pgfqpoint{1.254614in}{1.485326in}}%
\pgfpathcurveto{\pgfqpoint{1.254614in}{1.505500in}}{\pgfqpoint{1.246599in}{1.524851in}}{\pgfqpoint{1.232333in}{1.539117in}}%
\pgfpathcurveto{\pgfqpoint{1.218068in}{1.553383in}}{\pgfqpoint{1.198716in}{1.561398in}}{\pgfqpoint{1.178542in}{1.561398in}}%
\pgfpathcurveto{\pgfqpoint{1.158367in}{1.561398in}}{\pgfqpoint{1.139016in}{1.553383in}}{\pgfqpoint{1.124750in}{1.539117in}}%
\pgfpathcurveto{\pgfqpoint{1.110485in}{1.524851in}}{\pgfqpoint{1.102469in}{1.505500in}}{\pgfqpoint{1.102469in}{1.485326in}}%
\pgfpathcurveto{\pgfqpoint{1.102469in}{1.465151in}}{\pgfqpoint{1.110485in}{1.445800in}}{\pgfqpoint{1.124750in}{1.431534in}}%
\pgfpathcurveto{\pgfqpoint{1.139016in}{1.417269in}}{\pgfqpoint{1.158367in}{1.409253in}}{\pgfqpoint{1.178542in}{1.409253in}}%
\pgfpathlineto{\pgfqpoint{1.178542in}{1.409253in}}%
\pgfpathclose%
\pgfusepath{stroke}%
\end{pgfscope}%
\begin{pgfscope}%
\pgfpathrectangle{\pgfqpoint{0.782823in}{0.175231in}}{\pgfqpoint{2.766359in}{2.855141in}}%
\pgfusepath{clip}%
\pgfsetbuttcap%
\pgfsetroundjoin%
\pgfsetlinewidth{2.007500pt}%
\definecolor{currentstroke}{rgb}{1.000000,1.000000,1.000000}%
\pgfsetstrokecolor{currentstroke}%
\pgfsetdash{}{0pt}%
\pgfpathmoveto{\pgfqpoint{2.291214in}{2.813919in}}%
\pgfpathcurveto{\pgfqpoint{2.311389in}{2.813919in}}{\pgfqpoint{2.330740in}{2.821935in}}{\pgfqpoint{2.345006in}{2.836200in}}%
\pgfpathcurveto{\pgfqpoint{2.359271in}{2.850466in}}{\pgfqpoint{2.367287in}{2.869817in}}{\pgfqpoint{2.367287in}{2.889992in}}%
\pgfpathcurveto{\pgfqpoint{2.367287in}{2.910167in}}{\pgfqpoint{2.359271in}{2.929518in}}{\pgfqpoint{2.345006in}{2.943783in}}%
\pgfpathcurveto{\pgfqpoint{2.330740in}{2.958049in}}{\pgfqpoint{2.311389in}{2.966064in}}{\pgfqpoint{2.291214in}{2.966064in}}%
\pgfpathcurveto{\pgfqpoint{2.271040in}{2.966064in}}{\pgfqpoint{2.251689in}{2.958049in}}{\pgfqpoint{2.237423in}{2.943783in}}%
\pgfpathcurveto{\pgfqpoint{2.223157in}{2.929518in}}{\pgfqpoint{2.215142in}{2.910167in}}{\pgfqpoint{2.215142in}{2.889992in}}%
\pgfpathcurveto{\pgfqpoint{2.215142in}{2.869817in}}{\pgfqpoint{2.223157in}{2.850466in}}{\pgfqpoint{2.237423in}{2.836200in}}%
\pgfpathcurveto{\pgfqpoint{2.251689in}{2.821935in}}{\pgfqpoint{2.271040in}{2.813919in}}{\pgfqpoint{2.291214in}{2.813919in}}%
\pgfpathlineto{\pgfqpoint{2.291214in}{2.813919in}}%
\pgfpathclose%
\pgfusepath{stroke}%
\end{pgfscope}%
\begin{pgfscope}%
\pgfpathrectangle{\pgfqpoint{0.782823in}{0.175231in}}{\pgfqpoint{2.766359in}{2.855141in}}%
\pgfusepath{clip}%
\pgfsetbuttcap%
\pgfsetroundjoin%
\pgfsetlinewidth{2.007500pt}%
\definecolor{currentstroke}{rgb}{1.000000,1.000000,1.000000}%
\pgfsetstrokecolor{currentstroke}%
\pgfsetdash{}{0pt}%
\pgfpathmoveto{\pgfqpoint{1.154749in}{1.341354in}}%
\pgfpathcurveto{\pgfqpoint{1.174923in}{1.341354in}}{\pgfqpoint{1.194274in}{1.349369in}}{\pgfqpoint{1.208540in}{1.363635in}}%
\pgfpathcurveto{\pgfqpoint{1.222806in}{1.377900in}}{\pgfqpoint{1.230821in}{1.397252in}}{\pgfqpoint{1.230821in}{1.417426in}}%
\pgfpathcurveto{\pgfqpoint{1.230821in}{1.437601in}}{\pgfqpoint{1.222806in}{1.456952in}}{\pgfqpoint{1.208540in}{1.471218in}}%
\pgfpathcurveto{\pgfqpoint{1.194274in}{1.485483in}}{\pgfqpoint{1.174923in}{1.493499in}}{\pgfqpoint{1.154749in}{1.493499in}}%
\pgfpathcurveto{\pgfqpoint{1.134574in}{1.493499in}}{\pgfqpoint{1.115223in}{1.485483in}}{\pgfqpoint{1.100957in}{1.471218in}}%
\pgfpathcurveto{\pgfqpoint{1.086691in}{1.456952in}}{\pgfqpoint{1.078676in}{1.437601in}}{\pgfqpoint{1.078676in}{1.417426in}}%
\pgfpathcurveto{\pgfqpoint{1.078676in}{1.397252in}}{\pgfqpoint{1.086691in}{1.377900in}}{\pgfqpoint{1.100957in}{1.363635in}}%
\pgfpathcurveto{\pgfqpoint{1.115223in}{1.349369in}}{\pgfqpoint{1.134574in}{1.341354in}}{\pgfqpoint{1.154749in}{1.341354in}}%
\pgfpathlineto{\pgfqpoint{1.154749in}{1.341354in}}%
\pgfpathclose%
\pgfusepath{stroke}%
\end{pgfscope}%
\begin{pgfscope}%
\pgfpathrectangle{\pgfqpoint{0.782823in}{0.175231in}}{\pgfqpoint{2.766359in}{2.855141in}}%
\pgfusepath{clip}%
\pgfsetbuttcap%
\pgfsetroundjoin%
\pgfsetlinewidth{2.007500pt}%
\definecolor{currentstroke}{rgb}{1.000000,1.000000,1.000000}%
\pgfsetstrokecolor{currentstroke}%
\pgfsetdash{}{0pt}%
\pgfpathmoveto{\pgfqpoint{3.405081in}{0.302426in}}%
\pgfpathcurveto{\pgfqpoint{3.425256in}{0.302426in}}{\pgfqpoint{3.444607in}{0.310441in}}{\pgfqpoint{3.458872in}{0.324707in}}%
\pgfpathcurveto{\pgfqpoint{3.473138in}{0.338973in}}{\pgfqpoint{3.481153in}{0.358324in}}{\pgfqpoint{3.481153in}{0.378499in}}%
\pgfpathcurveto{\pgfqpoint{3.481153in}{0.398673in}}{\pgfqpoint{3.473138in}{0.418024in}}{\pgfqpoint{3.458872in}{0.432290in}}%
\pgfpathcurveto{\pgfqpoint{3.444607in}{0.446556in}}{\pgfqpoint{3.425256in}{0.454571in}}{\pgfqpoint{3.405081in}{0.454571in}}%
\pgfpathcurveto{\pgfqpoint{3.384906in}{0.454571in}}{\pgfqpoint{3.365555in}{0.446556in}}{\pgfqpoint{3.351289in}{0.432290in}}%
\pgfpathcurveto{\pgfqpoint{3.337024in}{0.418024in}}{\pgfqpoint{3.329008in}{0.398673in}}{\pgfqpoint{3.329008in}{0.378499in}}%
\pgfpathcurveto{\pgfqpoint{3.329008in}{0.358324in}}{\pgfqpoint{3.337024in}{0.338973in}}{\pgfqpoint{3.351289in}{0.324707in}}%
\pgfpathcurveto{\pgfqpoint{3.365555in}{0.310441in}}{\pgfqpoint{3.384906in}{0.302426in}}{\pgfqpoint{3.405081in}{0.302426in}}%
\pgfpathlineto{\pgfqpoint{3.405081in}{0.302426in}}%
\pgfpathclose%
\pgfusepath{stroke}%
\end{pgfscope}%
\begin{pgfscope}%
\pgfsetbuttcap%
\pgfsetroundjoin%
\definecolor{currentfill}{rgb}{0.000000,0.000000,0.000000}%
\pgfsetfillcolor{currentfill}%
\pgfsetlinewidth{0.803000pt}%
\definecolor{currentstroke}{rgb}{0.000000,0.000000,0.000000}%
\pgfsetstrokecolor{currentstroke}%
\pgfsetdash{}{0pt}%
\pgfsys@defobject{currentmarker}{\pgfqpoint{0.000000in}{0.000000in}}{\pgfqpoint{0.000000in}{0.048611in}}{%
\pgfpathmoveto{\pgfqpoint{0.000000in}{0.000000in}}%
\pgfpathlineto{\pgfqpoint{0.000000in}{0.048611in}}%
\pgfusepath{stroke,fill}%
}%
\begin{pgfscope}%
\pgfsys@transformshift{1.085394in}{3.030372in}%
\pgfsys@useobject{currentmarker}{}%
\end{pgfscope}%
\end{pgfscope}%
\begin{pgfscope}%
\definecolor{textcolor}{rgb}{0.000000,0.000000,0.000000}%
\pgfsetstrokecolor{textcolor}%
\pgfsetfillcolor{textcolor}%
\pgftext[x=1.085394in,y=3.127594in,,bottom]{\color{textcolor}{\rmfamily\fontsize{16.000000}{18.00000}\selectfont\catcode`\^=\active\def^{\ifmmode\sp\else\^{}\fi}\catcode`\%=\active\def
\end{pgfscope}%
\begin{pgfscope}%
\pgfsetbuttcap%
\pgfsetroundjoin%
\definecolor{currentfill}{rgb}{0.000000,0.000000,0.000000}%
\pgfsetfillcolor{currentfill}%
\pgfsetlinewidth{0.803000pt}%
\definecolor{currentstroke}{rgb}{0.000000,0.000000,0.000000}%
\pgfsetstrokecolor{currentstroke}%
\pgfsetdash{}{0pt}%
\pgfsys@defobject{currentmarker}{\pgfqpoint{0.000000in}{0.000000in}}{\pgfqpoint{0.000000in}{0.048611in}}{%
\pgfpathmoveto{\pgfqpoint{0.000000in}{0.000000in}}%
\pgfpathlineto{\pgfqpoint{0.000000in}{0.048611in}}%
\pgfusepath{stroke,fill}%
}%
\begin{pgfscope}%
\pgfsys@transformshift{1.625698in}{3.030372in}%
\pgfsys@useobject{currentmarker}{}%
\end{pgfscope}%
\end{pgfscope}%
\begin{pgfscope}%
\pgfsetbuttcap%
\pgfsetroundjoin%
\definecolor{currentfill}{rgb}{0.000000,0.000000,0.000000}%
\pgfsetfillcolor{currentfill}%
\pgfsetlinewidth{0.803000pt}%
\definecolor{currentstroke}{rgb}{0.000000,0.000000,0.000000}%
\pgfsetstrokecolor{currentstroke}%
\pgfsetdash{}{0pt}%
\pgfsys@defobject{currentmarker}{\pgfqpoint{0.000000in}{0.000000in}}{\pgfqpoint{0.000000in}{0.048611in}}{%
\pgfpathmoveto{\pgfqpoint{0.000000in}{0.000000in}}%
\pgfpathlineto{\pgfqpoint{0.000000in}{0.048611in}}%
\pgfusepath{stroke,fill}%
}%
\begin{pgfscope}%
\pgfsys@transformshift{2.166003in}{3.030372in}%
\pgfsys@useobject{currentmarker}{}%
\end{pgfscope}%
\end{pgfscope}%
\begin{pgfscope}%
\pgfsetbuttcap%
\pgfsetroundjoin%
\definecolor{currentfill}{rgb}{0.000000,0.000000,0.000000}%
\pgfsetfillcolor{currentfill}%
\pgfsetlinewidth{0.803000pt}%
\definecolor{currentstroke}{rgb}{0.000000,0.000000,0.000000}%
\pgfsetstrokecolor{currentstroke}%
\pgfsetdash{}{0pt}%
\pgfsys@defobject{currentmarker}{\pgfqpoint{0.000000in}{0.000000in}}{\pgfqpoint{0.000000in}{0.048611in}}{%
\pgfpathmoveto{\pgfqpoint{0.000000in}{0.000000in}}%
\pgfpathlineto{\pgfqpoint{0.000000in}{0.048611in}}%
\pgfusepath{stroke,fill}%
}%
\begin{pgfscope}%
\pgfsys@transformshift{2.706307in}{3.030372in}%
\pgfsys@useobject{currentmarker}{}%
\end{pgfscope}%
\end{pgfscope}%
\begin{pgfscope}%
\pgfsetbuttcap%
\pgfsetroundjoin%
\definecolor{currentfill}{rgb}{0.000000,0.000000,0.000000}%
\pgfsetfillcolor{currentfill}%
\pgfsetlinewidth{0.803000pt}%
\definecolor{currentstroke}{rgb}{0.000000,0.000000,0.000000}%
\pgfsetstrokecolor{currentstroke}%
\pgfsetdash{}{0pt}%
\pgfsys@defobject{currentmarker}{\pgfqpoint{0.000000in}{0.000000in}}{\pgfqpoint{0.000000in}{0.048611in}}{%
\pgfpathmoveto{\pgfqpoint{0.000000in}{0.000000in}}%
\pgfpathlineto{\pgfqpoint{0.000000in}{0.048611in}}%
\pgfusepath{stroke,fill}%
}%
\begin{pgfscope}%
\pgfsys@transformshift{3.246612in}{3.030372in}%
\pgfsys@useobject{currentmarker}{}%
\end{pgfscope}%
\end{pgfscope}%
\begin{pgfscope}%
\definecolor{textcolor}{rgb}{0.000000,0.000000,0.000000}%
\pgfsetstrokecolor{textcolor}%
\pgfsetfillcolor{textcolor}%
\pgftext[x=3.246612in,y=3.127594in,,bottom]{\color{textcolor}{\rmfamily\fontsize{16.000000}{18.00000}\selectfont\catcode`\^=\active\def^{\ifmmode\sp\else\^{}\fi}\catcode`\%=\active\def
\end{pgfscope}%
\begin{pgfscope}%
\definecolor{textcolor}{rgb}{0.000000,0.000000,0.000000}%
\pgfsetstrokecolor{textcolor}%
\pgfsetfillcolor{textcolor}%
\pgftext[x=2.166003in,y=3.266483in,,base]{\color{textcolor}{\rmfamily\fontsize{16.000000}{18.00000}\selectfont\catcode`\^=\active\def^{\ifmmode\sp\else\^{}\fi}\catcode`\%=\active\def
\end{pgfscope}%
\begin{pgfscope}%
\pgfsetbuttcap%
\pgfsetroundjoin%
\definecolor{currentfill}{rgb}{0.000000,0.000000,0.000000}%
\pgfsetfillcolor{currentfill}%
\pgfsetlinewidth{0.803000pt}%
\definecolor{currentstroke}{rgb}{0.000000,0.000000,0.000000}%
\pgfsetstrokecolor{currentstroke}%
\pgfsetdash{}{0pt}%
\pgfsys@defobject{currentmarker}{\pgfqpoint{-0.048611in}{0.000000in}}{\pgfqpoint{-0.000000in}{0.000000in}}{%
\pgfpathmoveto{\pgfqpoint{-0.000000in}{0.000000in}}%
\pgfpathlineto{\pgfqpoint{-0.048611in}{0.000000in}}%
\pgfusepath{stroke,fill}%
}%
\begin{pgfscope}%
\pgfsys@transformshift{0.782823in}{0.175231in}%
\pgfsys@useobject{currentmarker}{}%
\end{pgfscope}%
\end{pgfscope}%
\begin{pgfscope}%
\definecolor{textcolor}{rgb}{0.000000,0.000000,0.000000}%
\pgfsetstrokecolor{textcolor}%
\pgfsetfillcolor{textcolor}%
\pgftext[x=0.414767in, y=0.127006in, left, base]{\color{textcolor}{\rmfamily\fontsize{16.000000}{18.00000}\selectfont\catcode`\^=\active\def^{\ifmmode\sp\else\^{}\fi}\catcode`\%=\active\def
\end{pgfscope}%
\begin{pgfscope}%
\pgfsetbuttcap%
\pgfsetroundjoin%
\definecolor{currentfill}{rgb}{0.000000,0.000000,0.000000}%
\pgfsetfillcolor{currentfill}%
\pgfsetlinewidth{0.803000pt}%
\definecolor{currentstroke}{rgb}{0.000000,0.000000,0.000000}%
\pgfsetstrokecolor{currentstroke}%
\pgfsetdash{}{0pt}%
\pgfsys@defobject{currentmarker}{\pgfqpoint{-0.048611in}{0.000000in}}{\pgfqpoint{-0.000000in}{0.000000in}}{%
\pgfpathmoveto{\pgfqpoint{-0.000000in}{0.000000in}}%
\pgfpathlineto{\pgfqpoint{-0.048611in}{0.000000in}}%
\pgfusepath{stroke,fill}%
}%
\begin{pgfscope}%
\pgfsys@transformshift{0.782823in}{0.732876in}%
\pgfsys@useobject{currentmarker}{}%
\end{pgfscope}%
\end{pgfscope}%
\begin{pgfscope}%
\definecolor{textcolor}{rgb}{0.000000,0.000000,0.000000}%
\pgfsetstrokecolor{textcolor}%
\pgfsetfillcolor{textcolor}%
\pgftext[x=0.169445in, y=0.680793in, left, base]{\color{textcolor}{\rmfamily\fontsize{16.000000}{18.00000}\selectfont\catcode`\^=\active\def^{\ifmmode\sp\else\^{}\fi}\catcode`\%=\active\def
\end{pgfscope}%
\begin{pgfscope}%
\pgfsetbuttcap%
\pgfsetroundjoin%
\definecolor{currentfill}{rgb}{0.000000,0.000000,0.000000}%
\pgfsetfillcolor{currentfill}%
\pgfsetlinewidth{0.803000pt}%
\definecolor{currentstroke}{rgb}{0.000000,0.000000,0.000000}%
\pgfsetstrokecolor{currentstroke}%
\pgfsetdash{}{0pt}%
\pgfsys@defobject{currentmarker}{\pgfqpoint{-0.048611in}{0.000000in}}{\pgfqpoint{-0.000000in}{0.000000in}}{%
\pgfpathmoveto{\pgfqpoint{-0.000000in}{0.000000in}}%
\pgfpathlineto{\pgfqpoint{-0.048611in}{0.000000in}}%
\pgfusepath{stroke,fill}%
}%
\begin{pgfscope}%
\pgfsys@transformshift{0.782823in}{1.290521in}%
\pgfsys@useobject{currentmarker}{}%
\end{pgfscope}%
\end{pgfscope}%
\begin{pgfscope}%
\pgfsetbuttcap%
\pgfsetroundjoin%
\definecolor{currentfill}{rgb}{0.000000,0.000000,0.000000}%
\pgfsetfillcolor{currentfill}%
\pgfsetlinewidth{0.803000pt}%
\definecolor{currentstroke}{rgb}{0.000000,0.000000,0.000000}%
\pgfsetstrokecolor{currentstroke}%
\pgfsetdash{}{0pt}%
\pgfsys@defobject{currentmarker}{\pgfqpoint{-0.048611in}{0.000000in}}{\pgfqpoint{-0.000000in}{0.000000in}}{%
\pgfpathmoveto{\pgfqpoint{-0.000000in}{0.000000in}}%
\pgfpathlineto{\pgfqpoint{-0.048611in}{0.000000in}}%
\pgfusepath{stroke,fill}%
}%
\begin{pgfscope}%
\pgfsys@transformshift{0.782823in}{1.848165in}%
\pgfsys@useobject{currentmarker}{}%
\end{pgfscope}%
\end{pgfscope}%
\begin{pgfscope}%
\pgfsetbuttcap%
\pgfsetroundjoin%
\definecolor{currentfill}{rgb}{0.000000,0.000000,0.000000}%
\pgfsetfillcolor{currentfill}%
\pgfsetlinewidth{0.803000pt}%
\definecolor{currentstroke}{rgb}{0.000000,0.000000,0.000000}%
\pgfsetstrokecolor{currentstroke}%
\pgfsetdash{}{0pt}%
\pgfsys@defobject{currentmarker}{\pgfqpoint{-0.048611in}{0.000000in}}{\pgfqpoint{-0.000000in}{0.000000in}}{%
\pgfpathmoveto{\pgfqpoint{-0.000000in}{0.000000in}}%
\pgfpathlineto{\pgfqpoint{-0.048611in}{0.000000in}}%
\pgfusepath{stroke,fill}%
}%
\begin{pgfscope}%
\pgfsys@transformshift{0.782823in}{2.405810in}%
\pgfsys@useobject{currentmarker}{}%
\end{pgfscope}%
\end{pgfscope}%
\begin{pgfscope}%
\definecolor{textcolor}{rgb}{0.000000,0.000000,0.000000}%
\pgfsetstrokecolor{textcolor}%
\pgfsetfillcolor{textcolor}%
\pgftext[x=0.100000in, y=2.353727in, left, base]{\color{textcolor}{\rmfamily\fontsize{16.000000}{18.00000}\selectfont\catcode`\^=\active\def^{\ifmmode\sp\else\^{}\fi}\catcode`\%=\active\def
\end{pgfscope}%
\begin{pgfscope}%
\definecolor{textcolor}{rgb}{0.000000,0.000000,0.000000}%
\pgfsetstrokecolor{textcolor}%
\pgfsetfillcolor{textcolor}%
\pgftext[x=0.405556in,y=1.602802in,,bottom,rotate=90.000000]{\color{textcolor}{\rmfamily\fontsize{16.000000}{18.00000}\selectfont\catcode`\^=\active\def^{\ifmmode\sp\else\^{}\fi}\catcode`\%=\active\def
\end{pgfscope}%
\begin{pgfscope}%
\pgfsetrectcap%
\pgfsetmiterjoin%
\pgfsetlinewidth{0.803000pt}%
\definecolor{currentstroke}{rgb}{0.000000,0.000000,0.000000}%
\pgfsetstrokecolor{currentstroke}%
\pgfsetdash{}{0pt}%
\pgfpathmoveto{\pgfqpoint{0.782823in}{0.175231in}}%
\pgfpathlineto{\pgfqpoint{0.782823in}{3.030372in}}%
\pgfusepath{stroke}%
\end{pgfscope}%
\begin{pgfscope}%
\pgfsetrectcap%
\pgfsetmiterjoin%
\pgfsetlinewidth{0.803000pt}%
\definecolor{currentstroke}{rgb}{0.000000,0.000000,0.000000}%
\pgfsetstrokecolor{currentstroke}%
\pgfsetdash{}{0pt}%
\pgfpathmoveto{\pgfqpoint{3.549183in}{0.175231in}}%
\pgfpathlineto{\pgfqpoint{3.549183in}{3.030372in}}%
\pgfusepath{stroke}%
\end{pgfscope}%
\begin{pgfscope}%
\pgfsetrectcap%
\pgfsetmiterjoin%
\pgfsetlinewidth{0.803000pt}%
\definecolor{currentstroke}{rgb}{0.000000,0.000000,0.000000}%
\pgfsetstrokecolor{currentstroke}%
\pgfsetdash{}{0pt}%
\pgfpathmoveto{\pgfqpoint{0.782823in}{0.175231in}}%
\pgfpathlineto{\pgfqpoint{3.549183in}{0.175231in}}%
\pgfusepath{stroke}%
\end{pgfscope}%
\begin{pgfscope}%
\pgfsetrectcap%
\pgfsetmiterjoin%
\pgfsetlinewidth{0.803000pt}%
\definecolor{currentstroke}{rgb}{0.000000,0.000000,0.000000}%
\pgfsetstrokecolor{currentstroke}%
\pgfsetdash{}{0pt}%
\pgfpathmoveto{\pgfqpoint{0.782823in}{3.030372in}}%
\pgfpathlineto{\pgfqpoint{3.549183in}{3.030372in}}%
\pgfusepath{stroke}%
\end{pgfscope}%
\end{pgfpicture}%
\makeatother%
\endgroup%

%% file: figures/inference/snr=-20.pgf
\begingroup%
\makeatletter%
\begin{pgfpicture}%
\pgfpathrectangle{\pgfpointorigin}{\pgfqpoint{3.470650in}{3.470650in}}%
\pgfusepath{use as bounding box, clip}%
\begin{pgfscope}%
\pgfsetbuttcap%
\pgfsetmiterjoin%
\definecolor{currentfill}{rgb}{1.000000,1.000000,1.000000}%
\pgfsetfillcolor{currentfill}%
\pgfsetlinewidth{0.000000pt}%
\definecolor{currentstroke}{rgb}{1.000000,1.000000,1.000000}%
\pgfsetstrokecolor{currentstroke}%
\pgfsetdash{}{0pt}%
\pgfpathmoveto{\pgfqpoint{0.000000in}{0.000000in}}%
\pgfpathlineto{\pgfqpoint{3.470650in}{0.000000in}}%
\pgfpathlineto{\pgfqpoint{3.470650in}{3.470650in}}%
\pgfpathlineto{\pgfqpoint{0.000000in}{3.470650in}}%
\pgfpathlineto{\pgfqpoint{0.000000in}{0.000000in}}%
\pgfpathclose%
\pgfusepath{fill}%
\end{pgfscope}%
\begin{pgfscope}%
\pgfsetbuttcap%
\pgfsetmiterjoin%
\definecolor{currentfill}{rgb}{1.000000,1.000000,1.000000}%
\pgfsetfillcolor{currentfill}%
\pgfsetlinewidth{0.000000pt}%
\definecolor{currentstroke}{rgb}{0.000000,0.000000,0.000000}%
\pgfsetstrokecolor{currentstroke}%
\pgfsetstrokeopacity{0.000000}%
\pgfsetdash{}{0pt}%
\pgfpathmoveto{\pgfqpoint{0.266667in}{0.175231in}}%
\pgfpathlineto{\pgfqpoint{3.370650in}{0.175231in}}%
\pgfpathlineto{\pgfqpoint{3.370650in}{3.030372in}}%
\pgfpathlineto{\pgfqpoint{0.266667in}{3.030372in}}%
\pgfpathlineto{\pgfqpoint{0.266667in}{0.175231in}}%
\pgfpathclose%
\pgfusepath{fill}%
\end{pgfscope}%
\begin{pgfscope}%
\pgfpathrectangle{\pgfqpoint{0.266667in}{0.175231in}}{\pgfqpoint{3.103983in}{2.855141in}}%
\pgfusepath{clip}%
\pgfsys@transformshift{0.266667in}{0.175231in}%
\pgftext[left,bottom]{\includegraphics[interpolate=true,width=3.110000in,height=2.860000in]{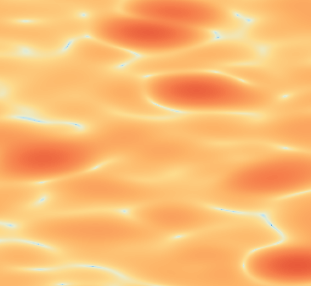}}%
\end{pgfscope}%
\begin{pgfscope}%
\pgfpathrectangle{\pgfqpoint{0.266667in}{0.175231in}}{\pgfqpoint{3.103983in}{2.855141in}}%
\pgfusepath{clip}%
\pgfsetbuttcap%
\pgfsetroundjoin%
\pgfsetlinewidth{2.007500pt}%
\definecolor{currentstroke}{rgb}{1.000000,1.000000,1.000000}%
\pgfsetstrokecolor{currentstroke}%
\pgfsetdash{}{0pt}%
\pgfpathmoveto{\pgfqpoint{1.737590in}{2.630829in}}%
\pgfpathcurveto{\pgfqpoint{1.757765in}{2.630829in}}{\pgfqpoint{1.777116in}{2.638845in}}{\pgfqpoint{1.791382in}{2.653110in}}%
\pgfpathcurveto{\pgfqpoint{1.805648in}{2.667376in}}{\pgfqpoint{1.813663in}{2.686727in}}{\pgfqpoint{1.813663in}{2.706902in}}%
\pgfpathcurveto{\pgfqpoint{1.813663in}{2.727076in}}{\pgfqpoint{1.805648in}{2.746428in}}{\pgfqpoint{1.791382in}{2.760693in}}%
\pgfpathcurveto{\pgfqpoint{1.777116in}{2.774959in}}{\pgfqpoint{1.757765in}{2.782974in}}{\pgfqpoint{1.737590in}{2.782974in}}%
\pgfpathcurveto{\pgfqpoint{1.717416in}{2.782974in}}{\pgfqpoint{1.698065in}{2.774959in}}{\pgfqpoint{1.683799in}{2.760693in}}%
\pgfpathcurveto{\pgfqpoint{1.669533in}{2.746428in}}{\pgfqpoint{1.661518in}{2.727076in}}{\pgfqpoint{1.661518in}{2.706902in}}%
\pgfpathcurveto{\pgfqpoint{1.661518in}{2.686727in}}{\pgfqpoint{1.669533in}{2.667376in}}{\pgfqpoint{1.683799in}{2.653110in}}%
\pgfpathcurveto{\pgfqpoint{1.698065in}{2.638845in}}{\pgfqpoint{1.717416in}{2.630829in}}{\pgfqpoint{1.737590in}{2.630829in}}%
\pgfpathlineto{\pgfqpoint{1.737590in}{2.630829in}}%
\pgfpathclose%
\pgfusepath{stroke}%
\end{pgfscope}%
\begin{pgfscope}%
\pgfpathrectangle{\pgfqpoint{0.266667in}{0.175231in}}{\pgfqpoint{3.103983in}{2.855141in}}%
\pgfusepath{clip}%
\pgfsetbuttcap%
\pgfsetroundjoin%
\pgfsetlinewidth{2.007500pt}%
\definecolor{currentstroke}{rgb}{1.000000,1.000000,1.000000}%
\pgfsetstrokecolor{currentstroke}%
\pgfsetdash{}{0pt}%
\pgfpathmoveto{\pgfqpoint{1.914370in}{0.752429in}}%
\pgfpathcurveto{\pgfqpoint{1.934545in}{0.752429in}}{\pgfqpoint{1.953896in}{0.760445in}}{\pgfqpoint{1.968162in}{0.774710in}}%
\pgfpathcurveto{\pgfqpoint{1.982427in}{0.788976in}}{\pgfqpoint{1.990443in}{0.808327in}}{\pgfqpoint{1.990443in}{0.828502in}}%
\pgfpathcurveto{\pgfqpoint{1.990443in}{0.848676in}}{\pgfqpoint{1.982427in}{0.868027in}}{\pgfqpoint{1.968162in}{0.882293in}}%
\pgfpathcurveto{\pgfqpoint{1.953896in}{0.896559in}}{\pgfqpoint{1.934545in}{0.904574in}}{\pgfqpoint{1.914370in}{0.904574in}}%
\pgfpathcurveto{\pgfqpoint{1.894195in}{0.904574in}}{\pgfqpoint{1.874844in}{0.896559in}}{\pgfqpoint{1.860579in}{0.882293in}}%
\pgfpathcurveto{\pgfqpoint{1.846313in}{0.868027in}}{\pgfqpoint{1.838298in}{0.848676in}}{\pgfqpoint{1.838298in}{0.828502in}}%
\pgfpathcurveto{\pgfqpoint{1.838298in}{0.808327in}}{\pgfqpoint{1.846313in}{0.788976in}}{\pgfqpoint{1.860579in}{0.774710in}}%
\pgfpathcurveto{\pgfqpoint{1.874844in}{0.760445in}}{\pgfqpoint{1.894195in}{0.752429in}}{\pgfqpoint{1.914370in}{0.752429in}}%
\pgfpathlineto{\pgfqpoint{1.914370in}{0.752429in}}%
\pgfpathclose%
\pgfusepath{stroke}%
\end{pgfscope}%
\begin{pgfscope}%
\pgfpathrectangle{\pgfqpoint{0.266667in}{0.175231in}}{\pgfqpoint{3.103983in}{2.855141in}}%
\pgfusepath{clip}%
\pgfsetbuttcap%
\pgfsetroundjoin%
\pgfsetlinewidth{2.007500pt}%
\definecolor{currentstroke}{rgb}{1.000000,1.000000,1.000000}%
\pgfsetstrokecolor{currentstroke}%
\pgfsetdash{}{0pt}%
\pgfpathmoveto{\pgfqpoint{2.916507in}{1.161008in}}%
\pgfpathcurveto{\pgfqpoint{2.936682in}{1.161008in}}{\pgfqpoint{2.956033in}{1.169023in}}{\pgfqpoint{2.970299in}{1.183289in}}%
\pgfpathcurveto{\pgfqpoint{2.984564in}{1.197554in}}{\pgfqpoint{2.992580in}{1.216906in}}{\pgfqpoint{2.992580in}{1.237080in}}%
\pgfpathcurveto{\pgfqpoint{2.992580in}{1.257255in}}{\pgfqpoint{2.984564in}{1.276606in}}{\pgfqpoint{2.970299in}{1.290872in}}%
\pgfpathcurveto{\pgfqpoint{2.956033in}{1.305137in}}{\pgfqpoint{2.936682in}{1.313153in}}{\pgfqpoint{2.916507in}{1.313153in}}%
\pgfpathcurveto{\pgfqpoint{2.896333in}{1.313153in}}{\pgfqpoint{2.876981in}{1.305137in}}{\pgfqpoint{2.862716in}{1.290872in}}%
\pgfpathcurveto{\pgfqpoint{2.848450in}{1.276606in}}{\pgfqpoint{2.840435in}{1.257255in}}{\pgfqpoint{2.840435in}{1.237080in}}%
\pgfpathcurveto{\pgfqpoint{2.840435in}{1.216906in}}{\pgfqpoint{2.848450in}{1.197554in}}{\pgfqpoint{2.862716in}{1.183289in}}%
\pgfpathcurveto{\pgfqpoint{2.876981in}{1.169023in}}{\pgfqpoint{2.896333in}{1.161008in}}{\pgfqpoint{2.916507in}{1.161008in}}%
\pgfpathlineto{\pgfqpoint{2.916507in}{1.161008in}}%
\pgfpathclose%
\pgfusepath{stroke}%
\end{pgfscope}%
\begin{pgfscope}%
\pgfpathrectangle{\pgfqpoint{0.266667in}{0.175231in}}{\pgfqpoint{3.103983in}{2.855141in}}%
\pgfusepath{clip}%
\pgfsetbuttcap%
\pgfsetroundjoin%
\pgfsetlinewidth{2.007500pt}%
\definecolor{currentstroke}{rgb}{1.000000,1.000000,1.000000}%
\pgfsetstrokecolor{currentstroke}%
\pgfsetdash{}{0pt}%
\pgfpathmoveto{\pgfqpoint{2.211039in}{2.044105in}}%
\pgfpathcurveto{\pgfqpoint{2.231214in}{2.044105in}}{\pgfqpoint{2.250565in}{2.052121in}}{\pgfqpoint{2.264831in}{2.066386in}}%
\pgfpathcurveto{\pgfqpoint{2.279096in}{2.080652in}}{\pgfqpoint{2.287112in}{2.100003in}}{\pgfqpoint{2.287112in}{2.120178in}}%
\pgfpathcurveto{\pgfqpoint{2.287112in}{2.140352in}}{\pgfqpoint{2.279096in}{2.159703in}}{\pgfqpoint{2.264831in}{2.173969in}}%
\pgfpathcurveto{\pgfqpoint{2.250565in}{2.188235in}}{\pgfqpoint{2.231214in}{2.196250in}}{\pgfqpoint{2.211039in}{2.196250in}}%
\pgfpathcurveto{\pgfqpoint{2.190864in}{2.196250in}}{\pgfqpoint{2.171513in}{2.188235in}}{\pgfqpoint{2.157248in}{2.173969in}}%
\pgfpathcurveto{\pgfqpoint{2.142982in}{2.159703in}}{\pgfqpoint{2.134967in}{2.140352in}}{\pgfqpoint{2.134967in}{2.120178in}}%
\pgfpathcurveto{\pgfqpoint{2.134967in}{2.100003in}}{\pgfqpoint{2.142982in}{2.080652in}}{\pgfqpoint{2.157248in}{2.066386in}}%
\pgfpathcurveto{\pgfqpoint{2.171513in}{2.052121in}}{\pgfqpoint{2.190864in}{2.044105in}}{\pgfqpoint{2.211039in}{2.044105in}}%
\pgfpathlineto{\pgfqpoint{2.211039in}{2.044105in}}%
\pgfpathclose%
\pgfusepath{stroke}%
\end{pgfscope}%
\begin{pgfscope}%
\pgfpathrectangle{\pgfqpoint{0.266667in}{0.175231in}}{\pgfqpoint{3.103983in}{2.855141in}}%
\pgfusepath{clip}%
\pgfsetbuttcap%
\pgfsetroundjoin%
\pgfsetlinewidth{2.007500pt}%
\definecolor{currentstroke}{rgb}{1.000000,1.000000,1.000000}%
\pgfsetstrokecolor{currentstroke}%
\pgfsetdash{}{0pt}%
\pgfpathmoveto{\pgfqpoint{0.710681in}{1.409253in}}%
\pgfpathcurveto{\pgfqpoint{0.730856in}{1.409253in}}{\pgfqpoint{0.750207in}{1.417269in}}{\pgfqpoint{0.764473in}{1.431534in}}%
\pgfpathcurveto{\pgfqpoint{0.778738in}{1.445800in}}{\pgfqpoint{0.786754in}{1.465151in}}{\pgfqpoint{0.786754in}{1.485326in}}%
\pgfpathcurveto{\pgfqpoint{0.786754in}{1.505500in}}{\pgfqpoint{0.778738in}{1.524851in}}{\pgfqpoint{0.764473in}{1.539117in}}%
\pgfpathcurveto{\pgfqpoint{0.750207in}{1.553383in}}{\pgfqpoint{0.730856in}{1.561398in}}{\pgfqpoint{0.710681in}{1.561398in}}%
\pgfpathcurveto{\pgfqpoint{0.690507in}{1.561398in}}{\pgfqpoint{0.671156in}{1.553383in}}{\pgfqpoint{0.656890in}{1.539117in}}%
\pgfpathcurveto{\pgfqpoint{0.642624in}{1.524851in}}{\pgfqpoint{0.634609in}{1.505500in}}{\pgfqpoint{0.634609in}{1.485326in}}%
\pgfpathcurveto{\pgfqpoint{0.634609in}{1.465151in}}{\pgfqpoint{0.642624in}{1.445800in}}{\pgfqpoint{0.656890in}{1.431534in}}%
\pgfpathcurveto{\pgfqpoint{0.671156in}{1.417269in}}{\pgfqpoint{0.690507in}{1.409253in}}{\pgfqpoint{0.710681in}{1.409253in}}%
\pgfpathlineto{\pgfqpoint{0.710681in}{1.409253in}}%
\pgfpathclose%
\pgfusepath{stroke}%
\end{pgfscope}%
\begin{pgfscope}%
\pgfpathrectangle{\pgfqpoint{0.266667in}{0.175231in}}{\pgfqpoint{3.103983in}{2.855141in}}%
\pgfusepath{clip}%
\pgfsetbuttcap%
\pgfsetroundjoin%
\pgfsetlinewidth{2.007500pt}%
\definecolor{currentstroke}{rgb}{1.000000,1.000000,1.000000}%
\pgfsetstrokecolor{currentstroke}%
\pgfsetdash{}{0pt}%
\pgfpathmoveto{\pgfqpoint{1.959151in}{2.813919in}}%
\pgfpathcurveto{\pgfqpoint{1.979326in}{2.813919in}}{\pgfqpoint{1.998677in}{2.821935in}}{\pgfqpoint{2.012943in}{2.836200in}}%
\pgfpathcurveto{\pgfqpoint{2.027209in}{2.850466in}}{\pgfqpoint{2.035224in}{2.869817in}}{\pgfqpoint{2.035224in}{2.889992in}}%
\pgfpathcurveto{\pgfqpoint{2.035224in}{2.910167in}}{\pgfqpoint{2.027209in}{2.929518in}}{\pgfqpoint{2.012943in}{2.943783in}}%
\pgfpathcurveto{\pgfqpoint{1.998677in}{2.958049in}}{\pgfqpoint{1.979326in}{2.966064in}}{\pgfqpoint{1.959151in}{2.966064in}}%
\pgfpathcurveto{\pgfqpoint{1.938977in}{2.966064in}}{\pgfqpoint{1.919626in}{2.958049in}}{\pgfqpoint{1.905360in}{2.943783in}}%
\pgfpathcurveto{\pgfqpoint{1.891094in}{2.929518in}}{\pgfqpoint{1.883079in}{2.910167in}}{\pgfqpoint{1.883079in}{2.889992in}}%
\pgfpathcurveto{\pgfqpoint{1.883079in}{2.869817in}}{\pgfqpoint{1.891094in}{2.850466in}}{\pgfqpoint{1.905360in}{2.836200in}}%
\pgfpathcurveto{\pgfqpoint{1.919626in}{2.821935in}}{\pgfqpoint{1.938977in}{2.813919in}}{\pgfqpoint{1.959151in}{2.813919in}}%
\pgfpathlineto{\pgfqpoint{1.959151in}{2.813919in}}%
\pgfpathclose%
\pgfusepath{stroke}%
\end{pgfscope}%
\begin{pgfscope}%
\pgfpathrectangle{\pgfqpoint{0.266667in}{0.175231in}}{\pgfqpoint{3.103983in}{2.855141in}}%
\pgfusepath{clip}%
\pgfsetbuttcap%
\pgfsetroundjoin%
\pgfsetlinewidth{2.007500pt}%
\definecolor{currentstroke}{rgb}{1.000000,1.000000,1.000000}%
\pgfsetstrokecolor{currentstroke}%
\pgfsetdash{}{0pt}%
\pgfpathmoveto{\pgfqpoint{0.683984in}{1.341354in}}%
\pgfpathcurveto{\pgfqpoint{0.704159in}{1.341354in}}{\pgfqpoint{0.723510in}{1.349369in}}{\pgfqpoint{0.737776in}{1.363635in}}%
\pgfpathcurveto{\pgfqpoint{0.752041in}{1.377900in}}{\pgfqpoint{0.760057in}{1.397252in}}{\pgfqpoint{0.760057in}{1.417426in}}%
\pgfpathcurveto{\pgfqpoint{0.760057in}{1.437601in}}{\pgfqpoint{0.752041in}{1.456952in}}{\pgfqpoint{0.737776in}{1.471218in}}%
\pgfpathcurveto{\pgfqpoint{0.723510in}{1.485483in}}{\pgfqpoint{0.704159in}{1.493499in}}{\pgfqpoint{0.683984in}{1.493499in}}%
\pgfpathcurveto{\pgfqpoint{0.663810in}{1.493499in}}{\pgfqpoint{0.644459in}{1.485483in}}{\pgfqpoint{0.630193in}{1.471218in}}%
\pgfpathcurveto{\pgfqpoint{0.615927in}{1.456952in}}{\pgfqpoint{0.607912in}{1.437601in}}{\pgfqpoint{0.607912in}{1.417426in}}%
\pgfpathcurveto{\pgfqpoint{0.607912in}{1.397252in}}{\pgfqpoint{0.615927in}{1.377900in}}{\pgfqpoint{0.630193in}{1.363635in}}%
\pgfpathcurveto{\pgfqpoint{0.644459in}{1.349369in}}{\pgfqpoint{0.663810in}{1.341354in}}{\pgfqpoint{0.683984in}{1.341354in}}%
\pgfpathlineto{\pgfqpoint{0.683984in}{1.341354in}}%
\pgfpathclose%
\pgfusepath{stroke}%
\end{pgfscope}%
\begin{pgfscope}%
\pgfpathrectangle{\pgfqpoint{0.266667in}{0.175231in}}{\pgfqpoint{3.103983in}{2.855141in}}%
\pgfusepath{clip}%
\pgfsetbuttcap%
\pgfsetroundjoin%
\pgfsetlinewidth{2.007500pt}%
\definecolor{currentstroke}{rgb}{1.000000,1.000000,1.000000}%
\pgfsetstrokecolor{currentstroke}%
\pgfsetdash{}{0pt}%
\pgfpathmoveto{\pgfqpoint{3.208961in}{0.302426in}}%
\pgfpathcurveto{\pgfqpoint{3.229136in}{0.302426in}}{\pgfqpoint{3.248487in}{0.310441in}}{\pgfqpoint{3.262753in}{0.324707in}}%
\pgfpathcurveto{\pgfqpoint{3.277018in}{0.338973in}}{\pgfqpoint{3.285034in}{0.358324in}}{\pgfqpoint{3.285034in}{0.378499in}}%
\pgfpathcurveto{\pgfqpoint{3.285034in}{0.398673in}}{\pgfqpoint{3.277018in}{0.418024in}}{\pgfqpoint{3.262753in}{0.432290in}}%
\pgfpathcurveto{\pgfqpoint{3.248487in}{0.446556in}}{\pgfqpoint{3.229136in}{0.454571in}}{\pgfqpoint{3.208961in}{0.454571in}}%
\pgfpathcurveto{\pgfqpoint{3.188787in}{0.454571in}}{\pgfqpoint{3.169435in}{0.446556in}}{\pgfqpoint{3.155170in}{0.432290in}}%
\pgfpathcurveto{\pgfqpoint{3.140904in}{0.418024in}}{\pgfqpoint{3.132889in}{0.398673in}}{\pgfqpoint{3.132889in}{0.378499in}}%
\pgfpathcurveto{\pgfqpoint{3.132889in}{0.358324in}}{\pgfqpoint{3.140904in}{0.338973in}}{\pgfqpoint{3.155170in}{0.324707in}}%
\pgfpathcurveto{\pgfqpoint{3.169435in}{0.310441in}}{\pgfqpoint{3.188787in}{0.302426in}}{\pgfqpoint{3.208961in}{0.302426in}}%
\pgfpathlineto{\pgfqpoint{3.208961in}{0.302426in}}%
\pgfpathclose%
\pgfusepath{stroke}%
\end{pgfscope}%
\begin{pgfscope}%
\pgfpathrectangle{\pgfqpoint{0.266667in}{0.175231in}}{\pgfqpoint{3.103983in}{2.855141in}}%
\pgfusepath{clip}%
\pgfsetbuttcap%
\pgfsetroundjoin%
\pgfsetlinewidth{2.007500pt}%
\definecolor{currentstroke}{rgb}{0.000000,0.000000,0.000000}%
\pgfsetstrokecolor{currentstroke}%
\pgfsetdash{}{0pt}%
\pgfpathmoveto{\pgfqpoint{1.718597in}{2.644640in}}%
\pgfpathcurveto{\pgfqpoint{1.738771in}{2.644640in}}{\pgfqpoint{1.758122in}{2.652655in}}{\pgfqpoint{1.772388in}{2.666921in}}%
\pgfpathcurveto{\pgfqpoint{1.786654in}{2.681186in}}{\pgfqpoint{1.794669in}{2.700537in}}{\pgfqpoint{1.794669in}{2.720712in}}%
\pgfpathcurveto{\pgfqpoint{1.794669in}{2.740887in}}{\pgfqpoint{1.786654in}{2.760238in}}{\pgfqpoint{1.772388in}{2.774504in}}%
\pgfpathcurveto{\pgfqpoint{1.758122in}{2.788769in}}{\pgfqpoint{1.738771in}{2.796785in}}{\pgfqpoint{1.718597in}{2.796785in}}%
\pgfpathcurveto{\pgfqpoint{1.698422in}{2.796785in}}{\pgfqpoint{1.679071in}{2.788769in}}{\pgfqpoint{1.664805in}{2.774504in}}%
\pgfpathcurveto{\pgfqpoint{1.650539in}{2.760238in}}{\pgfqpoint{1.642524in}{2.740887in}}{\pgfqpoint{1.642524in}{2.720712in}}%
\pgfpathcurveto{\pgfqpoint{1.642524in}{2.700537in}}{\pgfqpoint{1.650539in}{2.681186in}}{\pgfqpoint{1.664805in}{2.666921in}}%
\pgfpathcurveto{\pgfqpoint{1.679071in}{2.652655in}}{\pgfqpoint{1.698422in}{2.644640in}}{\pgfqpoint{1.718597in}{2.644640in}}%
\pgfpathlineto{\pgfqpoint{1.718597in}{2.644640in}}%
\pgfpathclose%
\pgfusepath{stroke}%
\end{pgfscope}%
\begin{pgfscope}%
\pgfpathrectangle{\pgfqpoint{0.266667in}{0.175231in}}{\pgfqpoint{3.103983in}{2.855141in}}%
\pgfusepath{clip}%
\pgfsetbuttcap%
\pgfsetroundjoin%
\pgfsetlinewidth{2.007500pt}%
\definecolor{currentstroke}{rgb}{0.000000,0.000000,0.000000}%
\pgfsetstrokecolor{currentstroke}%
\pgfsetdash{}{0pt}%
\pgfpathmoveto{\pgfqpoint{3.227998in}{0.303339in}}%
\pgfpathcurveto{\pgfqpoint{3.248173in}{0.303339in}}{\pgfqpoint{3.267524in}{0.311355in}}{\pgfqpoint{3.281789in}{0.325620in}}%
\pgfpathcurveto{\pgfqpoint{3.296055in}{0.339886in}}{\pgfqpoint{3.304071in}{0.359237in}}{\pgfqpoint{3.304071in}{0.379412in}}%
\pgfpathcurveto{\pgfqpoint{3.304071in}{0.399587in}}{\pgfqpoint{3.296055in}{0.418938in}}{\pgfqpoint{3.281789in}{0.433203in}}%
\pgfpathcurveto{\pgfqpoint{3.267524in}{0.447469in}}{\pgfqpoint{3.248173in}{0.455484in}}{\pgfqpoint{3.227998in}{0.455484in}}%
\pgfpathcurveto{\pgfqpoint{3.207823in}{0.455484in}}{\pgfqpoint{3.188472in}{0.447469in}}{\pgfqpoint{3.174207in}{0.433203in}}%
\pgfpathcurveto{\pgfqpoint{3.159941in}{0.418938in}}{\pgfqpoint{3.151925in}{0.399587in}}{\pgfqpoint{3.151925in}{0.379412in}}%
\pgfpathcurveto{\pgfqpoint{3.151925in}{0.359237in}}{\pgfqpoint{3.159941in}{0.339886in}}{\pgfqpoint{3.174207in}{0.325620in}}%
\pgfpathcurveto{\pgfqpoint{3.188472in}{0.311355in}}{\pgfqpoint{3.207823in}{0.303339in}}{\pgfqpoint{3.227998in}{0.303339in}}%
\pgfpathlineto{\pgfqpoint{3.227998in}{0.303339in}}%
\pgfpathclose%
\pgfusepath{stroke}%
\end{pgfscope}%
\begin{pgfscope}%
\pgfpathrectangle{\pgfqpoint{0.266667in}{0.175231in}}{\pgfqpoint{3.103983in}{2.855141in}}%
\pgfusepath{clip}%
\pgfsetbuttcap%
\pgfsetroundjoin%
\pgfsetlinewidth{2.007500pt}%
\definecolor{currentstroke}{rgb}{0.000000,0.000000,0.000000}%
\pgfsetstrokecolor{currentstroke}%
\pgfsetdash{}{0pt}%
\pgfpathmoveto{\pgfqpoint{1.960190in}{2.831173in}}%
\pgfpathcurveto{\pgfqpoint{1.980365in}{2.831173in}}{\pgfqpoint{1.999716in}{2.839188in}}{\pgfqpoint{2.013982in}{2.853454in}}%
\pgfpathcurveto{\pgfqpoint{2.028248in}{2.867720in}}{\pgfqpoint{2.036263in}{2.887071in}}{\pgfqpoint{2.036263in}{2.907245in}}%
\pgfpathcurveto{\pgfqpoint{2.036263in}{2.927420in}}{\pgfqpoint{2.028248in}{2.946771in}}{\pgfqpoint{2.013982in}{2.961037in}}%
\pgfpathcurveto{\pgfqpoint{1.999716in}{2.975302in}}{\pgfqpoint{1.980365in}{2.983318in}}{\pgfqpoint{1.960190in}{2.983318in}}%
\pgfpathcurveto{\pgfqpoint{1.940016in}{2.983318in}}{\pgfqpoint{1.920665in}{2.975302in}}{\pgfqpoint{1.906399in}{2.961037in}}%
\pgfpathcurveto{\pgfqpoint{1.892133in}{2.946771in}}{\pgfqpoint{1.884118in}{2.927420in}}{\pgfqpoint{1.884118in}{2.907245in}}%
\pgfpathcurveto{\pgfqpoint{1.884118in}{2.887071in}}{\pgfqpoint{1.892133in}{2.867720in}}{\pgfqpoint{1.906399in}{2.853454in}}%
\pgfpathcurveto{\pgfqpoint{1.920665in}{2.839188in}}{\pgfqpoint{1.940016in}{2.831173in}}{\pgfqpoint{1.960190in}{2.831173in}}%
\pgfpathlineto{\pgfqpoint{1.960190in}{2.831173in}}%
\pgfpathclose%
\pgfusepath{stroke}%
\end{pgfscope}%
\begin{pgfscope}%
\pgfpathrectangle{\pgfqpoint{0.266667in}{0.175231in}}{\pgfqpoint{3.103983in}{2.855141in}}%
\pgfusepath{clip}%
\pgfsetbuttcap%
\pgfsetroundjoin%
\pgfsetlinewidth{2.007500pt}%
\definecolor{currentstroke}{rgb}{0.000000,0.000000,0.000000}%
\pgfsetstrokecolor{currentstroke}%
\pgfsetdash{}{0pt}%
\pgfpathmoveto{\pgfqpoint{2.934050in}{1.164557in}}%
\pgfpathcurveto{\pgfqpoint{2.954225in}{1.164557in}}{\pgfqpoint{2.973576in}{1.172573in}}{\pgfqpoint{2.987841in}{1.186838in}}%
\pgfpathcurveto{\pgfqpoint{3.002107in}{1.201104in}}{\pgfqpoint{3.010123in}{1.220455in}}{\pgfqpoint{3.010123in}{1.240630in}}%
\pgfpathcurveto{\pgfqpoint{3.010123in}{1.260805in}}{\pgfqpoint{3.002107in}{1.280156in}}{\pgfqpoint{2.987841in}{1.294421in}}%
\pgfpathcurveto{\pgfqpoint{2.973576in}{1.308687in}}{\pgfqpoint{2.954225in}{1.316702in}}{\pgfqpoint{2.934050in}{1.316702in}}%
\pgfpathcurveto{\pgfqpoint{2.913875in}{1.316702in}}{\pgfqpoint{2.894524in}{1.308687in}}{\pgfqpoint{2.880259in}{1.294421in}}%
\pgfpathcurveto{\pgfqpoint{2.865993in}{1.280156in}}{\pgfqpoint{2.857977in}{1.260805in}}{\pgfqpoint{2.857977in}{1.240630in}}%
\pgfpathcurveto{\pgfqpoint{2.857977in}{1.220455in}}{\pgfqpoint{2.865993in}{1.201104in}}{\pgfqpoint{2.880259in}{1.186838in}}%
\pgfpathcurveto{\pgfqpoint{2.894524in}{1.172573in}}{\pgfqpoint{2.913875in}{1.164557in}}{\pgfqpoint{2.934050in}{1.164557in}}%
\pgfpathlineto{\pgfqpoint{2.934050in}{1.164557in}}%
\pgfpathclose%
\pgfusepath{stroke}%
\end{pgfscope}%
\begin{pgfscope}%
\pgfpathrectangle{\pgfqpoint{0.266667in}{0.175231in}}{\pgfqpoint{3.103983in}{2.855141in}}%
\pgfusepath{clip}%
\pgfsetbuttcap%
\pgfsetroundjoin%
\pgfsetlinewidth{2.007500pt}%
\definecolor{currentstroke}{rgb}{0.000000,0.000000,0.000000}%
\pgfsetstrokecolor{currentstroke}%
\pgfsetdash{}{0pt}%
\pgfpathmoveto{\pgfqpoint{2.216798in}{2.042734in}}%
\pgfpathcurveto{\pgfqpoint{2.236973in}{2.042734in}}{\pgfqpoint{2.256324in}{2.050750in}}{\pgfqpoint{2.270590in}{2.065015in}}%
\pgfpathcurveto{\pgfqpoint{2.284855in}{2.079281in}}{\pgfqpoint{2.292871in}{2.098632in}}{\pgfqpoint{2.292871in}{2.118807in}}%
\pgfpathcurveto{\pgfqpoint{2.292871in}{2.138981in}}{\pgfqpoint{2.284855in}{2.158332in}}{\pgfqpoint{2.270590in}{2.172598in}}%
\pgfpathcurveto{\pgfqpoint{2.256324in}{2.186864in}}{\pgfqpoint{2.236973in}{2.194879in}}{\pgfqpoint{2.216798in}{2.194879in}}%
\pgfpathcurveto{\pgfqpoint{2.196624in}{2.194879in}}{\pgfqpoint{2.177272in}{2.186864in}}{\pgfqpoint{2.163007in}{2.172598in}}%
\pgfpathcurveto{\pgfqpoint{2.148741in}{2.158332in}}{\pgfqpoint{2.140726in}{2.138981in}}{\pgfqpoint{2.140726in}{2.118807in}}%
\pgfpathcurveto{\pgfqpoint{2.140726in}{2.098632in}}{\pgfqpoint{2.148741in}{2.079281in}}{\pgfqpoint{2.163007in}{2.065015in}}%
\pgfpathcurveto{\pgfqpoint{2.177272in}{2.050750in}}{\pgfqpoint{2.196624in}{2.042734in}}{\pgfqpoint{2.216798in}{2.042734in}}%
\pgfpathlineto{\pgfqpoint{2.216798in}{2.042734in}}%
\pgfpathclose%
\pgfusepath{stroke}%
\end{pgfscope}%
\begin{pgfscope}%
\pgfpathrectangle{\pgfqpoint{0.266667in}{0.175231in}}{\pgfqpoint{3.103983in}{2.855141in}}%
\pgfusepath{clip}%
\pgfsetbuttcap%
\pgfsetroundjoin%
\pgfsetlinewidth{2.007500pt}%
\definecolor{currentstroke}{rgb}{0.000000,0.000000,0.000000}%
\pgfsetstrokecolor{currentstroke}%
\pgfsetdash{}{0pt}%
\pgfpathmoveto{\pgfqpoint{0.691549in}{1.363119in}}%
\pgfpathcurveto{\pgfqpoint{0.711724in}{1.363119in}}{\pgfqpoint{0.731075in}{1.371134in}}{\pgfqpoint{0.745340in}{1.385400in}}%
\pgfpathcurveto{\pgfqpoint{0.759606in}{1.399666in}}{\pgfqpoint{0.767621in}{1.419017in}}{\pgfqpoint{0.767621in}{1.439192in}}%
\pgfpathcurveto{\pgfqpoint{0.767621in}{1.459366in}}{\pgfqpoint{0.759606in}{1.478717in}}{\pgfqpoint{0.745340in}{1.492983in}}%
\pgfpathcurveto{\pgfqpoint{0.731075in}{1.507249in}}{\pgfqpoint{0.711724in}{1.515264in}}{\pgfqpoint{0.691549in}{1.515264in}}%
\pgfpathcurveto{\pgfqpoint{0.671374in}{1.515264in}}{\pgfqpoint{0.652023in}{1.507249in}}{\pgfqpoint{0.637757in}{1.492983in}}%
\pgfpathcurveto{\pgfqpoint{0.623492in}{1.478717in}}{\pgfqpoint{0.615476in}{1.459366in}}{\pgfqpoint{0.615476in}{1.439192in}}%
\pgfpathcurveto{\pgfqpoint{0.615476in}{1.419017in}}{\pgfqpoint{0.623492in}{1.399666in}}{\pgfqpoint{0.637757in}{1.385400in}}%
\pgfpathcurveto{\pgfqpoint{0.652023in}{1.371134in}}{\pgfqpoint{0.671374in}{1.363119in}}{\pgfqpoint{0.691549in}{1.363119in}}%
\pgfpathlineto{\pgfqpoint{0.691549in}{1.363119in}}%
\pgfpathclose%
\pgfusepath{stroke}%
\end{pgfscope}%
\begin{pgfscope}%
\pgfsetbuttcap%
\pgfsetroundjoin%
\definecolor{currentfill}{rgb}{0.000000,0.000000,0.000000}%
\pgfsetfillcolor{currentfill}%
\pgfsetlinewidth{0.803000pt}%
\definecolor{currentstroke}{rgb}{0.000000,0.000000,0.000000}%
\pgfsetstrokecolor{currentstroke}%
\pgfsetdash{}{0pt}%
\pgfsys@defobject{currentmarker}{\pgfqpoint{0.000000in}{0.000000in}}{\pgfqpoint{0.000000in}{0.048611in}}{%
\pgfpathmoveto{\pgfqpoint{0.000000in}{0.000000in}}%
\pgfpathlineto{\pgfqpoint{0.000000in}{0.048611in}}%
\pgfusepath{stroke,fill}%
}%
\begin{pgfscope}%
\pgfsys@transformshift{0.606165in}{3.030372in}%
\pgfsys@useobject{currentmarker}{}%
\end{pgfscope}%
\end{pgfscope}%
\begin{pgfscope}%
\definecolor{textcolor}{rgb}{0.000000,0.000000,0.000000}%
\pgfsetstrokecolor{textcolor}%
\pgfsetfillcolor{textcolor}%
\pgftext[x=0.606165in,y=3.127594in,,bottom]{\color{textcolor}{\rmfamily\fontsize{16.000000}{18.00000}\selectfont\catcode`\^=\active\def^{\ifmmode\sp\else\^{}\fi}\catcode`\%=\active\def
\end{pgfscope}%
\begin{pgfscope}%
\pgfsetbuttcap%
\pgfsetroundjoin%
\definecolor{currentfill}{rgb}{0.000000,0.000000,0.000000}%
\pgfsetfillcolor{currentfill}%
\pgfsetlinewidth{0.803000pt}%
\definecolor{currentstroke}{rgb}{0.000000,0.000000,0.000000}%
\pgfsetstrokecolor{currentstroke}%
\pgfsetdash{}{0pt}%
\pgfsys@defobject{currentmarker}{\pgfqpoint{0.000000in}{0.000000in}}{\pgfqpoint{0.000000in}{0.048611in}}{%
\pgfpathmoveto{\pgfqpoint{0.000000in}{0.000000in}}%
\pgfpathlineto{\pgfqpoint{0.000000in}{0.048611in}}%
\pgfusepath{stroke,fill}%
}%
\begin{pgfscope}%
\pgfsys@transformshift{1.212412in}{3.030372in}%
\pgfsys@useobject{currentmarker}{}%
\end{pgfscope}%
\end{pgfscope}%
\begin{pgfscope}%
\pgfsetbuttcap%
\pgfsetroundjoin%
\definecolor{currentfill}{rgb}{0.000000,0.000000,0.000000}%
\pgfsetfillcolor{currentfill}%
\pgfsetlinewidth{0.803000pt}%
\definecolor{currentstroke}{rgb}{0.000000,0.000000,0.000000}%
\pgfsetstrokecolor{currentstroke}%
\pgfsetdash{}{0pt}%
\pgfsys@defobject{currentmarker}{\pgfqpoint{0.000000in}{0.000000in}}{\pgfqpoint{0.000000in}{0.048611in}}{%
\pgfpathmoveto{\pgfqpoint{0.000000in}{0.000000in}}%
\pgfpathlineto{\pgfqpoint{0.000000in}{0.048611in}}%
\pgfusepath{stroke,fill}%
}%
\begin{pgfscope}%
\pgfsys@transformshift{1.818658in}{3.030372in}%
\pgfsys@useobject{currentmarker}{}%
\end{pgfscope}%
\end{pgfscope}%
\begin{pgfscope}%
\pgfsetbuttcap%
\pgfsetroundjoin%
\definecolor{currentfill}{rgb}{0.000000,0.000000,0.000000}%
\pgfsetfillcolor{currentfill}%
\pgfsetlinewidth{0.803000pt}%
\definecolor{currentstroke}{rgb}{0.000000,0.000000,0.000000}%
\pgfsetstrokecolor{currentstroke}%
\pgfsetdash{}{0pt}%
\pgfsys@defobject{currentmarker}{\pgfqpoint{0.000000in}{0.000000in}}{\pgfqpoint{0.000000in}{0.048611in}}{%
\pgfpathmoveto{\pgfqpoint{0.000000in}{0.000000in}}%
\pgfpathlineto{\pgfqpoint{0.000000in}{0.048611in}}%
\pgfusepath{stroke,fill}%
}%
\begin{pgfscope}%
\pgfsys@transformshift{2.424905in}{3.030372in}%
\pgfsys@useobject{currentmarker}{}%
\end{pgfscope}%
\end{pgfscope}%
\begin{pgfscope}%
\pgfsetbuttcap%
\pgfsetroundjoin%
\definecolor{currentfill}{rgb}{0.000000,0.000000,0.000000}%
\pgfsetfillcolor{currentfill}%
\pgfsetlinewidth{0.803000pt}%
\definecolor{currentstroke}{rgb}{0.000000,0.000000,0.000000}%
\pgfsetstrokecolor{currentstroke}%
\pgfsetdash{}{0pt}%
\pgfsys@defobject{currentmarker}{\pgfqpoint{0.000000in}{0.000000in}}{\pgfqpoint{0.000000in}{0.048611in}}{%
\pgfpathmoveto{\pgfqpoint{0.000000in}{0.000000in}}%
\pgfpathlineto{\pgfqpoint{0.000000in}{0.048611in}}%
\pgfusepath{stroke,fill}%
}%
\begin{pgfscope}%
\pgfsys@transformshift{3.031152in}{3.030372in}%
\pgfsys@useobject{currentmarker}{}%
\end{pgfscope}%
\end{pgfscope}%
\begin{pgfscope}%
\definecolor{textcolor}{rgb}{0.000000,0.000000,0.000000}%
\pgfsetstrokecolor{textcolor}%
\pgfsetfillcolor{textcolor}%
\pgftext[x=3.031152in,y=3.127594in,,bottom]{\color{textcolor}{\rmfamily\fontsize{16.000000}{18.00000}\selectfont\catcode`\^=\active\def^{\ifmmode\sp\else\^{}\fi}\catcode`\%=\active\def
\end{pgfscope}%
\begin{pgfscope}%
\definecolor{textcolor}{rgb}{0.000000,0.000000,0.000000}%
\pgfsetstrokecolor{textcolor}%
\pgfsetfillcolor{textcolor}%
\pgftext[x=1.818658in,y=3.266483in,,base]{\color{textcolor}{\rmfamily\fontsize{16.000000}{18.00000}\selectfont\catcode`\^=\active\def^{\ifmmode\sp\else\^{}\fi}\catcode`\%=\active\def
\end{pgfscope}%
\begin{pgfscope}%
\pgfsetbuttcap%
\pgfsetroundjoin%
\definecolor{currentfill}{rgb}{0.000000,0.000000,0.000000}%
\pgfsetfillcolor{currentfill}%
\pgfsetlinewidth{0.803000pt}%
\definecolor{currentstroke}{rgb}{0.000000,0.000000,0.000000}%
\pgfsetstrokecolor{currentstroke}%
\pgfsetdash{}{0pt}%
\pgfsys@defobject{currentmarker}{\pgfqpoint{-0.048611in}{0.000000in}}{\pgfqpoint{-0.000000in}{0.000000in}}{%
\pgfpathmoveto{\pgfqpoint{-0.000000in}{0.000000in}}%
\pgfpathlineto{\pgfqpoint{-0.048611in}{0.000000in}}%
\pgfusepath{stroke,fill}%
}%
\begin{pgfscope}%
\pgfsys@transformshift{0.266667in}{0.175231in}%
\pgfsys@useobject{currentmarker}{}%
\end{pgfscope}%
\end{pgfscope}%
\begin{pgfscope}%
\definecolor{textcolor}{rgb}{1.000000,1.000000,1.000000}%
\pgfsetstrokecolor{textcolor}%
\pgfsetfillcolor{textcolor}%
\pgftext[x=0.100000in, y=0.127006in, left, base]{\color{textcolor}{\rmfamily\fontsize{16.000000}{18.00000}\selectfont\catcode`\^=\active\def^{\ifmmode\sp\else\^{}\fi}\catcode`\%=\active\def
\end{pgfscope}%
\begin{pgfscope}%
\pgfsetbuttcap%
\pgfsetroundjoin%
\definecolor{currentfill}{rgb}{0.000000,0.000000,0.000000}%
\pgfsetfillcolor{currentfill}%
\pgfsetlinewidth{0.803000pt}%
\definecolor{currentstroke}{rgb}{0.000000,0.000000,0.000000}%
\pgfsetstrokecolor{currentstroke}%
\pgfsetdash{}{0pt}%
\pgfsys@defobject{currentmarker}{\pgfqpoint{-0.048611in}{0.000000in}}{\pgfqpoint{-0.000000in}{0.000000in}}{%
\pgfpathmoveto{\pgfqpoint{-0.000000in}{0.000000in}}%
\pgfpathlineto{\pgfqpoint{-0.048611in}{0.000000in}}%
\pgfusepath{stroke,fill}%
}%
\begin{pgfscope}%
\pgfsys@transformshift{0.266667in}{0.732876in}%
\pgfsys@useobject{currentmarker}{}%
\end{pgfscope}%
\end{pgfscope}%
\begin{pgfscope}%
\pgfsetbuttcap%
\pgfsetroundjoin%
\definecolor{currentfill}{rgb}{0.000000,0.000000,0.000000}%
\pgfsetfillcolor{currentfill}%
\pgfsetlinewidth{0.803000pt}%
\definecolor{currentstroke}{rgb}{0.000000,0.000000,0.000000}%
\pgfsetstrokecolor{currentstroke}%
\pgfsetdash{}{0pt}%
\pgfsys@defobject{currentmarker}{\pgfqpoint{-0.048611in}{0.000000in}}{\pgfqpoint{-0.000000in}{0.000000in}}{%
\pgfpathmoveto{\pgfqpoint{-0.000000in}{0.000000in}}%
\pgfpathlineto{\pgfqpoint{-0.048611in}{0.000000in}}%
\pgfusepath{stroke,fill}%
}%
\begin{pgfscope}%
\pgfsys@transformshift{0.266667in}{1.290521in}%
\pgfsys@useobject{currentmarker}{}%
\end{pgfscope}%
\end{pgfscope}%
\begin{pgfscope}%
\pgfsetbuttcap%
\pgfsetroundjoin%
\definecolor{currentfill}{rgb}{0.000000,0.000000,0.000000}%
\pgfsetfillcolor{currentfill}%
\pgfsetlinewidth{0.803000pt}%
\definecolor{currentstroke}{rgb}{0.000000,0.000000,0.000000}%
\pgfsetstrokecolor{currentstroke}%
\pgfsetdash{}{0pt}%
\pgfsys@defobject{currentmarker}{\pgfqpoint{-0.048611in}{0.000000in}}{\pgfqpoint{-0.000000in}{0.000000in}}{%
\pgfpathmoveto{\pgfqpoint{-0.000000in}{0.000000in}}%
\pgfpathlineto{\pgfqpoint{-0.048611in}{0.000000in}}%
\pgfusepath{stroke,fill}%
}%
\begin{pgfscope}%
\pgfsys@transformshift{0.266667in}{1.848165in}%
\pgfsys@useobject{currentmarker}{}%
\end{pgfscope}%
\end{pgfscope}%
\begin{pgfscope}%
\pgfsetbuttcap%
\pgfsetroundjoin%
\definecolor{currentfill}{rgb}{0.000000,0.000000,0.000000}%
\pgfsetfillcolor{currentfill}%
\pgfsetlinewidth{0.803000pt}%
\definecolor{currentstroke}{rgb}{0.000000,0.000000,0.000000}%
\pgfsetstrokecolor{currentstroke}%
\pgfsetdash{}{0pt}%
\pgfsys@defobject{currentmarker}{\pgfqpoint{-0.048611in}{0.000000in}}{\pgfqpoint{-0.000000in}{0.000000in}}{%
\pgfpathmoveto{\pgfqpoint{-0.000000in}{0.000000in}}%
\pgfpathlineto{\pgfqpoint{-0.048611in}{0.000000in}}%
\pgfusepath{stroke,fill}%
}%
\begin{pgfscope}%
\pgfsys@transformshift{0.266667in}{2.405810in}%
\pgfsys@useobject{currentmarker}{}%
\end{pgfscope}%
\end{pgfscope}%
\begin{pgfscope}%
\pgfsetrectcap%
\pgfsetmiterjoin%
\pgfsetlinewidth{0.803000pt}%
\definecolor{currentstroke}{rgb}{0.000000,0.000000,0.000000}%
\pgfsetstrokecolor{currentstroke}%
\pgfsetdash{}{0pt}%
\pgfpathmoveto{\pgfqpoint{0.266667in}{0.175231in}}%
\pgfpathlineto{\pgfqpoint{0.266667in}{3.030372in}}%
\pgfusepath{stroke}%
\end{pgfscope}%
\begin{pgfscope}%
\pgfsetrectcap%
\pgfsetmiterjoin%
\pgfsetlinewidth{0.803000pt}%
\definecolor{currentstroke}{rgb}{0.000000,0.000000,0.000000}%
\pgfsetstrokecolor{currentstroke}%
\pgfsetdash{}{0pt}%
\pgfpathmoveto{\pgfqpoint{3.370650in}{0.175231in}}%
\pgfpathlineto{\pgfqpoint{3.370650in}{3.030372in}}%
\pgfusepath{stroke}%
\end{pgfscope}%
\begin{pgfscope}%
\pgfsetrectcap%
\pgfsetmiterjoin%
\pgfsetlinewidth{0.803000pt}%
\definecolor{currentstroke}{rgb}{0.000000,0.000000,0.000000}%
\pgfsetstrokecolor{currentstroke}%
\pgfsetdash{}{0pt}%
\pgfpathmoveto{\pgfqpoint{0.266667in}{0.175231in}}%
\pgfpathlineto{\pgfqpoint{3.370650in}{0.175231in}}%
\pgfusepath{stroke}%
\end{pgfscope}%
\begin{pgfscope}%
\pgfsetrectcap%
\pgfsetmiterjoin%
\pgfsetlinewidth{0.803000pt}%
\definecolor{currentstroke}{rgb}{0.000000,0.000000,0.000000}%
\pgfsetstrokecolor{currentstroke}%
\pgfsetdash{}{0pt}%
\pgfpathmoveto{\pgfqpoint{0.266667in}{3.030372in}}%
\pgfpathlineto{\pgfqpoint{3.370650in}{3.030372in}}%
\pgfusepath{stroke}%
\end{pgfscope}%
\end{pgfpicture}%
\makeatother%
\endgroup%

%% file: figures/inference/snr=0.pgf
\begingroup%
\makeatletter%
\begin{pgfpicture}%
\pgfpathrectangle{\pgfpointorigin}{\pgfqpoint{3.470650in}{3.470650in}}%
\pgfusepath{use as bounding box, clip}%
\begin{pgfscope}%
\pgfsetbuttcap%
\pgfsetmiterjoin%
\definecolor{currentfill}{rgb}{1.000000,1.000000,1.000000}%
\pgfsetfillcolor{currentfill}%
\pgfsetlinewidth{0.000000pt}%
\definecolor{currentstroke}{rgb}{1.000000,1.000000,1.000000}%
\pgfsetstrokecolor{currentstroke}%
\pgfsetdash{}{0pt}%
\pgfpathmoveto{\pgfqpoint{0.000000in}{0.000000in}}%
\pgfpathlineto{\pgfqpoint{3.470650in}{0.000000in}}%
\pgfpathlineto{\pgfqpoint{3.470650in}{3.470650in}}%
\pgfpathlineto{\pgfqpoint{0.000000in}{3.470650in}}%
\pgfpathlineto{\pgfqpoint{0.000000in}{0.000000in}}%
\pgfpathclose%
\pgfusepath{fill}%
\end{pgfscope}%
\begin{pgfscope}%
\pgfsetbuttcap%
\pgfsetmiterjoin%
\definecolor{currentfill}{rgb}{1.000000,1.000000,1.000000}%
\pgfsetfillcolor{currentfill}%
\pgfsetlinewidth{0.000000pt}%
\definecolor{currentstroke}{rgb}{0.000000,0.000000,0.000000}%
\pgfsetstrokecolor{currentstroke}%
\pgfsetstrokeopacity{0.000000}%
\pgfsetdash{}{0pt}%
\pgfpathmoveto{\pgfqpoint{0.266667in}{0.175231in}}%
\pgfpathlineto{\pgfqpoint{3.370650in}{0.175231in}}%
\pgfpathlineto{\pgfqpoint{3.370650in}{3.030372in}}%
\pgfpathlineto{\pgfqpoint{0.266667in}{3.030372in}}%
\pgfpathlineto{\pgfqpoint{0.266667in}{0.175231in}}%
\pgfpathclose%
\pgfusepath{fill}%
\end{pgfscope}%
\begin{pgfscope}%
\pgfpathrectangle{\pgfqpoint{0.266667in}{0.175231in}}{\pgfqpoint{3.103983in}{2.855141in}}%
\pgfusepath{clip}%
\pgfsys@transformshift{0.266667in}{0.175231in}%
\pgftext[left,bottom]{\includegraphics[interpolate=true,width=3.110000in,height=2.860000in]{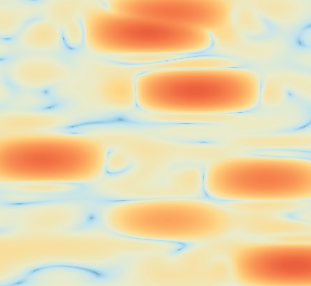}}%
\end{pgfscope}%
\begin{pgfscope}%
\pgfpathrectangle{\pgfqpoint{0.266667in}{0.175231in}}{\pgfqpoint{3.103983in}{2.855141in}}%
\pgfusepath{clip}%
\pgfsetbuttcap%
\pgfsetroundjoin%
\pgfsetlinewidth{2.007500pt}%
\definecolor{currentstroke}{rgb}{1.000000,1.000000,1.000000}%
\pgfsetstrokecolor{currentstroke}%
\pgfsetdash{}{0pt}%
\pgfpathmoveto{\pgfqpoint{1.737590in}{2.630829in}}%
\pgfpathcurveto{\pgfqpoint{1.757765in}{2.630829in}}{\pgfqpoint{1.777116in}{2.638845in}}{\pgfqpoint{1.791382in}{2.653110in}}%
\pgfpathcurveto{\pgfqpoint{1.805648in}{2.667376in}}{\pgfqpoint{1.813663in}{2.686727in}}{\pgfqpoint{1.813663in}{2.706902in}}%
\pgfpathcurveto{\pgfqpoint{1.813663in}{2.727076in}}{\pgfqpoint{1.805648in}{2.746428in}}{\pgfqpoint{1.791382in}{2.760693in}}%
\pgfpathcurveto{\pgfqpoint{1.777116in}{2.774959in}}{\pgfqpoint{1.757765in}{2.782974in}}{\pgfqpoint{1.737590in}{2.782974in}}%
\pgfpathcurveto{\pgfqpoint{1.717416in}{2.782974in}}{\pgfqpoint{1.698065in}{2.774959in}}{\pgfqpoint{1.683799in}{2.760693in}}%
\pgfpathcurveto{\pgfqpoint{1.669533in}{2.746428in}}{\pgfqpoint{1.661518in}{2.727076in}}{\pgfqpoint{1.661518in}{2.706902in}}%
\pgfpathcurveto{\pgfqpoint{1.661518in}{2.686727in}}{\pgfqpoint{1.669533in}{2.667376in}}{\pgfqpoint{1.683799in}{2.653110in}}%
\pgfpathcurveto{\pgfqpoint{1.698065in}{2.638845in}}{\pgfqpoint{1.717416in}{2.630829in}}{\pgfqpoint{1.737590in}{2.630829in}}%
\pgfpathlineto{\pgfqpoint{1.737590in}{2.630829in}}%
\pgfpathclose%
\pgfusepath{stroke}%
\end{pgfscope}%
\begin{pgfscope}%
\pgfpathrectangle{\pgfqpoint{0.266667in}{0.175231in}}{\pgfqpoint{3.103983in}{2.855141in}}%
\pgfusepath{clip}%
\pgfsetbuttcap%
\pgfsetroundjoin%
\pgfsetlinewidth{2.007500pt}%
\definecolor{currentstroke}{rgb}{1.000000,1.000000,1.000000}%
\pgfsetstrokecolor{currentstroke}%
\pgfsetdash{}{0pt}%
\pgfpathmoveto{\pgfqpoint{1.914370in}{0.752429in}}%
\pgfpathcurveto{\pgfqpoint{1.934545in}{0.752429in}}{\pgfqpoint{1.953896in}{0.760445in}}{\pgfqpoint{1.968162in}{0.774710in}}%
\pgfpathcurveto{\pgfqpoint{1.982427in}{0.788976in}}{\pgfqpoint{1.990443in}{0.808327in}}{\pgfqpoint{1.990443in}{0.828502in}}%
\pgfpathcurveto{\pgfqpoint{1.990443in}{0.848676in}}{\pgfqpoint{1.982427in}{0.868027in}}{\pgfqpoint{1.968162in}{0.882293in}}%
\pgfpathcurveto{\pgfqpoint{1.953896in}{0.896559in}}{\pgfqpoint{1.934545in}{0.904574in}}{\pgfqpoint{1.914370in}{0.904574in}}%
\pgfpathcurveto{\pgfqpoint{1.894195in}{0.904574in}}{\pgfqpoint{1.874844in}{0.896559in}}{\pgfqpoint{1.860579in}{0.882293in}}%
\pgfpathcurveto{\pgfqpoint{1.846313in}{0.868027in}}{\pgfqpoint{1.838298in}{0.848676in}}{\pgfqpoint{1.838298in}{0.828502in}}%
\pgfpathcurveto{\pgfqpoint{1.838298in}{0.808327in}}{\pgfqpoint{1.846313in}{0.788976in}}{\pgfqpoint{1.860579in}{0.774710in}}%
\pgfpathcurveto{\pgfqpoint{1.874844in}{0.760445in}}{\pgfqpoint{1.894195in}{0.752429in}}{\pgfqpoint{1.914370in}{0.752429in}}%
\pgfpathlineto{\pgfqpoint{1.914370in}{0.752429in}}%
\pgfpathclose%
\pgfusepath{stroke}%
\end{pgfscope}%
\begin{pgfscope}%
\pgfpathrectangle{\pgfqpoint{0.266667in}{0.175231in}}{\pgfqpoint{3.103983in}{2.855141in}}%
\pgfusepath{clip}%
\pgfsetbuttcap%
\pgfsetroundjoin%
\pgfsetlinewidth{2.007500pt}%
\definecolor{currentstroke}{rgb}{1.000000,1.000000,1.000000}%
\pgfsetstrokecolor{currentstroke}%
\pgfsetdash{}{0pt}%
\pgfpathmoveto{\pgfqpoint{2.916507in}{1.161008in}}%
\pgfpathcurveto{\pgfqpoint{2.936682in}{1.161008in}}{\pgfqpoint{2.956033in}{1.169023in}}{\pgfqpoint{2.970299in}{1.183289in}}%
\pgfpathcurveto{\pgfqpoint{2.984564in}{1.197554in}}{\pgfqpoint{2.992580in}{1.216906in}}{\pgfqpoint{2.992580in}{1.237080in}}%
\pgfpathcurveto{\pgfqpoint{2.992580in}{1.257255in}}{\pgfqpoint{2.984564in}{1.276606in}}{\pgfqpoint{2.970299in}{1.290872in}}%
\pgfpathcurveto{\pgfqpoint{2.956033in}{1.305137in}}{\pgfqpoint{2.936682in}{1.313153in}}{\pgfqpoint{2.916507in}{1.313153in}}%
\pgfpathcurveto{\pgfqpoint{2.896333in}{1.313153in}}{\pgfqpoint{2.876981in}{1.305137in}}{\pgfqpoint{2.862716in}{1.290872in}}%
\pgfpathcurveto{\pgfqpoint{2.848450in}{1.276606in}}{\pgfqpoint{2.840435in}{1.257255in}}{\pgfqpoint{2.840435in}{1.237080in}}%
\pgfpathcurveto{\pgfqpoint{2.840435in}{1.216906in}}{\pgfqpoint{2.848450in}{1.197554in}}{\pgfqpoint{2.862716in}{1.183289in}}%
\pgfpathcurveto{\pgfqpoint{2.876981in}{1.169023in}}{\pgfqpoint{2.896333in}{1.161008in}}{\pgfqpoint{2.916507in}{1.161008in}}%
\pgfpathlineto{\pgfqpoint{2.916507in}{1.161008in}}%
\pgfpathclose%
\pgfusepath{stroke}%
\end{pgfscope}%
\begin{pgfscope}%
\pgfpathrectangle{\pgfqpoint{0.266667in}{0.175231in}}{\pgfqpoint{3.103983in}{2.855141in}}%
\pgfusepath{clip}%
\pgfsetbuttcap%
\pgfsetroundjoin%
\pgfsetlinewidth{2.007500pt}%
\definecolor{currentstroke}{rgb}{1.000000,1.000000,1.000000}%
\pgfsetstrokecolor{currentstroke}%
\pgfsetdash{}{0pt}%
\pgfpathmoveto{\pgfqpoint{2.211039in}{2.044105in}}%
\pgfpathcurveto{\pgfqpoint{2.231214in}{2.044105in}}{\pgfqpoint{2.250565in}{2.052121in}}{\pgfqpoint{2.264831in}{2.066386in}}%
\pgfpathcurveto{\pgfqpoint{2.279096in}{2.080652in}}{\pgfqpoint{2.287112in}{2.100003in}}{\pgfqpoint{2.287112in}{2.120178in}}%
\pgfpathcurveto{\pgfqpoint{2.287112in}{2.140352in}}{\pgfqpoint{2.279096in}{2.159703in}}{\pgfqpoint{2.264831in}{2.173969in}}%
\pgfpathcurveto{\pgfqpoint{2.250565in}{2.188235in}}{\pgfqpoint{2.231214in}{2.196250in}}{\pgfqpoint{2.211039in}{2.196250in}}%
\pgfpathcurveto{\pgfqpoint{2.190864in}{2.196250in}}{\pgfqpoint{2.171513in}{2.188235in}}{\pgfqpoint{2.157248in}{2.173969in}}%
\pgfpathcurveto{\pgfqpoint{2.142982in}{2.159703in}}{\pgfqpoint{2.134967in}{2.140352in}}{\pgfqpoint{2.134967in}{2.120178in}}%
\pgfpathcurveto{\pgfqpoint{2.134967in}{2.100003in}}{\pgfqpoint{2.142982in}{2.080652in}}{\pgfqpoint{2.157248in}{2.066386in}}%
\pgfpathcurveto{\pgfqpoint{2.171513in}{2.052121in}}{\pgfqpoint{2.190864in}{2.044105in}}{\pgfqpoint{2.211039in}{2.044105in}}%
\pgfpathlineto{\pgfqpoint{2.211039in}{2.044105in}}%
\pgfpathclose%
\pgfusepath{stroke}%
\end{pgfscope}%
\begin{pgfscope}%
\pgfpathrectangle{\pgfqpoint{0.266667in}{0.175231in}}{\pgfqpoint{3.103983in}{2.855141in}}%
\pgfusepath{clip}%
\pgfsetbuttcap%
\pgfsetroundjoin%
\pgfsetlinewidth{2.007500pt}%
\definecolor{currentstroke}{rgb}{1.000000,1.000000,1.000000}%
\pgfsetstrokecolor{currentstroke}%
\pgfsetdash{}{0pt}%
\pgfpathmoveto{\pgfqpoint{0.710681in}{1.409253in}}%
\pgfpathcurveto{\pgfqpoint{0.730856in}{1.409253in}}{\pgfqpoint{0.750207in}{1.417269in}}{\pgfqpoint{0.764473in}{1.431534in}}%
\pgfpathcurveto{\pgfqpoint{0.778738in}{1.445800in}}{\pgfqpoint{0.786754in}{1.465151in}}{\pgfqpoint{0.786754in}{1.485326in}}%
\pgfpathcurveto{\pgfqpoint{0.786754in}{1.505500in}}{\pgfqpoint{0.778738in}{1.524851in}}{\pgfqpoint{0.764473in}{1.539117in}}%
\pgfpathcurveto{\pgfqpoint{0.750207in}{1.553383in}}{\pgfqpoint{0.730856in}{1.561398in}}{\pgfqpoint{0.710681in}{1.561398in}}%
\pgfpathcurveto{\pgfqpoint{0.690507in}{1.561398in}}{\pgfqpoint{0.671156in}{1.553383in}}{\pgfqpoint{0.656890in}{1.539117in}}%
\pgfpathcurveto{\pgfqpoint{0.642624in}{1.524851in}}{\pgfqpoint{0.634609in}{1.505500in}}{\pgfqpoint{0.634609in}{1.485326in}}%
\pgfpathcurveto{\pgfqpoint{0.634609in}{1.465151in}}{\pgfqpoint{0.642624in}{1.445800in}}{\pgfqpoint{0.656890in}{1.431534in}}%
\pgfpathcurveto{\pgfqpoint{0.671156in}{1.417269in}}{\pgfqpoint{0.690507in}{1.409253in}}{\pgfqpoint{0.710681in}{1.409253in}}%
\pgfpathlineto{\pgfqpoint{0.710681in}{1.409253in}}%
\pgfpathclose%
\pgfusepath{stroke}%
\end{pgfscope}%
\begin{pgfscope}%
\pgfpathrectangle{\pgfqpoint{0.266667in}{0.175231in}}{\pgfqpoint{3.103983in}{2.855141in}}%
\pgfusepath{clip}%
\pgfsetbuttcap%
\pgfsetroundjoin%
\pgfsetlinewidth{2.007500pt}%
\definecolor{currentstroke}{rgb}{1.000000,1.000000,1.000000}%
\pgfsetstrokecolor{currentstroke}%
\pgfsetdash{}{0pt}%
\pgfpathmoveto{\pgfqpoint{1.959151in}{2.813919in}}%
\pgfpathcurveto{\pgfqpoint{1.979326in}{2.813919in}}{\pgfqpoint{1.998677in}{2.821935in}}{\pgfqpoint{2.012943in}{2.836200in}}%
\pgfpathcurveto{\pgfqpoint{2.027209in}{2.850466in}}{\pgfqpoint{2.035224in}{2.869817in}}{\pgfqpoint{2.035224in}{2.889992in}}%
\pgfpathcurveto{\pgfqpoint{2.035224in}{2.910167in}}{\pgfqpoint{2.027209in}{2.929518in}}{\pgfqpoint{2.012943in}{2.943783in}}%
\pgfpathcurveto{\pgfqpoint{1.998677in}{2.958049in}}{\pgfqpoint{1.979326in}{2.966064in}}{\pgfqpoint{1.959151in}{2.966064in}}%
\pgfpathcurveto{\pgfqpoint{1.938977in}{2.966064in}}{\pgfqpoint{1.919626in}{2.958049in}}{\pgfqpoint{1.905360in}{2.943783in}}%
\pgfpathcurveto{\pgfqpoint{1.891094in}{2.929518in}}{\pgfqpoint{1.883079in}{2.910167in}}{\pgfqpoint{1.883079in}{2.889992in}}%
\pgfpathcurveto{\pgfqpoint{1.883079in}{2.869817in}}{\pgfqpoint{1.891094in}{2.850466in}}{\pgfqpoint{1.905360in}{2.836200in}}%
\pgfpathcurveto{\pgfqpoint{1.919626in}{2.821935in}}{\pgfqpoint{1.938977in}{2.813919in}}{\pgfqpoint{1.959151in}{2.813919in}}%
\pgfpathlineto{\pgfqpoint{1.959151in}{2.813919in}}%
\pgfpathclose%
\pgfusepath{stroke}%
\end{pgfscope}%
\begin{pgfscope}%
\pgfpathrectangle{\pgfqpoint{0.266667in}{0.175231in}}{\pgfqpoint{3.103983in}{2.855141in}}%
\pgfusepath{clip}%
\pgfsetbuttcap%
\pgfsetroundjoin%
\pgfsetlinewidth{2.007500pt}%
\definecolor{currentstroke}{rgb}{1.000000,1.000000,1.000000}%
\pgfsetstrokecolor{currentstroke}%
\pgfsetdash{}{0pt}%
\pgfpathmoveto{\pgfqpoint{0.683984in}{1.341354in}}%
\pgfpathcurveto{\pgfqpoint{0.704159in}{1.341354in}}{\pgfqpoint{0.723510in}{1.349369in}}{\pgfqpoint{0.737776in}{1.363635in}}%
\pgfpathcurveto{\pgfqpoint{0.752041in}{1.377900in}}{\pgfqpoint{0.760057in}{1.397252in}}{\pgfqpoint{0.760057in}{1.417426in}}%
\pgfpathcurveto{\pgfqpoint{0.760057in}{1.437601in}}{\pgfqpoint{0.752041in}{1.456952in}}{\pgfqpoint{0.737776in}{1.471218in}}%
\pgfpathcurveto{\pgfqpoint{0.723510in}{1.485483in}}{\pgfqpoint{0.704159in}{1.493499in}}{\pgfqpoint{0.683984in}{1.493499in}}%
\pgfpathcurveto{\pgfqpoint{0.663810in}{1.493499in}}{\pgfqpoint{0.644459in}{1.485483in}}{\pgfqpoint{0.630193in}{1.471218in}}%
\pgfpathcurveto{\pgfqpoint{0.615927in}{1.456952in}}{\pgfqpoint{0.607912in}{1.437601in}}{\pgfqpoint{0.607912in}{1.417426in}}%
\pgfpathcurveto{\pgfqpoint{0.607912in}{1.397252in}}{\pgfqpoint{0.615927in}{1.377900in}}{\pgfqpoint{0.630193in}{1.363635in}}%
\pgfpathcurveto{\pgfqpoint{0.644459in}{1.349369in}}{\pgfqpoint{0.663810in}{1.341354in}}{\pgfqpoint{0.683984in}{1.341354in}}%
\pgfpathlineto{\pgfqpoint{0.683984in}{1.341354in}}%
\pgfpathclose%
\pgfusepath{stroke}%
\end{pgfscope}%
\begin{pgfscope}%
\pgfpathrectangle{\pgfqpoint{0.266667in}{0.175231in}}{\pgfqpoint{3.103983in}{2.855141in}}%
\pgfusepath{clip}%
\pgfsetbuttcap%
\pgfsetroundjoin%
\pgfsetlinewidth{2.007500pt}%
\definecolor{currentstroke}{rgb}{1.000000,1.000000,1.000000}%
\pgfsetstrokecolor{currentstroke}%
\pgfsetdash{}{0pt}%
\pgfpathmoveto{\pgfqpoint{3.208961in}{0.302426in}}%
\pgfpathcurveto{\pgfqpoint{3.229136in}{0.302426in}}{\pgfqpoint{3.248487in}{0.310441in}}{\pgfqpoint{3.262753in}{0.324707in}}%
\pgfpathcurveto{\pgfqpoint{3.277018in}{0.338973in}}{\pgfqpoint{3.285034in}{0.358324in}}{\pgfqpoint{3.285034in}{0.378499in}}%
\pgfpathcurveto{\pgfqpoint{3.285034in}{0.398673in}}{\pgfqpoint{3.277018in}{0.418024in}}{\pgfqpoint{3.262753in}{0.432290in}}%
\pgfpathcurveto{\pgfqpoint{3.248487in}{0.446556in}}{\pgfqpoint{3.229136in}{0.454571in}}{\pgfqpoint{3.208961in}{0.454571in}}%
\pgfpathcurveto{\pgfqpoint{3.188787in}{0.454571in}}{\pgfqpoint{3.169435in}{0.446556in}}{\pgfqpoint{3.155170in}{0.432290in}}%
\pgfpathcurveto{\pgfqpoint{3.140904in}{0.418024in}}{\pgfqpoint{3.132889in}{0.398673in}}{\pgfqpoint{3.132889in}{0.378499in}}%
\pgfpathcurveto{\pgfqpoint{3.132889in}{0.358324in}}{\pgfqpoint{3.140904in}{0.338973in}}{\pgfqpoint{3.155170in}{0.324707in}}%
\pgfpathcurveto{\pgfqpoint{3.169435in}{0.310441in}}{\pgfqpoint{3.188787in}{0.302426in}}{\pgfqpoint{3.208961in}{0.302426in}}%
\pgfpathlineto{\pgfqpoint{3.208961in}{0.302426in}}%
\pgfpathclose%
\pgfusepath{stroke}%
\end{pgfscope}%
\begin{pgfscope}%
\pgfpathrectangle{\pgfqpoint{0.266667in}{0.175231in}}{\pgfqpoint{3.103983in}{2.855141in}}%
\pgfusepath{clip}%
\pgfsetbuttcap%
\pgfsetroundjoin%
\pgfsetlinewidth{2.007500pt}%
\definecolor{currentstroke}{rgb}{0.000000,0.000000,0.000000}%
\pgfsetstrokecolor{currentstroke}%
\pgfsetdash{}{0pt}%
\pgfpathmoveto{\pgfqpoint{1.735693in}{2.651742in}}%
\pgfpathcurveto{\pgfqpoint{1.755867in}{2.651742in}}{\pgfqpoint{1.775218in}{2.659758in}}{\pgfqpoint{1.789484in}{2.674023in}}%
\pgfpathcurveto{\pgfqpoint{1.803750in}{2.688289in}}{\pgfqpoint{1.811765in}{2.707640in}}{\pgfqpoint{1.811765in}{2.727815in}}%
\pgfpathcurveto{\pgfqpoint{1.811765in}{2.747989in}}{\pgfqpoint{1.803750in}{2.767341in}}{\pgfqpoint{1.789484in}{2.781606in}}%
\pgfpathcurveto{\pgfqpoint{1.775218in}{2.795872in}}{\pgfqpoint{1.755867in}{2.803887in}}{\pgfqpoint{1.735693in}{2.803887in}}%
\pgfpathcurveto{\pgfqpoint{1.715518in}{2.803887in}}{\pgfqpoint{1.696167in}{2.795872in}}{\pgfqpoint{1.681901in}{2.781606in}}%
\pgfpathcurveto{\pgfqpoint{1.667636in}{2.767341in}}{\pgfqpoint{1.659620in}{2.747989in}}{\pgfqpoint{1.659620in}{2.727815in}}%
\pgfpathcurveto{\pgfqpoint{1.659620in}{2.707640in}}{\pgfqpoint{1.667636in}{2.688289in}}{\pgfqpoint{1.681901in}{2.674023in}}%
\pgfpathcurveto{\pgfqpoint{1.696167in}{2.659758in}}{\pgfqpoint{1.715518in}{2.651742in}}{\pgfqpoint{1.735693in}{2.651742in}}%
\pgfpathlineto{\pgfqpoint{1.735693in}{2.651742in}}%
\pgfpathclose%
\pgfusepath{stroke}%
\end{pgfscope}%
\begin{pgfscope}%
\pgfpathrectangle{\pgfqpoint{0.266667in}{0.175231in}}{\pgfqpoint{3.103983in}{2.855141in}}%
\pgfusepath{clip}%
\pgfsetbuttcap%
\pgfsetroundjoin%
\pgfsetlinewidth{2.007500pt}%
\definecolor{currentstroke}{rgb}{0.000000,0.000000,0.000000}%
\pgfsetstrokecolor{currentstroke}%
\pgfsetdash{}{0pt}%
\pgfpathmoveto{\pgfqpoint{1.920759in}{0.752302in}}%
\pgfpathcurveto{\pgfqpoint{1.940933in}{0.752302in}}{\pgfqpoint{1.960284in}{0.760318in}}{\pgfqpoint{1.974550in}{0.774584in}}%
\pgfpathcurveto{\pgfqpoint{1.988816in}{0.788849in}}{\pgfqpoint{1.996831in}{0.808200in}}{\pgfqpoint{1.996831in}{0.828375in}}%
\pgfpathcurveto{\pgfqpoint{1.996831in}{0.848550in}}{\pgfqpoint{1.988816in}{0.867901in}}{\pgfqpoint{1.974550in}{0.882166in}}%
\pgfpathcurveto{\pgfqpoint{1.960284in}{0.896432in}}{\pgfqpoint{1.940933in}{0.904448in}}{\pgfqpoint{1.920759in}{0.904448in}}%
\pgfpathcurveto{\pgfqpoint{1.900584in}{0.904448in}}{\pgfqpoint{1.881233in}{0.896432in}}{\pgfqpoint{1.866967in}{0.882166in}}%
\pgfpathcurveto{\pgfqpoint{1.852701in}{0.867901in}}{\pgfqpoint{1.844686in}{0.848550in}}{\pgfqpoint{1.844686in}{0.828375in}}%
\pgfpathcurveto{\pgfqpoint{1.844686in}{0.808200in}}{\pgfqpoint{1.852701in}{0.788849in}}{\pgfqpoint{1.866967in}{0.774584in}}%
\pgfpathcurveto{\pgfqpoint{1.881233in}{0.760318in}}{\pgfqpoint{1.900584in}{0.752302in}}{\pgfqpoint{1.920759in}{0.752302in}}%
\pgfpathlineto{\pgfqpoint{1.920759in}{0.752302in}}%
\pgfpathclose%
\pgfusepath{stroke}%
\end{pgfscope}%
\begin{pgfscope}%
\pgfpathrectangle{\pgfqpoint{0.266667in}{0.175231in}}{\pgfqpoint{3.103983in}{2.855141in}}%
\pgfusepath{clip}%
\pgfsetbuttcap%
\pgfsetroundjoin%
\pgfsetlinewidth{2.007500pt}%
\definecolor{currentstroke}{rgb}{0.000000,0.000000,0.000000}%
\pgfsetstrokecolor{currentstroke}%
\pgfsetdash{}{0pt}%
\pgfpathmoveto{\pgfqpoint{1.952683in}{2.821781in}}%
\pgfpathcurveto{\pgfqpoint{1.972857in}{2.821781in}}{\pgfqpoint{1.992209in}{2.829796in}}{\pgfqpoint{2.006474in}{2.844062in}}%
\pgfpathcurveto{\pgfqpoint{2.020740in}{2.858328in}}{\pgfqpoint{2.028755in}{2.877679in}}{\pgfqpoint{2.028755in}{2.897854in}}%
\pgfpathcurveto{\pgfqpoint{2.028755in}{2.918028in}}{\pgfqpoint{2.020740in}{2.937379in}}{\pgfqpoint{2.006474in}{2.951645in}}%
\pgfpathcurveto{\pgfqpoint{1.992209in}{2.965911in}}{\pgfqpoint{1.972857in}{2.973926in}}{\pgfqpoint{1.952683in}{2.973926in}}%
\pgfpathcurveto{\pgfqpoint{1.932508in}{2.973926in}}{\pgfqpoint{1.913157in}{2.965911in}}{\pgfqpoint{1.898891in}{2.951645in}}%
\pgfpathcurveto{\pgfqpoint{1.884626in}{2.937379in}}{\pgfqpoint{1.876610in}{2.918028in}}{\pgfqpoint{1.876610in}{2.897854in}}%
\pgfpathcurveto{\pgfqpoint{1.876610in}{2.877679in}}{\pgfqpoint{1.884626in}{2.858328in}}{\pgfqpoint{1.898891in}{2.844062in}}%
\pgfpathcurveto{\pgfqpoint{1.913157in}{2.829796in}}{\pgfqpoint{1.932508in}{2.821781in}}{\pgfqpoint{1.952683in}{2.821781in}}%
\pgfpathlineto{\pgfqpoint{1.952683in}{2.821781in}}%
\pgfpathclose%
\pgfusepath{stroke}%
\end{pgfscope}%
\begin{pgfscope}%
\pgfpathrectangle{\pgfqpoint{0.266667in}{0.175231in}}{\pgfqpoint{3.103983in}{2.855141in}}%
\pgfusepath{clip}%
\pgfsetbuttcap%
\pgfsetroundjoin%
\pgfsetlinewidth{2.007500pt}%
\definecolor{currentstroke}{rgb}{0.000000,0.000000,0.000000}%
\pgfsetstrokecolor{currentstroke}%
\pgfsetdash{}{0pt}%
\pgfpathmoveto{\pgfqpoint{3.214807in}{0.298719in}}%
\pgfpathcurveto{\pgfqpoint{3.234982in}{0.298719in}}{\pgfqpoint{3.254333in}{0.306735in}}{\pgfqpoint{3.268598in}{0.321001in}}%
\pgfpathcurveto{\pgfqpoint{3.282864in}{0.335266in}}{\pgfqpoint{3.290879in}{0.354617in}}{\pgfqpoint{3.290879in}{0.374792in}}%
\pgfpathcurveto{\pgfqpoint{3.290879in}{0.394967in}}{\pgfqpoint{3.282864in}{0.414318in}}{\pgfqpoint{3.268598in}{0.428583in}}%
\pgfpathcurveto{\pgfqpoint{3.254333in}{0.442849in}}{\pgfqpoint{3.234982in}{0.450865in}}{\pgfqpoint{3.214807in}{0.450865in}}%
\pgfpathcurveto{\pgfqpoint{3.194632in}{0.450865in}}{\pgfqpoint{3.175281in}{0.442849in}}{\pgfqpoint{3.161015in}{0.428583in}}%
\pgfpathcurveto{\pgfqpoint{3.146750in}{0.414318in}}{\pgfqpoint{3.138734in}{0.394967in}}{\pgfqpoint{3.138734in}{0.374792in}}%
\pgfpathcurveto{\pgfqpoint{3.138734in}{0.354617in}}{\pgfqpoint{3.146750in}{0.335266in}}{\pgfqpoint{3.161015in}{0.321001in}}%
\pgfpathcurveto{\pgfqpoint{3.175281in}{0.306735in}}{\pgfqpoint{3.194632in}{0.298719in}}{\pgfqpoint{3.214807in}{0.298719in}}%
\pgfpathlineto{\pgfqpoint{3.214807in}{0.298719in}}%
\pgfpathclose%
\pgfusepath{stroke}%
\end{pgfscope}%
\begin{pgfscope}%
\pgfpathrectangle{\pgfqpoint{0.266667in}{0.175231in}}{\pgfqpoint{3.103983in}{2.855141in}}%
\pgfusepath{clip}%
\pgfsetbuttcap%
\pgfsetroundjoin%
\pgfsetlinewidth{2.007500pt}%
\definecolor{currentstroke}{rgb}{0.000000,0.000000,0.000000}%
\pgfsetstrokecolor{currentstroke}%
\pgfsetdash{}{0pt}%
\pgfpathmoveto{\pgfqpoint{2.922304in}{1.155878in}}%
\pgfpathcurveto{\pgfqpoint{2.942478in}{1.155878in}}{\pgfqpoint{2.961829in}{1.163893in}}{\pgfqpoint{2.976095in}{1.178159in}}%
\pgfpathcurveto{\pgfqpoint{2.990361in}{1.192424in}}{\pgfqpoint{2.998376in}{1.211775in}}{\pgfqpoint{2.998376in}{1.231950in}}%
\pgfpathcurveto{\pgfqpoint{2.998376in}{1.252125in}}{\pgfqpoint{2.990361in}{1.271476in}}{\pgfqpoint{2.976095in}{1.285742in}}%
\pgfpathcurveto{\pgfqpoint{2.961829in}{1.300007in}}{\pgfqpoint{2.942478in}{1.308023in}}{\pgfqpoint{2.922304in}{1.308023in}}%
\pgfpathcurveto{\pgfqpoint{2.902129in}{1.308023in}}{\pgfqpoint{2.882778in}{1.300007in}}{\pgfqpoint{2.868512in}{1.285742in}}%
\pgfpathcurveto{\pgfqpoint{2.854246in}{1.271476in}}{\pgfqpoint{2.846231in}{1.252125in}}{\pgfqpoint{2.846231in}{1.231950in}}%
\pgfpathcurveto{\pgfqpoint{2.846231in}{1.211775in}}{\pgfqpoint{2.854246in}{1.192424in}}{\pgfqpoint{2.868512in}{1.178159in}}%
\pgfpathcurveto{\pgfqpoint{2.882778in}{1.163893in}}{\pgfqpoint{2.902129in}{1.155878in}}{\pgfqpoint{2.922304in}{1.155878in}}%
\pgfpathlineto{\pgfqpoint{2.922304in}{1.155878in}}%
\pgfpathclose%
\pgfusepath{stroke}%
\end{pgfscope}%
\begin{pgfscope}%
\pgfpathrectangle{\pgfqpoint{0.266667in}{0.175231in}}{\pgfqpoint{3.103983in}{2.855141in}}%
\pgfusepath{clip}%
\pgfsetbuttcap%
\pgfsetroundjoin%
\pgfsetlinewidth{2.007500pt}%
\definecolor{currentstroke}{rgb}{0.000000,0.000000,0.000000}%
\pgfsetstrokecolor{currentstroke}%
\pgfsetdash{}{0pt}%
\pgfpathmoveto{\pgfqpoint{0.701637in}{1.431565in}}%
\pgfpathcurveto{\pgfqpoint{0.721811in}{1.431565in}}{\pgfqpoint{0.741162in}{1.439580in}}{\pgfqpoint{0.755428in}{1.453846in}}%
\pgfpathcurveto{\pgfqpoint{0.769694in}{1.468112in}}{\pgfqpoint{0.777709in}{1.487463in}}{\pgfqpoint{0.777709in}{1.507637in}}%
\pgfpathcurveto{\pgfqpoint{0.777709in}{1.527812in}}{\pgfqpoint{0.769694in}{1.547163in}}{\pgfqpoint{0.755428in}{1.561429in}}%
\pgfpathcurveto{\pgfqpoint{0.741162in}{1.575695in}}{\pgfqpoint{0.721811in}{1.583710in}}{\pgfqpoint{0.701637in}{1.583710in}}%
\pgfpathcurveto{\pgfqpoint{0.681462in}{1.583710in}}{\pgfqpoint{0.662111in}{1.575695in}}{\pgfqpoint{0.647845in}{1.561429in}}%
\pgfpathcurveto{\pgfqpoint{0.633579in}{1.547163in}}{\pgfqpoint{0.625564in}{1.527812in}}{\pgfqpoint{0.625564in}{1.507637in}}%
\pgfpathcurveto{\pgfqpoint{0.625564in}{1.487463in}}{\pgfqpoint{0.633579in}{1.468112in}}{\pgfqpoint{0.647845in}{1.453846in}}%
\pgfpathcurveto{\pgfqpoint{0.662111in}{1.439580in}}{\pgfqpoint{0.681462in}{1.431565in}}{\pgfqpoint{0.701637in}{1.431565in}}%
\pgfpathlineto{\pgfqpoint{0.701637in}{1.431565in}}%
\pgfpathclose%
\pgfusepath{stroke}%
\end{pgfscope}%
\begin{pgfscope}%
\pgfpathrectangle{\pgfqpoint{0.266667in}{0.175231in}}{\pgfqpoint{3.103983in}{2.855141in}}%
\pgfusepath{clip}%
\pgfsetbuttcap%
\pgfsetroundjoin%
\pgfsetlinewidth{2.007500pt}%
\definecolor{currentstroke}{rgb}{0.000000,0.000000,0.000000}%
\pgfsetstrokecolor{currentstroke}%
\pgfsetdash{}{0pt}%
\pgfpathmoveto{\pgfqpoint{2.214871in}{2.042137in}}%
\pgfpathcurveto{\pgfqpoint{2.235046in}{2.042137in}}{\pgfqpoint{2.254397in}{2.050152in}}{\pgfqpoint{2.268662in}{2.064418in}}%
\pgfpathcurveto{\pgfqpoint{2.282928in}{2.078684in}}{\pgfqpoint{2.290943in}{2.098035in}}{\pgfqpoint{2.290943in}{2.118209in}}%
\pgfpathcurveto{\pgfqpoint{2.290943in}{2.138384in}}{\pgfqpoint{2.282928in}{2.157735in}}{\pgfqpoint{2.268662in}{2.172001in}}%
\pgfpathcurveto{\pgfqpoint{2.254397in}{2.186266in}}{\pgfqpoint{2.235046in}{2.194282in}}{\pgfqpoint{2.214871in}{2.194282in}}%
\pgfpathcurveto{\pgfqpoint{2.194696in}{2.194282in}}{\pgfqpoint{2.175345in}{2.186266in}}{\pgfqpoint{2.161079in}{2.172001in}}%
\pgfpathcurveto{\pgfqpoint{2.146814in}{2.157735in}}{\pgfqpoint{2.138798in}{2.138384in}}{\pgfqpoint{2.138798in}{2.118209in}}%
\pgfpathcurveto{\pgfqpoint{2.138798in}{2.098035in}}{\pgfqpoint{2.146814in}{2.078684in}}{\pgfqpoint{2.161079in}{2.064418in}}%
\pgfpathcurveto{\pgfqpoint{2.175345in}{2.050152in}}{\pgfqpoint{2.194696in}{2.042137in}}{\pgfqpoint{2.214871in}{2.042137in}}%
\pgfpathlineto{\pgfqpoint{2.214871in}{2.042137in}}%
\pgfpathclose%
\pgfusepath{stroke}%
\end{pgfscope}%
\begin{pgfscope}%
\pgfpathrectangle{\pgfqpoint{0.266667in}{0.175231in}}{\pgfqpoint{3.103983in}{2.855141in}}%
\pgfusepath{clip}%
\pgfsetbuttcap%
\pgfsetroundjoin%
\pgfsetlinewidth{2.007500pt}%
\definecolor{currentstroke}{rgb}{0.000000,0.000000,0.000000}%
\pgfsetstrokecolor{currentstroke}%
\pgfsetdash{}{0pt}%
\pgfpathmoveto{\pgfqpoint{0.693346in}{1.331341in}}%
\pgfpathcurveto{\pgfqpoint{0.713521in}{1.331341in}}{\pgfqpoint{0.732872in}{1.339357in}}{\pgfqpoint{0.747138in}{1.353622in}}%
\pgfpathcurveto{\pgfqpoint{0.761403in}{1.367888in}}{\pgfqpoint{0.769419in}{1.387239in}}{\pgfqpoint{0.769419in}{1.407414in}}%
\pgfpathcurveto{\pgfqpoint{0.769419in}{1.427589in}}{\pgfqpoint{0.761403in}{1.446940in}}{\pgfqpoint{0.747138in}{1.461205in}}%
\pgfpathcurveto{\pgfqpoint{0.732872in}{1.475471in}}{\pgfqpoint{0.713521in}{1.483486in}}{\pgfqpoint{0.693346in}{1.483486in}}%
\pgfpathcurveto{\pgfqpoint{0.673172in}{1.483486in}}{\pgfqpoint{0.653820in}{1.475471in}}{\pgfqpoint{0.639555in}{1.461205in}}%
\pgfpathcurveto{\pgfqpoint{0.625289in}{1.446940in}}{\pgfqpoint{0.617274in}{1.427589in}}{\pgfqpoint{0.617274in}{1.407414in}}%
\pgfpathcurveto{\pgfqpoint{0.617274in}{1.387239in}}{\pgfqpoint{0.625289in}{1.367888in}}{\pgfqpoint{0.639555in}{1.353622in}}%
\pgfpathcurveto{\pgfqpoint{0.653820in}{1.339357in}}{\pgfqpoint{0.673172in}{1.331341in}}{\pgfqpoint{0.693346in}{1.331341in}}%
\pgfpathlineto{\pgfqpoint{0.693346in}{1.331341in}}%
\pgfpathclose%
\pgfusepath{stroke}%
\end{pgfscope}%
\begin{pgfscope}%
\pgfsetbuttcap%
\pgfsetroundjoin%
\definecolor{currentfill}{rgb}{0.000000,0.000000,0.000000}%
\pgfsetfillcolor{currentfill}%
\pgfsetlinewidth{0.803000pt}%
\definecolor{currentstroke}{rgb}{0.000000,0.000000,0.000000}%
\pgfsetstrokecolor{currentstroke}%
\pgfsetdash{}{0pt}%
\pgfsys@defobject{currentmarker}{\pgfqpoint{0.000000in}{0.000000in}}{\pgfqpoint{0.000000in}{0.048611in}}{%
\pgfpathmoveto{\pgfqpoint{0.000000in}{0.000000in}}%
\pgfpathlineto{\pgfqpoint{0.000000in}{0.048611in}}%
\pgfusepath{stroke,fill}%
}%
\begin{pgfscope}%
\pgfsys@transformshift{0.606165in}{3.030372in}%
\pgfsys@useobject{currentmarker}{}%
\end{pgfscope}%
\end{pgfscope}%
\begin{pgfscope}%
\definecolor{textcolor}{rgb}{0.000000,0.000000,0.000000}%
\pgfsetstrokecolor{textcolor}%
\pgfsetfillcolor{textcolor}%
\pgftext[x=0.606165in,y=3.127594in,,bottom]{\color{textcolor}{\rmfamily\fontsize{16.000000}{18.00000}\selectfont\catcode`\^=\active\def^{\ifmmode\sp\else\^{}\fi}\catcode`\%=\active\def
\end{pgfscope}%
\begin{pgfscope}%
\pgfsetbuttcap%
\pgfsetroundjoin%
\definecolor{currentfill}{rgb}{0.000000,0.000000,0.000000}%
\pgfsetfillcolor{currentfill}%
\pgfsetlinewidth{0.803000pt}%
\definecolor{currentstroke}{rgb}{0.000000,0.000000,0.000000}%
\pgfsetstrokecolor{currentstroke}%
\pgfsetdash{}{0pt}%
\pgfsys@defobject{currentmarker}{\pgfqpoint{0.000000in}{0.000000in}}{\pgfqpoint{0.000000in}{0.048611in}}{%
\pgfpathmoveto{\pgfqpoint{0.000000in}{0.000000in}}%
\pgfpathlineto{\pgfqpoint{0.000000in}{0.048611in}}%
\pgfusepath{stroke,fill}%
}%
\begin{pgfscope}%
\pgfsys@transformshift{1.212412in}{3.030372in}%
\pgfsys@useobject{currentmarker}{}%
\end{pgfscope}%
\end{pgfscope}%
\begin{pgfscope}%
\pgfsetbuttcap%
\pgfsetroundjoin%
\definecolor{currentfill}{rgb}{0.000000,0.000000,0.000000}%
\pgfsetfillcolor{currentfill}%
\pgfsetlinewidth{0.803000pt}%
\definecolor{currentstroke}{rgb}{0.000000,0.000000,0.000000}%
\pgfsetstrokecolor{currentstroke}%
\pgfsetdash{}{0pt}%
\pgfsys@defobject{currentmarker}{\pgfqpoint{0.000000in}{0.000000in}}{\pgfqpoint{0.000000in}{0.048611in}}{%
\pgfpathmoveto{\pgfqpoint{0.000000in}{0.000000in}}%
\pgfpathlineto{\pgfqpoint{0.000000in}{0.048611in}}%
\pgfusepath{stroke,fill}%
}%
\begin{pgfscope}%
\pgfsys@transformshift{1.818658in}{3.030372in}%
\pgfsys@useobject{currentmarker}{}%
\end{pgfscope}%
\end{pgfscope}%
\begin{pgfscope}%
\pgfsetbuttcap%
\pgfsetroundjoin%
\definecolor{currentfill}{rgb}{0.000000,0.000000,0.000000}%
\pgfsetfillcolor{currentfill}%
\pgfsetlinewidth{0.803000pt}%
\definecolor{currentstroke}{rgb}{0.000000,0.000000,0.000000}%
\pgfsetstrokecolor{currentstroke}%
\pgfsetdash{}{0pt}%
\pgfsys@defobject{currentmarker}{\pgfqpoint{0.000000in}{0.000000in}}{\pgfqpoint{0.000000in}{0.048611in}}{%
\pgfpathmoveto{\pgfqpoint{0.000000in}{0.000000in}}%
\pgfpathlineto{\pgfqpoint{0.000000in}{0.048611in}}%
\pgfusepath{stroke,fill}%
}%
\begin{pgfscope}%
\pgfsys@transformshift{2.424905in}{3.030372in}%
\pgfsys@useobject{currentmarker}{}%
\end{pgfscope}%
\end{pgfscope}%
\begin{pgfscope}%
\pgfsetbuttcap%
\pgfsetroundjoin%
\definecolor{currentfill}{rgb}{0.000000,0.000000,0.000000}%
\pgfsetfillcolor{currentfill}%
\pgfsetlinewidth{0.803000pt}%
\definecolor{currentstroke}{rgb}{0.000000,0.000000,0.000000}%
\pgfsetstrokecolor{currentstroke}%
\pgfsetdash{}{0pt}%
\pgfsys@defobject{currentmarker}{\pgfqpoint{0.000000in}{0.000000in}}{\pgfqpoint{0.000000in}{0.048611in}}{%
\pgfpathmoveto{\pgfqpoint{0.000000in}{0.000000in}}%
\pgfpathlineto{\pgfqpoint{0.000000in}{0.048611in}}%
\pgfusepath{stroke,fill}%
}%
\begin{pgfscope}%
\pgfsys@transformshift{3.031152in}{3.030372in}%
\pgfsys@useobject{currentmarker}{}%
\end{pgfscope}%
\end{pgfscope}%
\begin{pgfscope}%
\definecolor{textcolor}{rgb}{0.000000,0.000000,0.000000}%
\pgfsetstrokecolor{textcolor}%
\pgfsetfillcolor{textcolor}%
\pgftext[x=3.031152in,y=3.127594in,,bottom]{\color{textcolor}{\rmfamily\fontsize{16.000000}{18.00000}\selectfont\catcode`\^=\active\def^{\ifmmode\sp\else\^{}\fi}\catcode`\%=\active\def
\end{pgfscope}%
\begin{pgfscope}%
\definecolor{textcolor}{rgb}{0.000000,0.000000,0.000000}%
\pgfsetstrokecolor{textcolor}%
\pgfsetfillcolor{textcolor}%
\pgftext[x=1.818658in,y=3.266483in,,base]{\color{textcolor}{\rmfamily\fontsize{16.000000}{18.00000}\selectfont\catcode`\^=\active\def^{\ifmmode\sp\else\^{}\fi}\catcode`\%=\active\def
\end{pgfscope}%
\begin{pgfscope}%
\pgfsetbuttcap%
\pgfsetroundjoin%
\definecolor{currentfill}{rgb}{0.000000,0.000000,0.000000}%
\pgfsetfillcolor{currentfill}%
\pgfsetlinewidth{0.803000pt}%
\definecolor{currentstroke}{rgb}{0.000000,0.000000,0.000000}%
\pgfsetstrokecolor{currentstroke}%
\pgfsetdash{}{0pt}%
\pgfsys@defobject{currentmarker}{\pgfqpoint{-0.048611in}{0.000000in}}{\pgfqpoint{-0.000000in}{0.000000in}}{%
\pgfpathmoveto{\pgfqpoint{-0.000000in}{0.000000in}}%
\pgfpathlineto{\pgfqpoint{-0.048611in}{0.000000in}}%
\pgfusepath{stroke,fill}%
}%
\begin{pgfscope}%
\pgfsys@transformshift{0.266667in}{0.175231in}%
\pgfsys@useobject{currentmarker}{}%
\end{pgfscope}%
\end{pgfscope}%
\begin{pgfscope}%
\definecolor{textcolor}{rgb}{1.000000,1.000000,1.000000}%
\pgfsetstrokecolor{textcolor}%
\pgfsetfillcolor{textcolor}%
\pgftext[x=0.100000in, y=0.127006in, left, base]{\color{textcolor}{\rmfamily\fontsize{16.000000}{18.00000}\selectfont\catcode`\^=\active\def^{\ifmmode\sp\else\^{}\fi}\catcode`\%=\active\def
\end{pgfscope}%
\begin{pgfscope}%
\pgfsetbuttcap%
\pgfsetroundjoin%
\definecolor{currentfill}{rgb}{0.000000,0.000000,0.000000}%
\pgfsetfillcolor{currentfill}%
\pgfsetlinewidth{0.803000pt}%
\definecolor{currentstroke}{rgb}{0.000000,0.000000,0.000000}%
\pgfsetstrokecolor{currentstroke}%
\pgfsetdash{}{0pt}%
\pgfsys@defobject{currentmarker}{\pgfqpoint{-0.048611in}{0.000000in}}{\pgfqpoint{-0.000000in}{0.000000in}}{%
\pgfpathmoveto{\pgfqpoint{-0.000000in}{0.000000in}}%
\pgfpathlineto{\pgfqpoint{-0.048611in}{0.000000in}}%
\pgfusepath{stroke,fill}%
}%
\begin{pgfscope}%
\pgfsys@transformshift{0.266667in}{0.732876in}%
\pgfsys@useobject{currentmarker}{}%
\end{pgfscope}%
\end{pgfscope}%
\begin{pgfscope}%
\pgfsetbuttcap%
\pgfsetroundjoin%
\definecolor{currentfill}{rgb}{0.000000,0.000000,0.000000}%
\pgfsetfillcolor{currentfill}%
\pgfsetlinewidth{0.803000pt}%
\definecolor{currentstroke}{rgb}{0.000000,0.000000,0.000000}%
\pgfsetstrokecolor{currentstroke}%
\pgfsetdash{}{0pt}%
\pgfsys@defobject{currentmarker}{\pgfqpoint{-0.048611in}{0.000000in}}{\pgfqpoint{-0.000000in}{0.000000in}}{%
\pgfpathmoveto{\pgfqpoint{-0.000000in}{0.000000in}}%
\pgfpathlineto{\pgfqpoint{-0.048611in}{0.000000in}}%
\pgfusepath{stroke,fill}%
}%
\begin{pgfscope}%
\pgfsys@transformshift{0.266667in}{1.290521in}%
\pgfsys@useobject{currentmarker}{}%
\end{pgfscope}%
\end{pgfscope}%
\begin{pgfscope}%
\pgfsetbuttcap%
\pgfsetroundjoin%
\definecolor{currentfill}{rgb}{0.000000,0.000000,0.000000}%
\pgfsetfillcolor{currentfill}%
\pgfsetlinewidth{0.803000pt}%
\definecolor{currentstroke}{rgb}{0.000000,0.000000,0.000000}%
\pgfsetstrokecolor{currentstroke}%
\pgfsetdash{}{0pt}%
\pgfsys@defobject{currentmarker}{\pgfqpoint{-0.048611in}{0.000000in}}{\pgfqpoint{-0.000000in}{0.000000in}}{%
\pgfpathmoveto{\pgfqpoint{-0.000000in}{0.000000in}}%
\pgfpathlineto{\pgfqpoint{-0.048611in}{0.000000in}}%
\pgfusepath{stroke,fill}%
}%
\begin{pgfscope}%
\pgfsys@transformshift{0.266667in}{1.848165in}%
\pgfsys@useobject{currentmarker}{}%
\end{pgfscope}%
\end{pgfscope}%
\begin{pgfscope}%
\pgfsetbuttcap%
\pgfsetroundjoin%
\definecolor{currentfill}{rgb}{0.000000,0.000000,0.000000}%
\pgfsetfillcolor{currentfill}%
\pgfsetlinewidth{0.803000pt}%
\definecolor{currentstroke}{rgb}{0.000000,0.000000,0.000000}%
\pgfsetstrokecolor{currentstroke}%
\pgfsetdash{}{0pt}%
\pgfsys@defobject{currentmarker}{\pgfqpoint{-0.048611in}{0.000000in}}{\pgfqpoint{-0.000000in}{0.000000in}}{%
\pgfpathmoveto{\pgfqpoint{-0.000000in}{0.000000in}}%
\pgfpathlineto{\pgfqpoint{-0.048611in}{0.000000in}}%
\pgfusepath{stroke,fill}%
}%
\begin{pgfscope}%
\pgfsys@transformshift{0.266667in}{2.405810in}%
\pgfsys@useobject{currentmarker}{}%
\end{pgfscope}%
\end{pgfscope}%
\begin{pgfscope}%
\pgfsetrectcap%
\pgfsetmiterjoin%
\pgfsetlinewidth{0.803000pt}%
\definecolor{currentstroke}{rgb}{0.000000,0.000000,0.000000}%
\pgfsetstrokecolor{currentstroke}%
\pgfsetdash{}{0pt}%
\pgfpathmoveto{\pgfqpoint{0.266667in}{0.175231in}}%
\pgfpathlineto{\pgfqpoint{0.266667in}{3.030372in}}%
\pgfusepath{stroke}%
\end{pgfscope}%
\begin{pgfscope}%
\pgfsetrectcap%
\pgfsetmiterjoin%
\pgfsetlinewidth{0.803000pt}%
\definecolor{currentstroke}{rgb}{0.000000,0.000000,0.000000}%
\pgfsetstrokecolor{currentstroke}%
\pgfsetdash{}{0pt}%
\pgfpathmoveto{\pgfqpoint{3.370650in}{0.175231in}}%
\pgfpathlineto{\pgfqpoint{3.370650in}{3.030372in}}%
\pgfusepath{stroke}%
\end{pgfscope}%
\begin{pgfscope}%
\pgfsetrectcap%
\pgfsetmiterjoin%
\pgfsetlinewidth{0.803000pt}%
\definecolor{currentstroke}{rgb}{0.000000,0.000000,0.000000}%
\pgfsetstrokecolor{currentstroke}%
\pgfsetdash{}{0pt}%
\pgfpathmoveto{\pgfqpoint{0.266667in}{0.175231in}}%
\pgfpathlineto{\pgfqpoint{3.370650in}{0.175231in}}%
\pgfusepath{stroke}%
\end{pgfscope}%
\begin{pgfscope}%
\pgfsetrectcap%
\pgfsetmiterjoin%
\pgfsetlinewidth{0.803000pt}%
\definecolor{currentstroke}{rgb}{0.000000,0.000000,0.000000}%
\pgfsetstrokecolor{currentstroke}%
\pgfsetdash{}{0pt}%
\pgfpathmoveto{\pgfqpoint{0.266667in}{3.030372in}}%
\pgfpathlineto{\pgfqpoint{3.370650in}{3.030372in}}%
\pgfusepath{stroke}%
\end{pgfscope}%
\end{pgfpicture}%
\makeatother%
\endgroup%

%% file: figures/inference/snr=20.pgf
\begingroup%
\makeatletter%
\begin{pgfpicture}%
\pgfpathrectangle{\pgfpointorigin}{\pgfqpoint{4.006248in}{3.470650in}}%
\pgfusepath{use as bounding box, clip}%
\begin{pgfscope}%
\pgfsetbuttcap%
\pgfsetmiterjoin%
\definecolor{currentfill}{rgb}{1.000000,1.000000,1.000000}%
\pgfsetfillcolor{currentfill}%
\pgfsetlinewidth{0.000000pt}%
\definecolor{currentstroke}{rgb}{1.000000,1.000000,1.000000}%
\pgfsetstrokecolor{currentstroke}%
\pgfsetdash{}{0pt}%
\pgfpathmoveto{\pgfqpoint{0.000000in}{0.000000in}}%
\pgfpathlineto{\pgfqpoint{4.006248in}{0.000000in}}%
\pgfpathlineto{\pgfqpoint{4.006248in}{3.470650in}}%
\pgfpathlineto{\pgfqpoint{0.000000in}{3.470650in}}%
\pgfpathlineto{\pgfqpoint{0.000000in}{0.000000in}}%
\pgfpathclose%
\pgfusepath{fill}%
\end{pgfscope}%
\begin{pgfscope}%
\pgfsetbuttcap%
\pgfsetmiterjoin%
\definecolor{currentfill}{rgb}{1.000000,1.000000,1.000000}%
\pgfsetfillcolor{currentfill}%
\pgfsetlinewidth{0.000000pt}%
\definecolor{currentstroke}{rgb}{0.000000,0.000000,0.000000}%
\pgfsetstrokecolor{currentstroke}%
\pgfsetstrokeopacity{0.000000}%
\pgfsetdash{}{0pt}%
\pgfpathmoveto{\pgfqpoint{0.266667in}{0.175231in}}%
\pgfpathlineto{\pgfqpoint{3.208288in}{0.175231in}}%
\pgfpathlineto{\pgfqpoint{3.208288in}{3.030372in}}%
\pgfpathlineto{\pgfqpoint{0.266667in}{3.030372in}}%
\pgfpathlineto{\pgfqpoint{0.266667in}{0.175231in}}%
\pgfpathclose%
\pgfusepath{fill}%
\end{pgfscope}%
\begin{pgfscope}%
\pgfpathrectangle{\pgfqpoint{0.266667in}{0.175231in}}{\pgfqpoint{2.941621in}{2.855141in}}%
\pgfusepath{clip}%
\pgfsys@transformshift{0.266667in}{0.175231in}%
\pgftext[left,bottom]{\includegraphics[interpolate=true,width=2.950000in,height=2.860000in]{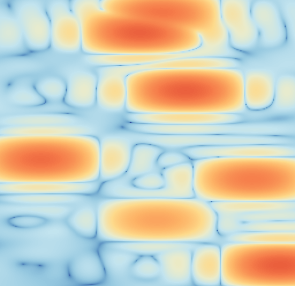}}%
\end{pgfscope}%
\begin{pgfscope}%
\pgfpathrectangle{\pgfqpoint{0.266667in}{0.175231in}}{\pgfqpoint{2.941621in}{2.855141in}}%
\pgfusepath{clip}%
\pgfsetbuttcap%
\pgfsetroundjoin%
\pgfsetlinewidth{2.007500pt}%
\definecolor{currentstroke}{rgb}{1.000000,1.000000,1.000000}%
\pgfsetstrokecolor{currentstroke}%
\pgfsetdash{}{0pt}%
\pgfpathmoveto{\pgfqpoint{1.660650in}{2.630829in}}%
\pgfpathcurveto{\pgfqpoint{1.680825in}{2.630829in}}{\pgfqpoint{1.700176in}{2.638845in}}{\pgfqpoint{1.714441in}{2.653110in}}%
\pgfpathcurveto{\pgfqpoint{1.728707in}{2.667376in}}{\pgfqpoint{1.736723in}{2.686727in}}{\pgfqpoint{1.736723in}{2.706902in}}%
\pgfpathcurveto{\pgfqpoint{1.736723in}{2.727076in}}{\pgfqpoint{1.728707in}{2.746428in}}{\pgfqpoint{1.714441in}{2.760693in}}%
\pgfpathcurveto{\pgfqpoint{1.700176in}{2.774959in}}{\pgfqpoint{1.680825in}{2.782974in}}{\pgfqpoint{1.660650in}{2.782974in}}%
\pgfpathcurveto{\pgfqpoint{1.640475in}{2.782974in}}{\pgfqpoint{1.621124in}{2.774959in}}{\pgfqpoint{1.606859in}{2.760693in}}%
\pgfpathcurveto{\pgfqpoint{1.592593in}{2.746428in}}{\pgfqpoint{1.584577in}{2.727076in}}{\pgfqpoint{1.584577in}{2.706902in}}%
\pgfpathcurveto{\pgfqpoint{1.584577in}{2.686727in}}{\pgfqpoint{1.592593in}{2.667376in}}{\pgfqpoint{1.606859in}{2.653110in}}%
\pgfpathcurveto{\pgfqpoint{1.621124in}{2.638845in}}{\pgfqpoint{1.640475in}{2.630829in}}{\pgfqpoint{1.660650in}{2.630829in}}%
\pgfpathlineto{\pgfqpoint{1.660650in}{2.630829in}}%
\pgfpathclose%
\pgfusepath{stroke}%
\end{pgfscope}%
\begin{pgfscope}%
\pgfpathrectangle{\pgfqpoint{0.266667in}{0.175231in}}{\pgfqpoint{2.941621in}{2.855141in}}%
\pgfusepath{clip}%
\pgfsetbuttcap%
\pgfsetroundjoin%
\pgfsetlinewidth{2.007500pt}%
\definecolor{currentstroke}{rgb}{1.000000,1.000000,1.000000}%
\pgfsetstrokecolor{currentstroke}%
\pgfsetdash{}{0pt}%
\pgfpathmoveto{\pgfqpoint{1.828183in}{0.752429in}}%
\pgfpathcurveto{\pgfqpoint{1.848357in}{0.752429in}}{\pgfqpoint{1.867709in}{0.760445in}}{\pgfqpoint{1.881974in}{0.774710in}}%
\pgfpathcurveto{\pgfqpoint{1.896240in}{0.788976in}}{\pgfqpoint{1.904255in}{0.808327in}}{\pgfqpoint{1.904255in}{0.828502in}}%
\pgfpathcurveto{\pgfqpoint{1.904255in}{0.848676in}}{\pgfqpoint{1.896240in}{0.868027in}}{\pgfqpoint{1.881974in}{0.882293in}}%
\pgfpathcurveto{\pgfqpoint{1.867709in}{0.896559in}}{\pgfqpoint{1.848357in}{0.904574in}}{\pgfqpoint{1.828183in}{0.904574in}}%
\pgfpathcurveto{\pgfqpoint{1.808008in}{0.904574in}}{\pgfqpoint{1.788657in}{0.896559in}}{\pgfqpoint{1.774391in}{0.882293in}}%
\pgfpathcurveto{\pgfqpoint{1.760126in}{0.868027in}}{\pgfqpoint{1.752110in}{0.848676in}}{\pgfqpoint{1.752110in}{0.828502in}}%
\pgfpathcurveto{\pgfqpoint{1.752110in}{0.808327in}}{\pgfqpoint{1.760126in}{0.788976in}}{\pgfqpoint{1.774391in}{0.774710in}}%
\pgfpathcurveto{\pgfqpoint{1.788657in}{0.760445in}}{\pgfqpoint{1.808008in}{0.752429in}}{\pgfqpoint{1.828183in}{0.752429in}}%
\pgfpathlineto{\pgfqpoint{1.828183in}{0.752429in}}%
\pgfpathclose%
\pgfusepath{stroke}%
\end{pgfscope}%
\begin{pgfscope}%
\pgfpathrectangle{\pgfqpoint{0.266667in}{0.175231in}}{\pgfqpoint{2.941621in}{2.855141in}}%
\pgfusepath{clip}%
\pgfsetbuttcap%
\pgfsetroundjoin%
\pgfsetlinewidth{2.007500pt}%
\definecolor{currentstroke}{rgb}{1.000000,1.000000,1.000000}%
\pgfsetstrokecolor{currentstroke}%
\pgfsetdash{}{0pt}%
\pgfpathmoveto{\pgfqpoint{2.777901in}{1.161008in}}%
\pgfpathcurveto{\pgfqpoint{2.798075in}{1.161008in}}{\pgfqpoint{2.817426in}{1.169023in}}{\pgfqpoint{2.831692in}{1.183289in}}%
\pgfpathcurveto{\pgfqpoint{2.845958in}{1.197554in}}{\pgfqpoint{2.853973in}{1.216906in}}{\pgfqpoint{2.853973in}{1.237080in}}%
\pgfpathcurveto{\pgfqpoint{2.853973in}{1.257255in}}{\pgfqpoint{2.845958in}{1.276606in}}{\pgfqpoint{2.831692in}{1.290872in}}%
\pgfpathcurveto{\pgfqpoint{2.817426in}{1.305137in}}{\pgfqpoint{2.798075in}{1.313153in}}{\pgfqpoint{2.777901in}{1.313153in}}%
\pgfpathcurveto{\pgfqpoint{2.757726in}{1.313153in}}{\pgfqpoint{2.738375in}{1.305137in}}{\pgfqpoint{2.724109in}{1.290872in}}%
\pgfpathcurveto{\pgfqpoint{2.709843in}{1.276606in}}{\pgfqpoint{2.701828in}{1.257255in}}{\pgfqpoint{2.701828in}{1.237080in}}%
\pgfpathcurveto{\pgfqpoint{2.701828in}{1.216906in}}{\pgfqpoint{2.709843in}{1.197554in}}{\pgfqpoint{2.724109in}{1.183289in}}%
\pgfpathcurveto{\pgfqpoint{2.738375in}{1.169023in}}{\pgfqpoint{2.757726in}{1.161008in}}{\pgfqpoint{2.777901in}{1.161008in}}%
\pgfpathlineto{\pgfqpoint{2.777901in}{1.161008in}}%
\pgfpathclose%
\pgfusepath{stroke}%
\end{pgfscope}%
\begin{pgfscope}%
\pgfpathrectangle{\pgfqpoint{0.266667in}{0.175231in}}{\pgfqpoint{2.941621in}{2.855141in}}%
\pgfusepath{clip}%
\pgfsetbuttcap%
\pgfsetroundjoin%
\pgfsetlinewidth{2.007500pt}%
\definecolor{currentstroke}{rgb}{1.000000,1.000000,1.000000}%
\pgfsetstrokecolor{currentstroke}%
\pgfsetdash{}{0pt}%
\pgfpathmoveto{\pgfqpoint{2.109334in}{2.044105in}}%
\pgfpathcurveto{\pgfqpoint{2.129509in}{2.044105in}}{\pgfqpoint{2.148860in}{2.052121in}}{\pgfqpoint{2.163125in}{2.066386in}}%
\pgfpathcurveto{\pgfqpoint{2.177391in}{2.080652in}}{\pgfqpoint{2.185406in}{2.100003in}}{\pgfqpoint{2.185406in}{2.120178in}}%
\pgfpathcurveto{\pgfqpoint{2.185406in}{2.140352in}}{\pgfqpoint{2.177391in}{2.159703in}}{\pgfqpoint{2.163125in}{2.173969in}}%
\pgfpathcurveto{\pgfqpoint{2.148860in}{2.188235in}}{\pgfqpoint{2.129509in}{2.196250in}}{\pgfqpoint{2.109334in}{2.196250in}}%
\pgfpathcurveto{\pgfqpoint{2.089159in}{2.196250in}}{\pgfqpoint{2.069808in}{2.188235in}}{\pgfqpoint{2.055542in}{2.173969in}}%
\pgfpathcurveto{\pgfqpoint{2.041277in}{2.159703in}}{\pgfqpoint{2.033261in}{2.140352in}}{\pgfqpoint{2.033261in}{2.120178in}}%
\pgfpathcurveto{\pgfqpoint{2.033261in}{2.100003in}}{\pgfqpoint{2.041277in}{2.080652in}}{\pgfqpoint{2.055542in}{2.066386in}}%
\pgfpathcurveto{\pgfqpoint{2.069808in}{2.052121in}}{\pgfqpoint{2.089159in}{2.044105in}}{\pgfqpoint{2.109334in}{2.044105in}}%
\pgfpathlineto{\pgfqpoint{2.109334in}{2.044105in}}%
\pgfpathclose%
\pgfusepath{stroke}%
\end{pgfscope}%
\begin{pgfscope}%
\pgfpathrectangle{\pgfqpoint{0.266667in}{0.175231in}}{\pgfqpoint{2.941621in}{2.855141in}}%
\pgfusepath{clip}%
\pgfsetbuttcap%
\pgfsetroundjoin%
\pgfsetlinewidth{2.007500pt}%
\definecolor{currentstroke}{rgb}{1.000000,1.000000,1.000000}%
\pgfsetstrokecolor{currentstroke}%
\pgfsetdash{}{0pt}%
\pgfpathmoveto{\pgfqpoint{0.687456in}{1.409253in}}%
\pgfpathcurveto{\pgfqpoint{0.707631in}{1.409253in}}{\pgfqpoint{0.726982in}{1.417269in}}{\pgfqpoint{0.741248in}{1.431534in}}%
\pgfpathcurveto{\pgfqpoint{0.755513in}{1.445800in}}{\pgfqpoint{0.763529in}{1.465151in}}{\pgfqpoint{0.763529in}{1.485326in}}%
\pgfpathcurveto{\pgfqpoint{0.763529in}{1.505500in}}{\pgfqpoint{0.755513in}{1.524851in}}{\pgfqpoint{0.741248in}{1.539117in}}%
\pgfpathcurveto{\pgfqpoint{0.726982in}{1.553383in}}{\pgfqpoint{0.707631in}{1.561398in}}{\pgfqpoint{0.687456in}{1.561398in}}%
\pgfpathcurveto{\pgfqpoint{0.667281in}{1.561398in}}{\pgfqpoint{0.647930in}{1.553383in}}{\pgfqpoint{0.633665in}{1.539117in}}%
\pgfpathcurveto{\pgfqpoint{0.619399in}{1.524851in}}{\pgfqpoint{0.611384in}{1.505500in}}{\pgfqpoint{0.611384in}{1.485326in}}%
\pgfpathcurveto{\pgfqpoint{0.611384in}{1.465151in}}{\pgfqpoint{0.619399in}{1.445800in}}{\pgfqpoint{0.633665in}{1.431534in}}%
\pgfpathcurveto{\pgfqpoint{0.647930in}{1.417269in}}{\pgfqpoint{0.667281in}{1.409253in}}{\pgfqpoint{0.687456in}{1.409253in}}%
\pgfpathlineto{\pgfqpoint{0.687456in}{1.409253in}}%
\pgfpathclose%
\pgfusepath{stroke}%
\end{pgfscope}%
\begin{pgfscope}%
\pgfpathrectangle{\pgfqpoint{0.266667in}{0.175231in}}{\pgfqpoint{2.941621in}{2.855141in}}%
\pgfusepath{clip}%
\pgfsetbuttcap%
\pgfsetroundjoin%
\pgfsetlinewidth{2.007500pt}%
\definecolor{currentstroke}{rgb}{1.000000,1.000000,1.000000}%
\pgfsetstrokecolor{currentstroke}%
\pgfsetdash{}{0pt}%
\pgfpathmoveto{\pgfqpoint{1.870622in}{2.813919in}}%
\pgfpathcurveto{\pgfqpoint{1.890796in}{2.813919in}}{\pgfqpoint{1.910148in}{2.821935in}}{\pgfqpoint{1.924413in}{2.836200in}}%
\pgfpathcurveto{\pgfqpoint{1.938679in}{2.850466in}}{\pgfqpoint{1.946694in}{2.869817in}}{\pgfqpoint{1.946694in}{2.889992in}}%
\pgfpathcurveto{\pgfqpoint{1.946694in}{2.910167in}}{\pgfqpoint{1.938679in}{2.929518in}}{\pgfqpoint{1.924413in}{2.943783in}}%
\pgfpathcurveto{\pgfqpoint{1.910148in}{2.958049in}}{\pgfqpoint{1.890796in}{2.966064in}}{\pgfqpoint{1.870622in}{2.966064in}}%
\pgfpathcurveto{\pgfqpoint{1.850447in}{2.966064in}}{\pgfqpoint{1.831096in}{2.958049in}}{\pgfqpoint{1.816830in}{2.943783in}}%
\pgfpathcurveto{\pgfqpoint{1.802565in}{2.929518in}}{\pgfqpoint{1.794549in}{2.910167in}}{\pgfqpoint{1.794549in}{2.889992in}}%
\pgfpathcurveto{\pgfqpoint{1.794549in}{2.869817in}}{\pgfqpoint{1.802565in}{2.850466in}}{\pgfqpoint{1.816830in}{2.836200in}}%
\pgfpathcurveto{\pgfqpoint{1.831096in}{2.821935in}}{\pgfqpoint{1.850447in}{2.813919in}}{\pgfqpoint{1.870622in}{2.813919in}}%
\pgfpathlineto{\pgfqpoint{1.870622in}{2.813919in}}%
\pgfpathclose%
\pgfusepath{stroke}%
\end{pgfscope}%
\begin{pgfscope}%
\pgfpathrectangle{\pgfqpoint{0.266667in}{0.175231in}}{\pgfqpoint{2.941621in}{2.855141in}}%
\pgfusepath{clip}%
\pgfsetbuttcap%
\pgfsetroundjoin%
\pgfsetlinewidth{2.007500pt}%
\definecolor{currentstroke}{rgb}{1.000000,1.000000,1.000000}%
\pgfsetstrokecolor{currentstroke}%
\pgfsetdash{}{0pt}%
\pgfpathmoveto{\pgfqpoint{0.662155in}{1.341354in}}%
\pgfpathcurveto{\pgfqpoint{0.682330in}{1.341354in}}{\pgfqpoint{0.701681in}{1.349369in}}{\pgfqpoint{0.715947in}{1.363635in}}%
\pgfpathcurveto{\pgfqpoint{0.730213in}{1.377900in}}{\pgfqpoint{0.738228in}{1.397252in}}{\pgfqpoint{0.738228in}{1.417426in}}%
\pgfpathcurveto{\pgfqpoint{0.738228in}{1.437601in}}{\pgfqpoint{0.730213in}{1.456952in}}{\pgfqpoint{0.715947in}{1.471218in}}%
\pgfpathcurveto{\pgfqpoint{0.701681in}{1.485483in}}{\pgfqpoint{0.682330in}{1.493499in}}{\pgfqpoint{0.662155in}{1.493499in}}%
\pgfpathcurveto{\pgfqpoint{0.641981in}{1.493499in}}{\pgfqpoint{0.622630in}{1.485483in}}{\pgfqpoint{0.608364in}{1.471218in}}%
\pgfpathcurveto{\pgfqpoint{0.594098in}{1.456952in}}{\pgfqpoint{0.586083in}{1.437601in}}{\pgfqpoint{0.586083in}{1.417426in}}%
\pgfpathcurveto{\pgfqpoint{0.586083in}{1.397252in}}{\pgfqpoint{0.594098in}{1.377900in}}{\pgfqpoint{0.608364in}{1.363635in}}%
\pgfpathcurveto{\pgfqpoint{0.622630in}{1.349369in}}{\pgfqpoint{0.641981in}{1.341354in}}{\pgfqpoint{0.662155in}{1.341354in}}%
\pgfpathlineto{\pgfqpoint{0.662155in}{1.341354in}}%
\pgfpathclose%
\pgfusepath{stroke}%
\end{pgfscope}%
\begin{pgfscope}%
\pgfpathrectangle{\pgfqpoint{0.266667in}{0.175231in}}{\pgfqpoint{2.941621in}{2.855141in}}%
\pgfusepath{clip}%
\pgfsetbuttcap%
\pgfsetroundjoin%
\pgfsetlinewidth{2.007500pt}%
\definecolor{currentstroke}{rgb}{1.000000,1.000000,1.000000}%
\pgfsetstrokecolor{currentstroke}%
\pgfsetdash{}{0pt}%
\pgfpathmoveto{\pgfqpoint{3.055057in}{0.302426in}}%
\pgfpathcurveto{\pgfqpoint{3.075232in}{0.302426in}}{\pgfqpoint{3.094583in}{0.310441in}}{\pgfqpoint{3.108849in}{0.324707in}}%
\pgfpathcurveto{\pgfqpoint{3.123114in}{0.338973in}}{\pgfqpoint{3.131130in}{0.358324in}}{\pgfqpoint{3.131130in}{0.378499in}}%
\pgfpathcurveto{\pgfqpoint{3.131130in}{0.398673in}}{\pgfqpoint{3.123114in}{0.418024in}}{\pgfqpoint{3.108849in}{0.432290in}}%
\pgfpathcurveto{\pgfqpoint{3.094583in}{0.446556in}}{\pgfqpoint{3.075232in}{0.454571in}}{\pgfqpoint{3.055057in}{0.454571in}}%
\pgfpathcurveto{\pgfqpoint{3.034882in}{0.454571in}}{\pgfqpoint{3.015531in}{0.446556in}}{\pgfqpoint{3.001266in}{0.432290in}}%
\pgfpathcurveto{\pgfqpoint{2.987000in}{0.418024in}}{\pgfqpoint{2.978985in}{0.398673in}}{\pgfqpoint{2.978985in}{0.378499in}}%
\pgfpathcurveto{\pgfqpoint{2.978985in}{0.358324in}}{\pgfqpoint{2.987000in}{0.338973in}}{\pgfqpoint{3.001266in}{0.324707in}}%
\pgfpathcurveto{\pgfqpoint{3.015531in}{0.310441in}}{\pgfqpoint{3.034882in}{0.302426in}}{\pgfqpoint{3.055057in}{0.302426in}}%
\pgfpathlineto{\pgfqpoint{3.055057in}{0.302426in}}%
\pgfpathclose%
\pgfusepath{stroke}%
\end{pgfscope}%
\begin{pgfscope}%
\pgfpathrectangle{\pgfqpoint{0.266667in}{0.175231in}}{\pgfqpoint{2.941621in}{2.855141in}}%
\pgfusepath{clip}%
\pgfsetbuttcap%
\pgfsetroundjoin%
\pgfsetlinewidth{2.007500pt}%
\definecolor{currentstroke}{rgb}{0.000000,0.000000,0.000000}%
\pgfsetstrokecolor{currentstroke}%
\pgfsetdash{}{0pt}%
\pgfpathmoveto{\pgfqpoint{1.825626in}{0.749830in}}%
\pgfpathcurveto{\pgfqpoint{1.845801in}{0.749830in}}{\pgfqpoint{1.865152in}{0.757845in}}{\pgfqpoint{1.879418in}{0.772111in}}%
\pgfpathcurveto{\pgfqpoint{1.893684in}{0.786377in}}{\pgfqpoint{1.901699in}{0.805728in}}{\pgfqpoint{1.901699in}{0.825902in}}%
\pgfpathcurveto{\pgfqpoint{1.901699in}{0.846077in}}{\pgfqpoint{1.893684in}{0.865428in}}{\pgfqpoint{1.879418in}{0.879694in}}%
\pgfpathcurveto{\pgfqpoint{1.865152in}{0.893959in}}{\pgfqpoint{1.845801in}{0.901975in}}{\pgfqpoint{1.825626in}{0.901975in}}%
\pgfpathcurveto{\pgfqpoint{1.805452in}{0.901975in}}{\pgfqpoint{1.786101in}{0.893959in}}{\pgfqpoint{1.771835in}{0.879694in}}%
\pgfpathcurveto{\pgfqpoint{1.757569in}{0.865428in}}{\pgfqpoint{1.749554in}{0.846077in}}{\pgfqpoint{1.749554in}{0.825902in}}%
\pgfpathcurveto{\pgfqpoint{1.749554in}{0.805728in}}{\pgfqpoint{1.757569in}{0.786377in}}{\pgfqpoint{1.771835in}{0.772111in}}%
\pgfpathcurveto{\pgfqpoint{1.786101in}{0.757845in}}{\pgfqpoint{1.805452in}{0.749830in}}{\pgfqpoint{1.825626in}{0.749830in}}%
\pgfpathlineto{\pgfqpoint{1.825626in}{0.749830in}}%
\pgfpathclose%
\pgfusepath{stroke}%
\end{pgfscope}%
\begin{pgfscope}%
\pgfpathrectangle{\pgfqpoint{0.266667in}{0.175231in}}{\pgfqpoint{2.941621in}{2.855141in}}%
\pgfusepath{clip}%
\pgfsetbuttcap%
\pgfsetroundjoin%
\pgfsetlinewidth{2.007500pt}%
\definecolor{currentstroke}{rgb}{0.000000,0.000000,0.000000}%
\pgfsetstrokecolor{currentstroke}%
\pgfsetdash{}{0pt}%
\pgfpathmoveto{\pgfqpoint{2.779708in}{1.154302in}}%
\pgfpathcurveto{\pgfqpoint{2.799883in}{1.154302in}}{\pgfqpoint{2.819234in}{1.162318in}}{\pgfqpoint{2.833499in}{1.176584in}}%
\pgfpathcurveto{\pgfqpoint{2.847765in}{1.190849in}}{\pgfqpoint{2.855781in}{1.210200in}}{\pgfqpoint{2.855781in}{1.230375in}}%
\pgfpathcurveto{\pgfqpoint{2.855781in}{1.250550in}}{\pgfqpoint{2.847765in}{1.269901in}}{\pgfqpoint{2.833499in}{1.284166in}}%
\pgfpathcurveto{\pgfqpoint{2.819234in}{1.298432in}}{\pgfqpoint{2.799883in}{1.306448in}}{\pgfqpoint{2.779708in}{1.306448in}}%
\pgfpathcurveto{\pgfqpoint{2.759533in}{1.306448in}}{\pgfqpoint{2.740182in}{1.298432in}}{\pgfqpoint{2.725917in}{1.284166in}}%
\pgfpathcurveto{\pgfqpoint{2.711651in}{1.269901in}}{\pgfqpoint{2.703635in}{1.250550in}}{\pgfqpoint{2.703635in}{1.230375in}}%
\pgfpathcurveto{\pgfqpoint{2.703635in}{1.210200in}}{\pgfqpoint{2.711651in}{1.190849in}}{\pgfqpoint{2.725917in}{1.176584in}}%
\pgfpathcurveto{\pgfqpoint{2.740182in}{1.162318in}}{\pgfqpoint{2.759533in}{1.154302in}}{\pgfqpoint{2.779708in}{1.154302in}}%
\pgfpathlineto{\pgfqpoint{2.779708in}{1.154302in}}%
\pgfpathclose%
\pgfusepath{stroke}%
\end{pgfscope}%
\begin{pgfscope}%
\pgfpathrectangle{\pgfqpoint{0.266667in}{0.175231in}}{\pgfqpoint{2.941621in}{2.855141in}}%
\pgfusepath{clip}%
\pgfsetbuttcap%
\pgfsetroundjoin%
\pgfsetlinewidth{2.007500pt}%
\definecolor{currentstroke}{rgb}{0.000000,0.000000,0.000000}%
\pgfsetstrokecolor{currentstroke}%
\pgfsetdash{}{0pt}%
\pgfpathmoveto{\pgfqpoint{1.655878in}{2.655804in}}%
\pgfpathcurveto{\pgfqpoint{1.676052in}{2.655804in}}{\pgfqpoint{1.695403in}{2.663819in}}{\pgfqpoint{1.709669in}{2.678085in}}%
\pgfpathcurveto{\pgfqpoint{1.723935in}{2.692350in}}{\pgfqpoint{1.731950in}{2.711701in}}{\pgfqpoint{1.731950in}{2.731876in}}%
\pgfpathcurveto{\pgfqpoint{1.731950in}{2.752051in}}{\pgfqpoint{1.723935in}{2.771402in}}{\pgfqpoint{1.709669in}{2.785668in}}%
\pgfpathcurveto{\pgfqpoint{1.695403in}{2.799933in}}{\pgfqpoint{1.676052in}{2.807949in}}{\pgfqpoint{1.655878in}{2.807949in}}%
\pgfpathcurveto{\pgfqpoint{1.635703in}{2.807949in}}{\pgfqpoint{1.616352in}{2.799933in}}{\pgfqpoint{1.602086in}{2.785668in}}%
\pgfpathcurveto{\pgfqpoint{1.587821in}{2.771402in}}{\pgfqpoint{1.579805in}{2.752051in}}{\pgfqpoint{1.579805in}{2.731876in}}%
\pgfpathcurveto{\pgfqpoint{1.579805in}{2.711701in}}{\pgfqpoint{1.587821in}{2.692350in}}{\pgfqpoint{1.602086in}{2.678085in}}%
\pgfpathcurveto{\pgfqpoint{1.616352in}{2.663819in}}{\pgfqpoint{1.635703in}{2.655804in}}{\pgfqpoint{1.655878in}{2.655804in}}%
\pgfpathlineto{\pgfqpoint{1.655878in}{2.655804in}}%
\pgfpathclose%
\pgfusepath{stroke}%
\end{pgfscope}%
\begin{pgfscope}%
\pgfpathrectangle{\pgfqpoint{0.266667in}{0.175231in}}{\pgfqpoint{2.941621in}{2.855141in}}%
\pgfusepath{clip}%
\pgfsetbuttcap%
\pgfsetroundjoin%
\pgfsetlinewidth{2.007500pt}%
\definecolor{currentstroke}{rgb}{0.000000,0.000000,0.000000}%
\pgfsetstrokecolor{currentstroke}%
\pgfsetdash{}{0pt}%
\pgfpathmoveto{\pgfqpoint{3.059288in}{0.302928in}}%
\pgfpathcurveto{\pgfqpoint{3.079463in}{0.302928in}}{\pgfqpoint{3.098814in}{0.310943in}}{\pgfqpoint{3.113079in}{0.325209in}}%
\pgfpathcurveto{\pgfqpoint{3.127345in}{0.339474in}}{\pgfqpoint{3.135360in}{0.358825in}}{\pgfqpoint{3.135360in}{0.379000in}}%
\pgfpathcurveto{\pgfqpoint{3.135360in}{0.399175in}}{\pgfqpoint{3.127345in}{0.418526in}}{\pgfqpoint{3.113079in}{0.432792in}}%
\pgfpathcurveto{\pgfqpoint{3.098814in}{0.447057in}}{\pgfqpoint{3.079463in}{0.455073in}}{\pgfqpoint{3.059288in}{0.455073in}}%
\pgfpathcurveto{\pgfqpoint{3.039113in}{0.455073in}}{\pgfqpoint{3.019762in}{0.447057in}}{\pgfqpoint{3.005496in}{0.432792in}}%
\pgfpathcurveto{\pgfqpoint{2.991231in}{0.418526in}}{\pgfqpoint{2.983215in}{0.399175in}}{\pgfqpoint{2.983215in}{0.379000in}}%
\pgfpathcurveto{\pgfqpoint{2.983215in}{0.358825in}}{\pgfqpoint{2.991231in}{0.339474in}}{\pgfqpoint{3.005496in}{0.325209in}}%
\pgfpathcurveto{\pgfqpoint{3.019762in}{0.310943in}}{\pgfqpoint{3.039113in}{0.302928in}}{\pgfqpoint{3.059288in}{0.302928in}}%
\pgfpathlineto{\pgfqpoint{3.059288in}{0.302928in}}%
\pgfpathclose%
\pgfusepath{stroke}%
\end{pgfscope}%
\begin{pgfscope}%
\pgfpathrectangle{\pgfqpoint{0.266667in}{0.175231in}}{\pgfqpoint{2.941621in}{2.855141in}}%
\pgfusepath{clip}%
\pgfsetbuttcap%
\pgfsetroundjoin%
\pgfsetlinewidth{2.007500pt}%
\definecolor{currentstroke}{rgb}{0.000000,0.000000,0.000000}%
\pgfsetstrokecolor{currentstroke}%
\pgfsetdash{}{0pt}%
\pgfpathmoveto{\pgfqpoint{1.863964in}{2.817246in}}%
\pgfpathcurveto{\pgfqpoint{1.884138in}{2.817246in}}{\pgfqpoint{1.903489in}{2.825262in}}{\pgfqpoint{1.917755in}{2.839527in}}%
\pgfpathcurveto{\pgfqpoint{1.932021in}{2.853793in}}{\pgfqpoint{1.940036in}{2.873144in}}{\pgfqpoint{1.940036in}{2.893319in}}%
\pgfpathcurveto{\pgfqpoint{1.940036in}{2.913494in}}{\pgfqpoint{1.932021in}{2.932845in}}{\pgfqpoint{1.917755in}{2.947110in}}%
\pgfpathcurveto{\pgfqpoint{1.903489in}{2.961376in}}{\pgfqpoint{1.884138in}{2.969392in}}{\pgfqpoint{1.863964in}{2.969392in}}%
\pgfpathcurveto{\pgfqpoint{1.843789in}{2.969392in}}{\pgfqpoint{1.824438in}{2.961376in}}{\pgfqpoint{1.810172in}{2.947110in}}%
\pgfpathcurveto{\pgfqpoint{1.795907in}{2.932845in}}{\pgfqpoint{1.787891in}{2.913494in}}{\pgfqpoint{1.787891in}{2.893319in}}%
\pgfpathcurveto{\pgfqpoint{1.787891in}{2.873144in}}{\pgfqpoint{1.795907in}{2.853793in}}{\pgfqpoint{1.810172in}{2.839527in}}%
\pgfpathcurveto{\pgfqpoint{1.824438in}{2.825262in}}{\pgfqpoint{1.843789in}{2.817246in}}{\pgfqpoint{1.863964in}{2.817246in}}%
\pgfpathlineto{\pgfqpoint{1.863964in}{2.817246in}}%
\pgfpathclose%
\pgfusepath{stroke}%
\end{pgfscope}%
\begin{pgfscope}%
\pgfpathrectangle{\pgfqpoint{0.266667in}{0.175231in}}{\pgfqpoint{2.941621in}{2.855141in}}%
\pgfusepath{clip}%
\pgfsetbuttcap%
\pgfsetroundjoin%
\pgfsetlinewidth{2.007500pt}%
\definecolor{currentstroke}{rgb}{0.000000,0.000000,0.000000}%
\pgfsetstrokecolor{currentstroke}%
\pgfsetdash{}{0pt}%
\pgfpathmoveto{\pgfqpoint{0.681465in}{1.423523in}}%
\pgfpathcurveto{\pgfqpoint{0.701640in}{1.423523in}}{\pgfqpoint{0.720991in}{1.431539in}}{\pgfqpoint{0.735257in}{1.445804in}}%
\pgfpathcurveto{\pgfqpoint{0.749522in}{1.460070in}}{\pgfqpoint{0.757538in}{1.479421in}}{\pgfqpoint{0.757538in}{1.499596in}}%
\pgfpathcurveto{\pgfqpoint{0.757538in}{1.519770in}}{\pgfqpoint{0.749522in}{1.539122in}}{\pgfqpoint{0.735257in}{1.553387in}}%
\pgfpathcurveto{\pgfqpoint{0.720991in}{1.567653in}}{\pgfqpoint{0.701640in}{1.575668in}}{\pgfqpoint{0.681465in}{1.575668in}}%
\pgfpathcurveto{\pgfqpoint{0.661290in}{1.575668in}}{\pgfqpoint{0.641939in}{1.567653in}}{\pgfqpoint{0.627674in}{1.553387in}}%
\pgfpathcurveto{\pgfqpoint{0.613408in}{1.539122in}}{\pgfqpoint{0.605393in}{1.519770in}}{\pgfqpoint{0.605393in}{1.499596in}}%
\pgfpathcurveto{\pgfqpoint{0.605393in}{1.479421in}}{\pgfqpoint{0.613408in}{1.460070in}}{\pgfqpoint{0.627674in}{1.445804in}}%
\pgfpathcurveto{\pgfqpoint{0.641939in}{1.431539in}}{\pgfqpoint{0.661290in}{1.423523in}}{\pgfqpoint{0.681465in}{1.423523in}}%
\pgfpathlineto{\pgfqpoint{0.681465in}{1.423523in}}%
\pgfpathclose%
\pgfusepath{stroke}%
\end{pgfscope}%
\begin{pgfscope}%
\pgfpathrectangle{\pgfqpoint{0.266667in}{0.175231in}}{\pgfqpoint{2.941621in}{2.855141in}}%
\pgfusepath{clip}%
\pgfsetbuttcap%
\pgfsetroundjoin%
\pgfsetlinewidth{2.007500pt}%
\definecolor{currentstroke}{rgb}{0.000000,0.000000,0.000000}%
\pgfsetstrokecolor{currentstroke}%
\pgfsetdash{}{0pt}%
\pgfpathmoveto{\pgfqpoint{2.111895in}{2.046232in}}%
\pgfpathcurveto{\pgfqpoint{2.132070in}{2.046232in}}{\pgfqpoint{2.151421in}{2.054248in}}{\pgfqpoint{2.165687in}{2.068513in}}%
\pgfpathcurveto{\pgfqpoint{2.179953in}{2.082779in}}{\pgfqpoint{2.187968in}{2.102130in}}{\pgfqpoint{2.187968in}{2.122305in}}%
\pgfpathcurveto{\pgfqpoint{2.187968in}{2.142480in}}{\pgfqpoint{2.179953in}{2.161831in}}{\pgfqpoint{2.165687in}{2.176096in}}%
\pgfpathcurveto{\pgfqpoint{2.151421in}{2.190362in}}{\pgfqpoint{2.132070in}{2.198377in}}{\pgfqpoint{2.111895in}{2.198377in}}%
\pgfpathcurveto{\pgfqpoint{2.091721in}{2.198377in}}{\pgfqpoint{2.072370in}{2.190362in}}{\pgfqpoint{2.058104in}{2.176096in}}%
\pgfpathcurveto{\pgfqpoint{2.043838in}{2.161831in}}{\pgfqpoint{2.035823in}{2.142480in}}{\pgfqpoint{2.035823in}{2.122305in}}%
\pgfpathcurveto{\pgfqpoint{2.035823in}{2.102130in}}{\pgfqpoint{2.043838in}{2.082779in}}{\pgfqpoint{2.058104in}{2.068513in}}%
\pgfpathcurveto{\pgfqpoint{2.072370in}{2.054248in}}{\pgfqpoint{2.091721in}{2.046232in}}{\pgfqpoint{2.111895in}{2.046232in}}%
\pgfpathlineto{\pgfqpoint{2.111895in}{2.046232in}}%
\pgfpathclose%
\pgfusepath{stroke}%
\end{pgfscope}%
\begin{pgfscope}%
\pgfpathrectangle{\pgfqpoint{0.266667in}{0.175231in}}{\pgfqpoint{2.941621in}{2.855141in}}%
\pgfusepath{clip}%
\pgfsetbuttcap%
\pgfsetroundjoin%
\pgfsetlinewidth{2.007500pt}%
\definecolor{currentstroke}{rgb}{0.000000,0.000000,0.000000}%
\pgfsetstrokecolor{currentstroke}%
\pgfsetdash{}{0pt}%
\pgfpathmoveto{\pgfqpoint{0.652733in}{1.328005in}}%
\pgfpathcurveto{\pgfqpoint{0.672908in}{1.328005in}}{\pgfqpoint{0.692259in}{1.336020in}}{\pgfqpoint{0.706524in}{1.350286in}}%
\pgfpathcurveto{\pgfqpoint{0.720790in}{1.364551in}}{\pgfqpoint{0.728806in}{1.383903in}}{\pgfqpoint{0.728806in}{1.404077in}}%
\pgfpathcurveto{\pgfqpoint{0.728806in}{1.424252in}}{\pgfqpoint{0.720790in}{1.443603in}}{\pgfqpoint{0.706524in}{1.457869in}}%
\pgfpathcurveto{\pgfqpoint{0.692259in}{1.472134in}}{\pgfqpoint{0.672908in}{1.480150in}}{\pgfqpoint{0.652733in}{1.480150in}}%
\pgfpathcurveto{\pgfqpoint{0.632558in}{1.480150in}}{\pgfqpoint{0.613207in}{1.472134in}}{\pgfqpoint{0.598942in}{1.457869in}}%
\pgfpathcurveto{\pgfqpoint{0.584676in}{1.443603in}}{\pgfqpoint{0.576660in}{1.424252in}}{\pgfqpoint{0.576660in}{1.404077in}}%
\pgfpathcurveto{\pgfqpoint{0.576660in}{1.383903in}}{\pgfqpoint{0.584676in}{1.364551in}}{\pgfqpoint{0.598942in}{1.350286in}}%
\pgfpathcurveto{\pgfqpoint{0.613207in}{1.336020in}}{\pgfqpoint{0.632558in}{1.328005in}}{\pgfqpoint{0.652733in}{1.328005in}}%
\pgfpathlineto{\pgfqpoint{0.652733in}{1.328005in}}%
\pgfpathclose%
\pgfusepath{stroke}%
\end{pgfscope}%
\begin{pgfscope}%
\pgfsetbuttcap%
\pgfsetroundjoin%
\definecolor{currentfill}{rgb}{0.000000,0.000000,0.000000}%
\pgfsetfillcolor{currentfill}%
\pgfsetlinewidth{0.803000pt}%
\definecolor{currentstroke}{rgb}{0.000000,0.000000,0.000000}%
\pgfsetstrokecolor{currentstroke}%
\pgfsetdash{}{0pt}%
\pgfsys@defobject{currentmarker}{\pgfqpoint{0.000000in}{0.000000in}}{\pgfqpoint{0.000000in}{0.048611in}}{%
\pgfpathmoveto{\pgfqpoint{0.000000in}{0.000000in}}%
\pgfpathlineto{\pgfqpoint{0.000000in}{0.048611in}}%
\pgfusepath{stroke,fill}%
}%
\begin{pgfscope}%
\pgfsys@transformshift{0.588407in}{3.030372in}%
\pgfsys@useobject{currentmarker}{}%
\end{pgfscope}%
\end{pgfscope}%
\begin{pgfscope}%
\definecolor{textcolor}{rgb}{0.000000,0.000000,0.000000}%
\pgfsetstrokecolor{textcolor}%
\pgfsetfillcolor{textcolor}%
\pgftext[x=0.588407in,y=3.127594in,,bottom]{\color{textcolor}{\rmfamily\fontsize{16.000000}{18.00000}\selectfont\catcode`\^=\active\def^{\ifmmode\sp\else\^{}\fi}\catcode`\%=\active\def
\end{pgfscope}%
\begin{pgfscope}%
\pgfsetbuttcap%
\pgfsetroundjoin%
\definecolor{currentfill}{rgb}{0.000000,0.000000,0.000000}%
\pgfsetfillcolor{currentfill}%
\pgfsetlinewidth{0.803000pt}%
\definecolor{currentstroke}{rgb}{0.000000,0.000000,0.000000}%
\pgfsetstrokecolor{currentstroke}%
\pgfsetdash{}{0pt}%
\pgfsys@defobject{currentmarker}{\pgfqpoint{0.000000in}{0.000000in}}{\pgfqpoint{0.000000in}{0.048611in}}{%
\pgfpathmoveto{\pgfqpoint{0.000000in}{0.000000in}}%
\pgfpathlineto{\pgfqpoint{0.000000in}{0.048611in}}%
\pgfusepath{stroke,fill}%
}%
\begin{pgfscope}%
\pgfsys@transformshift{1.162942in}{3.030372in}%
\pgfsys@useobject{currentmarker}{}%
\end{pgfscope}%
\end{pgfscope}%
\begin{pgfscope}%
\pgfsetbuttcap%
\pgfsetroundjoin%
\definecolor{currentfill}{rgb}{0.000000,0.000000,0.000000}%
\pgfsetfillcolor{currentfill}%
\pgfsetlinewidth{0.803000pt}%
\definecolor{currentstroke}{rgb}{0.000000,0.000000,0.000000}%
\pgfsetstrokecolor{currentstroke}%
\pgfsetdash{}{0pt}%
\pgfsys@defobject{currentmarker}{\pgfqpoint{0.000000in}{0.000000in}}{\pgfqpoint{0.000000in}{0.048611in}}{%
\pgfpathmoveto{\pgfqpoint{0.000000in}{0.000000in}}%
\pgfpathlineto{\pgfqpoint{0.000000in}{0.048611in}}%
\pgfusepath{stroke,fill}%
}%
\begin{pgfscope}%
\pgfsys@transformshift{1.737478in}{3.030372in}%
\pgfsys@useobject{currentmarker}{}%
\end{pgfscope}%
\end{pgfscope}%
\begin{pgfscope}%
\pgfsetbuttcap%
\pgfsetroundjoin%
\definecolor{currentfill}{rgb}{0.000000,0.000000,0.000000}%
\pgfsetfillcolor{currentfill}%
\pgfsetlinewidth{0.803000pt}%
\definecolor{currentstroke}{rgb}{0.000000,0.000000,0.000000}%
\pgfsetstrokecolor{currentstroke}%
\pgfsetdash{}{0pt}%
\pgfsys@defobject{currentmarker}{\pgfqpoint{0.000000in}{0.000000in}}{\pgfqpoint{0.000000in}{0.048611in}}{%
\pgfpathmoveto{\pgfqpoint{0.000000in}{0.000000in}}%
\pgfpathlineto{\pgfqpoint{0.000000in}{0.048611in}}%
\pgfusepath{stroke,fill}%
}%
\begin{pgfscope}%
\pgfsys@transformshift{2.312013in}{3.030372in}%
\pgfsys@useobject{currentmarker}{}%
\end{pgfscope}%
\end{pgfscope}%
\begin{pgfscope}%
\pgfsetbuttcap%
\pgfsetroundjoin%
\definecolor{currentfill}{rgb}{0.000000,0.000000,0.000000}%
\pgfsetfillcolor{currentfill}%
\pgfsetlinewidth{0.803000pt}%
\definecolor{currentstroke}{rgb}{0.000000,0.000000,0.000000}%
\pgfsetstrokecolor{currentstroke}%
\pgfsetdash{}{0pt}%
\pgfsys@defobject{currentmarker}{\pgfqpoint{0.000000in}{0.000000in}}{\pgfqpoint{0.000000in}{0.048611in}}{%
\pgfpathmoveto{\pgfqpoint{0.000000in}{0.000000in}}%
\pgfpathlineto{\pgfqpoint{0.000000in}{0.048611in}}%
\pgfusepath{stroke,fill}%
}%
\begin{pgfscope}%
\pgfsys@transformshift{2.886548in}{3.030372in}%
\pgfsys@useobject{currentmarker}{}%
\end{pgfscope}%
\end{pgfscope}%
\begin{pgfscope}%
\definecolor{textcolor}{rgb}{0.000000,0.000000,0.000000}%
\pgfsetstrokecolor{textcolor}%
\pgfsetfillcolor{textcolor}%
\pgftext[x=2.886548in,y=3.127594in,,bottom]{\color{textcolor}{\rmfamily\fontsize{16.000000}{18.00000}\selectfont\catcode`\^=\active\def^{\ifmmode\sp\else\^{}\fi}\catcode`\%=\active\def
\end{pgfscope}%
\begin{pgfscope}%
\definecolor{textcolor}{rgb}{0.000000,0.000000,0.000000}%
\pgfsetstrokecolor{textcolor}%
\pgfsetfillcolor{textcolor}%
\pgftext[x=1.737478in,y=3.266483in,,base]{\color{textcolor}{\rmfamily\fontsize{16.000000}{18.00000}\selectfont\catcode`\^=\active\def^{\ifmmode\sp\else\^{}\fi}\catcode`\%=\active\def
\end{pgfscope}%
\begin{pgfscope}%
\pgfsetbuttcap%
\pgfsetroundjoin%
\definecolor{currentfill}{rgb}{0.000000,0.000000,0.000000}%
\pgfsetfillcolor{currentfill}%
\pgfsetlinewidth{0.803000pt}%
\definecolor{currentstroke}{rgb}{0.000000,0.000000,0.000000}%
\pgfsetstrokecolor{currentstroke}%
\pgfsetdash{}{0pt}%
\pgfsys@defobject{currentmarker}{\pgfqpoint{-0.048611in}{0.000000in}}{\pgfqpoint{-0.000000in}{0.000000in}}{%
\pgfpathmoveto{\pgfqpoint{-0.000000in}{0.000000in}}%
\pgfpathlineto{\pgfqpoint{-0.048611in}{0.000000in}}%
\pgfusepath{stroke,fill}%
}%
\begin{pgfscope}%
\pgfsys@transformshift{0.266667in}{0.175231in}%
\pgfsys@useobject{currentmarker}{}%
\end{pgfscope}%
\end{pgfscope}%
\begin{pgfscope}%
\definecolor{textcolor}{rgb}{1.000000,1.000000,1.000000}%
\pgfsetstrokecolor{textcolor}%
\pgfsetfillcolor{textcolor}%
\pgftext[x=0.100000in, y=0.127006in, left, base]{\color{textcolor}{\rmfamily\fontsize{16.000000}{18.00000}\selectfont\catcode`\^=\active\def^{\ifmmode\sp\else\^{}\fi}\catcode`\%=\active\def
\end{pgfscope}%
\begin{pgfscope}%
\pgfsetbuttcap%
\pgfsetroundjoin%
\definecolor{currentfill}{rgb}{0.000000,0.000000,0.000000}%
\pgfsetfillcolor{currentfill}%
\pgfsetlinewidth{0.803000pt}%
\definecolor{currentstroke}{rgb}{0.000000,0.000000,0.000000}%
\pgfsetstrokecolor{currentstroke}%
\pgfsetdash{}{0pt}%
\pgfsys@defobject{currentmarker}{\pgfqpoint{-0.048611in}{0.000000in}}{\pgfqpoint{-0.000000in}{0.000000in}}{%
\pgfpathmoveto{\pgfqpoint{-0.000000in}{0.000000in}}%
\pgfpathlineto{\pgfqpoint{-0.048611in}{0.000000in}}%
\pgfusepath{stroke,fill}%
}%
\begin{pgfscope}%
\pgfsys@transformshift{0.266667in}{0.732876in}%
\pgfsys@useobject{currentmarker}{}%
\end{pgfscope}%
\end{pgfscope}%
\begin{pgfscope}%
\pgfsetbuttcap%
\pgfsetroundjoin%
\definecolor{currentfill}{rgb}{0.000000,0.000000,0.000000}%
\pgfsetfillcolor{currentfill}%
\pgfsetlinewidth{0.803000pt}%
\definecolor{currentstroke}{rgb}{0.000000,0.000000,0.000000}%
\pgfsetstrokecolor{currentstroke}%
\pgfsetdash{}{0pt}%
\pgfsys@defobject{currentmarker}{\pgfqpoint{-0.048611in}{0.000000in}}{\pgfqpoint{-0.000000in}{0.000000in}}{%
\pgfpathmoveto{\pgfqpoint{-0.000000in}{0.000000in}}%
\pgfpathlineto{\pgfqpoint{-0.048611in}{0.000000in}}%
\pgfusepath{stroke,fill}%
}%
\begin{pgfscope}%
\pgfsys@transformshift{0.266667in}{1.290521in}%
\pgfsys@useobject{currentmarker}{}%
\end{pgfscope}%
\end{pgfscope}%
\begin{pgfscope}%
\pgfsetbuttcap%
\pgfsetroundjoin%
\definecolor{currentfill}{rgb}{0.000000,0.000000,0.000000}%
\pgfsetfillcolor{currentfill}%
\pgfsetlinewidth{0.803000pt}%
\definecolor{currentstroke}{rgb}{0.000000,0.000000,0.000000}%
\pgfsetstrokecolor{currentstroke}%
\pgfsetdash{}{0pt}%
\pgfsys@defobject{currentmarker}{\pgfqpoint{-0.048611in}{0.000000in}}{\pgfqpoint{-0.000000in}{0.000000in}}{%
\pgfpathmoveto{\pgfqpoint{-0.000000in}{0.000000in}}%
\pgfpathlineto{\pgfqpoint{-0.048611in}{0.000000in}}%
\pgfusepath{stroke,fill}%
}%
\begin{pgfscope}%
\pgfsys@transformshift{0.266667in}{1.848165in}%
\pgfsys@useobject{currentmarker}{}%
\end{pgfscope}%
\end{pgfscope}%
\begin{pgfscope}%
\pgfsetbuttcap%
\pgfsetroundjoin%
\definecolor{currentfill}{rgb}{0.000000,0.000000,0.000000}%
\pgfsetfillcolor{currentfill}%
\pgfsetlinewidth{0.803000pt}%
\definecolor{currentstroke}{rgb}{0.000000,0.000000,0.000000}%
\pgfsetstrokecolor{currentstroke}%
\pgfsetdash{}{0pt}%
\pgfsys@defobject{currentmarker}{\pgfqpoint{-0.048611in}{0.000000in}}{\pgfqpoint{-0.000000in}{0.000000in}}{%
\pgfpathmoveto{\pgfqpoint{-0.000000in}{0.000000in}}%
\pgfpathlineto{\pgfqpoint{-0.048611in}{0.000000in}}%
\pgfusepath{stroke,fill}%
}%
\begin{pgfscope}%
\pgfsys@transformshift{0.266667in}{2.405810in}%
\pgfsys@useobject{currentmarker}{}%
\end{pgfscope}%
\end{pgfscope}%
\begin{pgfscope}%
\pgfsetrectcap%
\pgfsetmiterjoin%
\pgfsetlinewidth{0.803000pt}%
\definecolor{currentstroke}{rgb}{0.000000,0.000000,0.000000}%
\pgfsetstrokecolor{currentstroke}%
\pgfsetdash{}{0pt}%
\pgfpathmoveto{\pgfqpoint{0.266667in}{0.175231in}}%
\pgfpathlineto{\pgfqpoint{0.266667in}{3.030372in}}%
\pgfusepath{stroke}%
\end{pgfscope}%
\begin{pgfscope}%
\pgfsetrectcap%
\pgfsetmiterjoin%
\pgfsetlinewidth{0.803000pt}%
\definecolor{currentstroke}{rgb}{0.000000,0.000000,0.000000}%
\pgfsetstrokecolor{currentstroke}%
\pgfsetdash{}{0pt}%
\pgfpathmoveto{\pgfqpoint{3.208288in}{0.175231in}}%
\pgfpathlineto{\pgfqpoint{3.208288in}{3.030372in}}%
\pgfusepath{stroke}%
\end{pgfscope}%
\begin{pgfscope}%
\pgfsetrectcap%
\pgfsetmiterjoin%
\pgfsetlinewidth{0.803000pt}%
\definecolor{currentstroke}{rgb}{0.000000,0.000000,0.000000}%
\pgfsetstrokecolor{currentstroke}%
\pgfsetdash{}{0pt}%
\pgfpathmoveto{\pgfqpoint{0.266667in}{0.175231in}}%
\pgfpathlineto{\pgfqpoint{3.208288in}{0.175231in}}%
\pgfusepath{stroke}%
\end{pgfscope}%
\begin{pgfscope}%
\pgfsetrectcap%
\pgfsetmiterjoin%
\pgfsetlinewidth{0.803000pt}%
\definecolor{currentstroke}{rgb}{0.000000,0.000000,0.000000}%
\pgfsetstrokecolor{currentstroke}%
\pgfsetdash{}{0pt}%
\pgfpathmoveto{\pgfqpoint{0.266667in}{3.030372in}}%
\pgfpathlineto{\pgfqpoint{3.208288in}{3.030372in}}%
\pgfusepath{stroke}%
\end{pgfscope}%
\begin{pgfscope}%
\pgfsetbuttcap%
\pgfsetmiterjoin%
\definecolor{currentfill}{rgb}{1.000000,1.000000,1.000000}%
\pgfsetfillcolor{currentfill}%
\pgfsetlinewidth{0.000000pt}%
\definecolor{currentstroke}{rgb}{0.000000,0.000000,0.000000}%
\pgfsetstrokecolor{currentstroke}%
\pgfsetstrokeopacity{0.000000}%
\pgfsetdash{}{0pt}%
\pgfpathmoveto{\pgfqpoint{3.258288in}{0.175231in}}%
\pgfpathlineto{\pgfqpoint{3.552450in}{0.175231in}}%
\pgfpathlineto{\pgfqpoint{3.552450in}{3.030372in}}%
\pgfpathlineto{\pgfqpoint{3.258288in}{3.030372in}}%
\pgfpathlineto{\pgfqpoint{3.258288in}{0.175231in}}%
\pgfpathclose%
\pgfusepath{fill}%
\end{pgfscope}%
\begin{pgfscope}%
\pgfsys@transformshift{3.260000in}{0.180650in}%
\pgftext[left,bottom]{\includegraphics[interpolate=true,width=0.290000in,height=2.850000in]{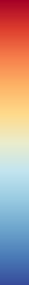}}%
\end{pgfscope}%
\begin{pgfscope}%
\pgfsetbuttcap%
\pgfsetroundjoin%
\definecolor{currentfill}{rgb}{0.000000,0.000000,0.000000}%
\pgfsetfillcolor{currentfill}%
\pgfsetlinewidth{0.803000pt}%
\definecolor{currentstroke}{rgb}{0.000000,0.000000,0.000000}%
\pgfsetstrokecolor{currentstroke}%
\pgfsetdash{}{0pt}%
\pgfsys@defobject{currentmarker}{\pgfqpoint{0.000000in}{0.000000in}}{\pgfqpoint{0.048611in}{0.000000in}}{%
\pgfpathmoveto{\pgfqpoint{0.000000in}{0.000000in}}%
\pgfpathlineto{\pgfqpoint{0.048611in}{0.000000in}}%
\pgfusepath{stroke,fill}%
}%
\begin{pgfscope}%
\pgfsys@transformshift{3.552450in}{3.030372in}%
\pgfsys@useobject{currentmarker}{}%
\end{pgfscope}%
\end{pgfscope}%
\begin{pgfscope}%
\definecolor{textcolor}{rgb}{0.000000,0.000000,0.000000}%
\pgfsetstrokecolor{textcolor}%
\pgfsetfillcolor{textcolor}%
\pgftext[x=3.649673in, y=2.982147in, left, base]{\color{textcolor}{\rmfamily\fontsize{16.000000}{18.00000}\selectfont\catcode`\^=\active\def^{\ifmmode\sp\else\^{}\fi}\catcode`\%=\active\def
\end{pgfscope}%
\begin{pgfscope}%
\pgfsetbuttcap%
\pgfsetroundjoin%
\definecolor{currentfill}{rgb}{0.000000,0.000000,0.000000}%
\pgfsetfillcolor{currentfill}%
\pgfsetlinewidth{0.803000pt}%
\definecolor{currentstroke}{rgb}{0.000000,0.000000,0.000000}%
\pgfsetstrokecolor{currentstroke}%
\pgfsetdash{}{0pt}%
\pgfsys@defobject{currentmarker}{\pgfqpoint{0.000000in}{0.000000in}}{\pgfqpoint{0.048611in}{0.000000in}}{%
\pgfpathmoveto{\pgfqpoint{0.000000in}{0.000000in}}%
\pgfpathlineto{\pgfqpoint{0.048611in}{0.000000in}}%
\pgfusepath{stroke,fill}%
}%
\begin{pgfscope}%
\pgfsys@transformshift{3.552450in}{2.554515in}%
\pgfsys@useobject{currentmarker}{}%
\end{pgfscope}%
\end{pgfscope}%
\begin{pgfscope}%
\pgfsetbuttcap%
\pgfsetroundjoin%
\definecolor{currentfill}{rgb}{0.000000,0.000000,0.000000}%
\pgfsetfillcolor{currentfill}%
\pgfsetlinewidth{0.803000pt}%
\definecolor{currentstroke}{rgb}{0.000000,0.000000,0.000000}%
\pgfsetstrokecolor{currentstroke}%
\pgfsetdash{}{0pt}%
\pgfsys@defobject{currentmarker}{\pgfqpoint{0.000000in}{0.000000in}}{\pgfqpoint{0.048611in}{0.000000in}}{%
\pgfpathmoveto{\pgfqpoint{0.000000in}{0.000000in}}%
\pgfpathlineto{\pgfqpoint{0.048611in}{0.000000in}}%
\pgfusepath{stroke,fill}%
}%
\begin{pgfscope}%
\pgfsys@transformshift{3.552450in}{2.078659in}%
\pgfsys@useobject{currentmarker}{}%
\end{pgfscope}%
\end{pgfscope}%
\begin{pgfscope}%
\pgfsetbuttcap%
\pgfsetroundjoin%
\definecolor{currentfill}{rgb}{0.000000,0.000000,0.000000}%
\pgfsetfillcolor{currentfill}%
\pgfsetlinewidth{0.803000pt}%
\definecolor{currentstroke}{rgb}{0.000000,0.000000,0.000000}%
\pgfsetstrokecolor{currentstroke}%
\pgfsetdash{}{0pt}%
\pgfsys@defobject{currentmarker}{\pgfqpoint{0.000000in}{0.000000in}}{\pgfqpoint{0.048611in}{0.000000in}}{%
\pgfpathmoveto{\pgfqpoint{0.000000in}{0.000000in}}%
\pgfpathlineto{\pgfqpoint{0.048611in}{0.000000in}}%
\pgfusepath{stroke,fill}%
}%
\begin{pgfscope}%
\pgfsys@transformshift{3.552450in}{1.602802in}%
\pgfsys@useobject{currentmarker}{}%
\end{pgfscope}%
\end{pgfscope}%
\begin{pgfscope}%
\pgfsetbuttcap%
\pgfsetroundjoin%
\definecolor{currentfill}{rgb}{0.000000,0.000000,0.000000}%
\pgfsetfillcolor{currentfill}%
\pgfsetlinewidth{0.803000pt}%
\definecolor{currentstroke}{rgb}{0.000000,0.000000,0.000000}%
\pgfsetstrokecolor{currentstroke}%
\pgfsetdash{}{0pt}%
\pgfsys@defobject{currentmarker}{\pgfqpoint{0.000000in}{0.000000in}}{\pgfqpoint{0.048611in}{0.000000in}}{%
\pgfpathmoveto{\pgfqpoint{0.000000in}{0.000000in}}%
\pgfpathlineto{\pgfqpoint{0.048611in}{0.000000in}}%
\pgfusepath{stroke,fill}%
}%
\begin{pgfscope}%
\pgfsys@transformshift{3.552450in}{1.126945in}%
\pgfsys@useobject{currentmarker}{}%
\end{pgfscope}%
\end{pgfscope}%
\begin{pgfscope}%
\pgfsetbuttcap%
\pgfsetroundjoin%
\definecolor{currentfill}{rgb}{0.000000,0.000000,0.000000}%
\pgfsetfillcolor{currentfill}%
\pgfsetlinewidth{0.803000pt}%
\definecolor{currentstroke}{rgb}{0.000000,0.000000,0.000000}%
\pgfsetstrokecolor{currentstroke}%
\pgfsetdash{}{0pt}%
\pgfsys@defobject{currentmarker}{\pgfqpoint{0.000000in}{0.000000in}}{\pgfqpoint{0.048611in}{0.000000in}}{%
\pgfpathmoveto{\pgfqpoint{0.000000in}{0.000000in}}%
\pgfpathlineto{\pgfqpoint{0.048611in}{0.000000in}}%
\pgfusepath{stroke,fill}%
}%
\begin{pgfscope}%
\pgfsys@transformshift{3.552450in}{0.651088in}%
\pgfsys@useobject{currentmarker}{}%
\end{pgfscope}%
\end{pgfscope}%
\begin{pgfscope}%
\pgfsetbuttcap%
\pgfsetroundjoin%
\definecolor{currentfill}{rgb}{0.000000,0.000000,0.000000}%
\pgfsetfillcolor{currentfill}%
\pgfsetlinewidth{0.803000pt}%
\definecolor{currentstroke}{rgb}{0.000000,0.000000,0.000000}%
\pgfsetstrokecolor{currentstroke}%
\pgfsetdash{}{0pt}%
\pgfsys@defobject{currentmarker}{\pgfqpoint{0.000000in}{0.000000in}}{\pgfqpoint{0.048611in}{0.000000in}}{%
\pgfpathmoveto{\pgfqpoint{0.000000in}{0.000000in}}%
\pgfpathlineto{\pgfqpoint{0.048611in}{0.000000in}}%
\pgfusepath{stroke,fill}%
}%
\begin{pgfscope}%
\pgfsys@transformshift{3.552450in}{0.175231in}%
\pgfsys@useobject{currentmarker}{}%
\end{pgfscope}%
\end{pgfscope}%
\begin{pgfscope}%
\definecolor{textcolor}{rgb}{0.000000,0.000000,0.000000}%
\pgfsetstrokecolor{textcolor}%
\pgfsetfillcolor{textcolor}%
\pgftext[x=3.649673in, y=0.127006in, left, base]{\color{textcolor}{\rmfamily\fontsize{16.000000}{18.00000}\selectfont\catcode`\^=\active\def^{\ifmmode\sp\else\^{}\fi}\catcode`\%=\active\def
\end{pgfscope}%
\begin{pgfscope}%
\definecolor{textcolor}{rgb}{0.000000,0.000000,0.000000}%
\pgfsetstrokecolor{textcolor}%
\pgfsetfillcolor{textcolor}%
\pgftext[x=3.751525in,y=1.602802in,,top,rotate=90.000000]{\color{textcolor}{\rmfamily\fontsize{16.000000}{18.00000}\selectfont\catcode`\^=\active\def^{\ifmmode\sp\else\^{}\fi}\catcode`\%=\active\def
\end{pgfscope}%
\begin{pgfscope}%
\pgfsetrectcap%
\pgfsetmiterjoin%
\pgfsetlinewidth{0.803000pt}%
\definecolor{currentstroke}{rgb}{0.000000,0.000000,0.000000}%
\pgfsetstrokecolor{currentstroke}%
\pgfsetdash{}{0pt}%
\pgfpathmoveto{\pgfqpoint{3.258288in}{0.175231in}}%
\pgfpathlineto{\pgfqpoint{3.405369in}{0.175231in}}%
\pgfpathlineto{\pgfqpoint{3.552450in}{0.175231in}}%
\pgfpathlineto{\pgfqpoint{3.552450in}{3.030372in}}%
\pgfpathlineto{\pgfqpoint{3.405369in}{3.030372in}}%
\pgfpathlineto{\pgfqpoint{3.258288in}{3.030372in}}%
\pgfpathlineto{\pgfqpoint{3.258288in}{0.175231in}}%
\pgfpathclose%
\pgfusepath{stroke}%
\end{pgfscope}%
\end{pgfpicture}%
\makeatother%
\endgroup%

%% file: tex/analysis.tex
We conduct a two-fold evaluation of our approach: first, by assessing its estimation performance compared to state-of-the-art algorithms and theoretic bounds, and second, by examining its applicability to real-world measurement data.
To assess the estimation performance of our approach, we perform an isolated study with respect to the \gls{mse} of the parameter estimates and the accuracy of the estimated model order.
As these comparisons mandate the availability of a groundtruth, we perform them on synthetic data from a test dataset.
Second, we apply the approach to previously published measurement data (see~\cite{Schwind2019}), captured in an anechoic chamber and compare the results to the available groundtruth where possible. 
\begin{figure}[t]
    \begin{center}
        \def\mathdefault#1{#1}
        \scalebox{0.9}{
            \input{figures/mse.pgf}
        }
    \end{center}
    \caption{\gls{mse} of different methods. With \num{10} iterations of \eqref{eq:gradient_steps}, the \gls{cnn} achieves similar performance to the iterative \gls{ml} method at significantly lower computation times.}
    \label{fig:mse}
\end{figure}
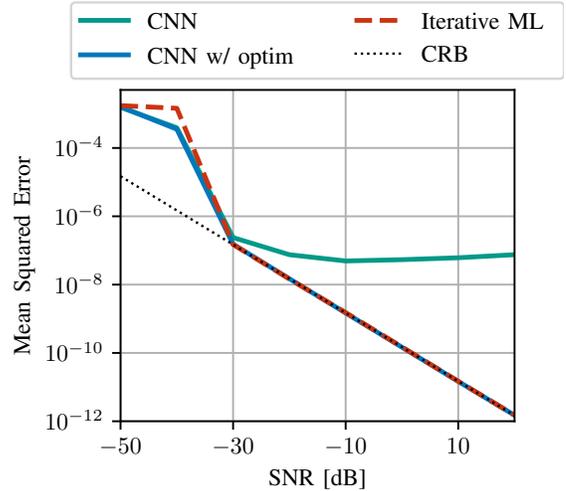
\begin{figure}[t]
    \begin{center}
        \def\mathdefault#1{#1}
        \scalebox{0.9}{
            \input{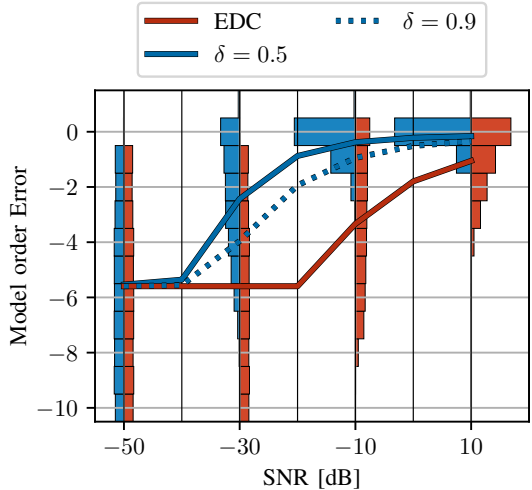}
        }
    \end{center}
    \caption{Model order error for different threshold $\delta$ and \gls{edc}, where negative values represent underestimation. Our approach works better at lower \glspl{snr}. Tuning the threshold enables finetuning without retraining.} 
    \label{fig:modelordererror}
\end{figure}
\subsection{Synthetic Data}\label{sec:analysis:synthetic}
To assess the estimation accuracy, we create a random single path scenario and evaluate it at different \glspl{snr} from \qtyrange{-50}{20}{\decibel}.
For each \gls{snr} we average the squared error in the parameter domain of the estimates from \num{10000} different noise realizations to obtain an estimate of the \gls{mse}.
The comparison includes the raw estimates obtained from the \gls{cnn} and those obtained with the additional gradient-based optimization.
For reference, we provide comparisons with an iterative \gls{ml} estimator from ~\cite{richter_estimation_2005} and the \gls{crb} as a theoretical lower bound for unbiased estimators.

The results in \Cref{fig:mse} show that our approach successfully estimates the delay-Doppler parameters, with the raw and optimized estimates showing similar performance in low \gls{snr}.
At higher \glspl{snr}, the raw estimates' accuracy of our approach starts to saturate around \qty{-20}{\decibel} \gls{snr}.
This indicates that the approach produces a bias in the estimates, whose source is currently unknown and is subject to future research.
However, it can be overcome with the additional model-based optimization defined by \cref{eq:gradient_steps}.
With \num{10} second-order Newton steps, the results from our estimator reach the \gls{crb}, removing the bias and demonstrating identical behavior compared to the iterative \gls{ml}. 
This is a highly promising feature of our approach, particularly for scenarios with multiple paths (shown in \cref{sec:analysis:meas}), since the \emph{joint} estimates are close enough to the global minimum of \eqref{eq:llf} such that \eqref{eq:gradient_steps} can converge to the true solution.
This procedure starkly contrasts iterative \gls{ml} approaches, which only add a single path in each step to the set of estimates and carry out iterative refinement between two newly added paths (also known as successive interference cancellation).
In our case, we initialize the iteration much more efficiently with a single, fixed-clock, forward pass through the \gls{cnn} but still accurate enough for the global optimization to converge.

Apart from the accuracy, the computational complexity of the algorithms is also of interest.
As the assessment of the computational complexity of the iterative \gls{ml} methods is not straightforward due to the use of involved numerical methods, we obtain the average runtime per sample on an identical system.
Averaging \num{1000} runs, our approach requires \qty{19}{\milli\second} without and \qty{60}{\milli\second} with the optional Gauss-Newton scheme \eqref{eq:gradient_steps}, whereas the iterative \gls{ml} algorithm requires around \SI{11.9}{\second}.
While these timing numbers are subject to implementation and must be interpreted accordingly, they still highlight that our approach can address applications requiring fast, fixed-clock estimates.
Additionally, the global optimization runtime is flexibly tune-able via the number of iterations, providing a runtime-accuracy trade-off for system design.

To assess the model order estimation capabilities of our approach, we study different thresholds $\delta$ and compare with the \gls{edc}~\cite{zhao_detection_1986}.
The results are illustrated in \Cref{fig:modelordererror} and reaffirm the findings of previous publications, such as~\cite{izacard_data-driven_2019, barthelme_machine_2021}, which show that \glspl{dnn} can reliably predict the model order in low \gls{snr}.
Similarly, our approach achieves good results across the \gls{snr} range and outperforms \gls{edc} over all threshold levels. 
The histogram results for $\delta=0.5$ and \gls{edc} illustrate the empirical distribution of the model order error and highlight that error probability decreases consistently with increasing \gls{snr}.
This result highlights the approach's model order estimation capabilities, particularly in the challenging low-\gls{snr} domain.
\begin{figure*}[t]
    \def\mathdefault#1{#1}
    \input{figures/meas/meas.tex}
\end{figure*}
\subsection{Measurement Data}\label{sec:analysis:meas}
So far, the analysis is limited to synthetic data, generated from the observation model in \cref{eq:observation_discrete}.
It allowed an assessment of the overall feasibility and theoretical performance, but not real-world performance.
Therefore, the question remains: If a model trained on such synthetic data, is it applicable to measurement data?
And if so, how is its performance affected by adverse effects, such as imperfect synchronization or additional phase noise?

To answer these questions, we utilize measurement data bi-static delay-Doppler emulation measurement in an anechoic chamber, previously presented in~\cite{Schwind2019} and evaluated in~\cite{dobereiner_joint_2019}.
The setup features two metallic spheres (\qty{12}{\centi\meter} diameter) mounted at the ends of a \qty{3}{\meter} beam rotating at \num{60} rpm.
They are illuminated in a bi-static setup by a \gls{tx} and \gls{rx} spaced \qty{2.24}{\meter} apart, schematically shown in \Cref{fig:analysis:setup}.
The \gls{ofdm}-like signal is transmitted at $f_c=\qty{5.9}{\giga\hertz}$ with a bandwidth of \qty{160}{\mega\hertz} and a symbol length of \qty{64}{\mu\second}.
It is modulated with a Newman sequence providing constant spectral amplitude and minimal Crest factor \cite{boyd1986}.
To provide a groundtruth for the performance evaluation, the sphere positions are recorded by an angular sensor in the motor.
From these positions and the known rotation speed, the path parameters $\bm\tau$ and $\bm\alpha$ of the spheres can be computed and evaluated against the estimates.

We divide the recorded data of one full rotation of the spherical targets into consecutive snapshots, each with $N_t=100$ successive measurements, to form the \num{2}D data according to \cref{eq:observation_discrete}.
The associated sampling processes result in a Rayleigh resolution of \qty{6.25}{\nano\second} in delay and \qty{156.25}{\hertz} in Doppler-shift.
While rotating, the spheres result in two paths with oscillating delay and Doppler parameters, illustrated by the black lines in \Cref{fig:analysis:delay} and \Cref{fig:analysis:doppler}.
Despite the anechoic environment, the \gls{los} at \qty{7.4}{\nano\second}, motor, and other objects (at \qty{37.5}{\nano\second}) contribute further static paths, creating a challenging task for the detection and estimation with our approach.\par
As before, we use our approach for the joint estimation of the delay and Doppler-shifts.
We utilized the same network as in \cref{sec:analysis:synthetic}, including the additional optimization with the Gauss-Newton scheme during the postprocessing and a threshold of $\delta=0.8$ for the model order estimation.
Most notably, this \gls{cnn} was trained solely on synthetic data and never exposed to any measurement data during its training.\par
The plots in \Cref{fig:analysis:delay} and \cref{fig:analysis:doppler} illustrate the estimation results for one full rotation of both spheres (from motor angle \qtyrange{0}{360}{\degree}).
It is immediately visible that our approach can accurately estimate the spheres' parameters.
A notable exception are motor rotation angles from \qtyrange{90}{135}{\degree}, where one sphere is masked by a strong beam reflection, visible by the linearly increasing estimates in \cref{fig:analysis:doppler} (labeled \textit{beam}).
Furthermore, other static paths, such as the motor reflection, are also estimated.
Using a groundtruth filter and matching the remaining estimates to the respective sphere parameters, we can compute the average \gls{rmse} for the parameter estimates of the spheres.
This comes out to \qty{1.1}{\nano\second} and \qty{30}{\hertz} in Doppler-shift, and hence, well below the inherent resolution limit of the original sampling process (illustrated by the y-grid lines in \Cref{fig:analysis:delay} and \Cref{fig:analysis:doppler}), showing super-resolution capabilities.
However, the most significant insight is that despite being trained solely on synthetic data, the \gls{cnn} generalized sufficiently well to be applicable even in measurement applications.

%% file: figures/mse.pgf
\begingroup%
\makeatletter%
\begin{pgfpicture}%
\pgfpathrectangle{\pgfpointorigin}{\pgfqpoint{3.347115in}{3.347115in}}%
\pgfusepath{use as bounding box, clip}%
\begin{pgfscope}%
\pgfsetbuttcap%
\pgfsetmiterjoin%
\definecolor{currentfill}{rgb}{1.000000,1.000000,1.000000}%
\pgfsetfillcolor{currentfill}%
\pgfsetlinewidth{0.000000pt}%
\definecolor{currentstroke}{rgb}{1.000000,1.000000,1.000000}%
\pgfsetstrokecolor{currentstroke}%
\pgfsetdash{}{0pt}%
\pgfpathmoveto{\pgfqpoint{0.000000in}{0.000000in}}%
\pgfpathlineto{\pgfqpoint{3.347115in}{0.000000in}}%
\pgfpathlineto{\pgfqpoint{3.347115in}{3.347115in}}%
\pgfpathlineto{\pgfqpoint{0.000000in}{3.347115in}}%
\pgfpathlineto{\pgfqpoint{0.000000in}{0.000000in}}%
\pgfpathclose%
\pgfusepath{fill}%
\end{pgfscope}%
\begin{pgfscope}%
\pgfsetbuttcap%
\pgfsetmiterjoin%
\definecolor{currentfill}{rgb}{1.000000,1.000000,1.000000}%
\pgfsetfillcolor{currentfill}%
\pgfsetlinewidth{0.000000pt}%
\definecolor{currentstroke}{rgb}{0.000000,0.000000,0.000000}%
\pgfsetstrokecolor{currentstroke}%
\pgfsetstrokeopacity{0.000000}%
\pgfsetdash{}{0pt}%
\pgfpathmoveto{\pgfqpoint{0.769600in}{0.565123in}}%
\pgfpathlineto{\pgfqpoint{3.060864in}{0.565123in}}%
\pgfpathlineto{\pgfqpoint{3.060864in}{2.493376in}}%
\pgfpathlineto{\pgfqpoint{0.769600in}{2.493376in}}%
\pgfpathlineto{\pgfqpoint{0.769600in}{0.565123in}}%
\pgfpathclose%
\pgfusepath{fill}%
\end{pgfscope}%
\begin{pgfscope}%
\pgfpathrectangle{\pgfqpoint{0.769600in}{0.565123in}}{\pgfqpoint{2.291265in}{1.928253in}}%
\pgfusepath{clip}%
\pgfsetrectcap%
\pgfsetroundjoin%
\pgfsetlinewidth{0.803000pt}%
\definecolor{currentstroke}{rgb}{0.690196,0.690196,0.690196}%
\pgfsetstrokecolor{currentstroke}%
\pgfsetdash{}{0pt}%
\pgfpathmoveto{\pgfqpoint{0.769600in}{0.565123in}}%
\pgfpathlineto{\pgfqpoint{0.769600in}{2.493376in}}%
\pgfusepath{stroke}%
\end{pgfscope}%
\begin{pgfscope}%
\pgfsetbuttcap%
\pgfsetroundjoin%
\definecolor{currentfill}{rgb}{0.000000,0.000000,0.000000}%
\pgfsetfillcolor{currentfill}%
\pgfsetlinewidth{0.803000pt}%
\definecolor{currentstroke}{rgb}{0.000000,0.000000,0.000000}%
\pgfsetstrokecolor{currentstroke}%
\pgfsetdash{}{0pt}%
\pgfsys@defobject{currentmarker}{\pgfqpoint{0.000000in}{-0.048611in}}{\pgfqpoint{0.000000in}{0.000000in}}{%
\pgfpathmoveto{\pgfqpoint{0.000000in}{0.000000in}}%
\pgfpathlineto{\pgfqpoint{0.000000in}{-0.048611in}}%
\pgfusepath{stroke,fill}%
}%
\begin{pgfscope}%
\pgfsys@transformshift{0.769600in}{0.565123in}%
\pgfsys@useobject{currentmarker}{}%
\end{pgfscope}%
\end{pgfscope}%
\begin{pgfscope}%
\definecolor{textcolor}{rgb}{0.000000,0.000000,0.000000}%
\pgfsetstrokecolor{textcolor}%
\pgfsetfillcolor{textcolor}%
\pgftext[x=0.769600in,y=0.467901in,,top]{\color{textcolor}{\rmfamily\fontsize{10.000000}{12.000000}\selectfont\catcode`\^=\active\def^{\ifmmode\sp\else\^{}\fi}\catcode`\%=\active\def
\end{pgfscope}%
\begin{pgfscope}%
\pgfpathrectangle{\pgfqpoint{0.769600in}{0.565123in}}{\pgfqpoint{2.291265in}{1.928253in}}%
\pgfusepath{clip}%
\pgfsetrectcap%
\pgfsetroundjoin%
\pgfsetlinewidth{0.803000pt}%
\definecolor{currentstroke}{rgb}{0.690196,0.690196,0.690196}%
\pgfsetstrokecolor{currentstroke}%
\pgfsetdash{}{0pt}%
\pgfpathmoveto{\pgfqpoint{1.424247in}{0.565123in}}%
\pgfpathlineto{\pgfqpoint{1.424247in}{2.493376in}}%
\pgfusepath{stroke}%
\end{pgfscope}%
\begin{pgfscope}%
\pgfsetbuttcap%
\pgfsetroundjoin%
\definecolor{currentfill}{rgb}{0.000000,0.000000,0.000000}%
\pgfsetfillcolor{currentfill}%
\pgfsetlinewidth{0.803000pt}%
\definecolor{currentstroke}{rgb}{0.000000,0.000000,0.000000}%
\pgfsetstrokecolor{currentstroke}%
\pgfsetdash{}{0pt}%
\pgfsys@defobject{currentmarker}{\pgfqpoint{0.000000in}{-0.048611in}}{\pgfqpoint{0.000000in}{0.000000in}}{%
\pgfpathmoveto{\pgfqpoint{0.000000in}{0.000000in}}%
\pgfpathlineto{\pgfqpoint{0.000000in}{-0.048611in}}%
\pgfusepath{stroke,fill}%
}%
\begin{pgfscope}%
\pgfsys@transformshift{1.424247in}{0.565123in}%
\pgfsys@useobject{currentmarker}{}%
\end{pgfscope}%
\end{pgfscope}%
\begin{pgfscope}%
\definecolor{textcolor}{rgb}{0.000000,0.000000,0.000000}%
\pgfsetstrokecolor{textcolor}%
\pgfsetfillcolor{textcolor}%
\pgftext[x=1.424247in,y=0.467901in,,top]{\color{textcolor}{\rmfamily\fontsize{10.000000}{12.000000}\selectfont\catcode`\^=\active\def^{\ifmmode\sp\else\^{}\fi}\catcode`\%=\active\def
\end{pgfscope}%
\begin{pgfscope}%
\pgfpathrectangle{\pgfqpoint{0.769600in}{0.565123in}}{\pgfqpoint{2.291265in}{1.928253in}}%
\pgfusepath{clip}%
\pgfsetrectcap%
\pgfsetroundjoin%
\pgfsetlinewidth{0.803000pt}%
\definecolor{currentstroke}{rgb}{0.690196,0.690196,0.690196}%
\pgfsetstrokecolor{currentstroke}%
\pgfsetdash{}{0pt}%
\pgfpathmoveto{\pgfqpoint{2.078894in}{0.565123in}}%
\pgfpathlineto{\pgfqpoint{2.078894in}{2.493376in}}%
\pgfusepath{stroke}%
\end{pgfscope}%
\begin{pgfscope}%
\pgfsetbuttcap%
\pgfsetroundjoin%
\definecolor{currentfill}{rgb}{0.000000,0.000000,0.000000}%
\pgfsetfillcolor{currentfill}%
\pgfsetlinewidth{0.803000pt}%
\definecolor{currentstroke}{rgb}{0.000000,0.000000,0.000000}%
\pgfsetstrokecolor{currentstroke}%
\pgfsetdash{}{0pt}%
\pgfsys@defobject{currentmarker}{\pgfqpoint{0.000000in}{-0.048611in}}{\pgfqpoint{0.000000in}{0.000000in}}{%
\pgfpathmoveto{\pgfqpoint{0.000000in}{0.000000in}}%
\pgfpathlineto{\pgfqpoint{0.000000in}{-0.048611in}}%
\pgfusepath{stroke,fill}%
}%
\begin{pgfscope}%
\pgfsys@transformshift{2.078894in}{0.565123in}%
\pgfsys@useobject{currentmarker}{}%
\end{pgfscope}%
\end{pgfscope}%
\begin{pgfscope}%
\definecolor{textcolor}{rgb}{0.000000,0.000000,0.000000}%
\pgfsetstrokecolor{textcolor}%
\pgfsetfillcolor{textcolor}%
\pgftext[x=2.078894in,y=0.467901in,,top]{\color{textcolor}{\rmfamily\fontsize{10.000000}{12.000000}\selectfont\catcode`\^=\active\def^{\ifmmode\sp\else\^{}\fi}\catcode`\%=\active\def
\end{pgfscope}%
\begin{pgfscope}%
\pgfpathrectangle{\pgfqpoint{0.769600in}{0.565123in}}{\pgfqpoint{2.291265in}{1.928253in}}%
\pgfusepath{clip}%
\pgfsetrectcap%
\pgfsetroundjoin%
\pgfsetlinewidth{0.803000pt}%
\definecolor{currentstroke}{rgb}{0.690196,0.690196,0.690196}%
\pgfsetstrokecolor{currentstroke}%
\pgfsetdash{}{0pt}%
\pgfpathmoveto{\pgfqpoint{2.733541in}{0.565123in}}%
\pgfpathlineto{\pgfqpoint{2.733541in}{2.493376in}}%
\pgfusepath{stroke}%
\end{pgfscope}%
\begin{pgfscope}%
\pgfsetbuttcap%
\pgfsetroundjoin%
\definecolor{currentfill}{rgb}{0.000000,0.000000,0.000000}%
\pgfsetfillcolor{currentfill}%
\pgfsetlinewidth{0.803000pt}%
\definecolor{currentstroke}{rgb}{0.000000,0.000000,0.000000}%
\pgfsetstrokecolor{currentstroke}%
\pgfsetdash{}{0pt}%
\pgfsys@defobject{currentmarker}{\pgfqpoint{0.000000in}{-0.048611in}}{\pgfqpoint{0.000000in}{0.000000in}}{%
\pgfpathmoveto{\pgfqpoint{0.000000in}{0.000000in}}%
\pgfpathlineto{\pgfqpoint{0.000000in}{-0.048611in}}%
\pgfusepath{stroke,fill}%
}%
\begin{pgfscope}%
\pgfsys@transformshift{2.733541in}{0.565123in}%
\pgfsys@useobject{currentmarker}{}%
\end{pgfscope}%
\end{pgfscope}%
\begin{pgfscope}%
\definecolor{textcolor}{rgb}{0.000000,0.000000,0.000000}%
\pgfsetstrokecolor{textcolor}%
\pgfsetfillcolor{textcolor}%
\pgftext[x=2.733541in,y=0.467901in,,top]{\color{textcolor}{\rmfamily\fontsize{10.000000}{12.000000}\selectfont\catcode`\^=\active\def^{\ifmmode\sp\else\^{}\fi}\catcode`\%=\active\def
\end{pgfscope}%
\begin{pgfscope}%
\definecolor{textcolor}{rgb}{0.000000,0.000000,0.000000}%
\pgfsetstrokecolor{textcolor}%
\pgfsetfillcolor{textcolor}%
\pgftext[x=1.915232in,y=0.288889in,,top]{\color{textcolor}{\rmfamily\fontsize{10.000000}{12.000000}\selectfont\catcode`\^=\active\def^{\ifmmode\sp\else\^{}\fi}\catcode`\%=\active\def
\end{pgfscope}%
\begin{pgfscope}%
\pgfpathrectangle{\pgfqpoint{0.769600in}{0.565123in}}{\pgfqpoint{2.291265in}{1.928253in}}%
\pgfusepath{clip}%
\pgfsetrectcap%
\pgfsetroundjoin%
\pgfsetlinewidth{0.803000pt}%
\definecolor{currentstroke}{rgb}{0.690196,0.690196,0.690196}%
\pgfsetstrokecolor{currentstroke}%
\pgfsetdash{}{0pt}%
\pgfpathmoveto{\pgfqpoint{0.769600in}{0.565123in}}%
\pgfpathlineto{\pgfqpoint{3.060864in}{0.565123in}}%
\pgfusepath{stroke}%
\end{pgfscope}%
\begin{pgfscope}%
\pgfsetbuttcap%
\pgfsetroundjoin%
\definecolor{currentfill}{rgb}{0.000000,0.000000,0.000000}%
\pgfsetfillcolor{currentfill}%
\pgfsetlinewidth{0.803000pt}%
\definecolor{currentstroke}{rgb}{0.000000,0.000000,0.000000}%
\pgfsetstrokecolor{currentstroke}%
\pgfsetdash{}{0pt}%
\pgfsys@defobject{currentmarker}{\pgfqpoint{-0.048611in}{0.000000in}}{\pgfqpoint{-0.000000in}{0.000000in}}{%
\pgfpathmoveto{\pgfqpoint{-0.000000in}{0.000000in}}%
\pgfpathlineto{\pgfqpoint{-0.048611in}{0.000000in}}%
\pgfusepath{stroke,fill}%
}%
\begin{pgfscope}%
\pgfsys@transformshift{0.769600in}{0.565123in}%
\pgfsys@useobject{currentmarker}{}%
\end{pgfscope}%
\end{pgfscope}%
\begin{pgfscope}%
\definecolor{textcolor}{rgb}{0.000000,0.000000,0.000000}%
\pgfsetstrokecolor{textcolor}%
\pgfsetfillcolor{textcolor}%
\pgftext[x=0.329012in, y=0.516898in, left, base]{\color{textcolor}{\rmfamily\fontsize{10.000000}{12.000000}\selectfont\catcode`\^=\active\def^{\ifmmode\sp\else\^{}\fi}\catcode`\%=\active\def
\end{pgfscope}%
\begin{pgfscope}%
\pgfpathrectangle{\pgfqpoint{0.769600in}{0.565123in}}{\pgfqpoint{2.291265in}{1.928253in}}%
\pgfusepath{clip}%
\pgfsetrectcap%
\pgfsetroundjoin%
\pgfsetlinewidth{0.803000pt}%
\definecolor{currentstroke}{rgb}{0.690196,0.690196,0.690196}%
\pgfsetstrokecolor{currentstroke}%
\pgfsetdash{}{0pt}%
\pgfpathmoveto{\pgfqpoint{0.769600in}{0.962743in}}%
\pgfpathlineto{\pgfqpoint{3.060864in}{0.962743in}}%
\pgfusepath{stroke}%
\end{pgfscope}%
\begin{pgfscope}%
\pgfsetbuttcap%
\pgfsetroundjoin%
\definecolor{currentfill}{rgb}{0.000000,0.000000,0.000000}%
\pgfsetfillcolor{currentfill}%
\pgfsetlinewidth{0.803000pt}%
\definecolor{currentstroke}{rgb}{0.000000,0.000000,0.000000}%
\pgfsetstrokecolor{currentstroke}%
\pgfsetdash{}{0pt}%
\pgfsys@defobject{currentmarker}{\pgfqpoint{-0.048611in}{0.000000in}}{\pgfqpoint{-0.000000in}{0.000000in}}{%
\pgfpathmoveto{\pgfqpoint{-0.000000in}{0.000000in}}%
\pgfpathlineto{\pgfqpoint{-0.048611in}{0.000000in}}%
\pgfusepath{stroke,fill}%
}%
\begin{pgfscope}%
\pgfsys@transformshift{0.769600in}{0.962743in}%
\pgfsys@useobject{currentmarker}{}%
\end{pgfscope}%
\end{pgfscope}%
\begin{pgfscope}%
\definecolor{textcolor}{rgb}{0.000000,0.000000,0.000000}%
\pgfsetstrokecolor{textcolor}%
\pgfsetfillcolor{textcolor}%
\pgftext[x=0.329012in, y=0.914518in, left, base]{\color{textcolor}{\rmfamily\fontsize{10.000000}{12.000000}\selectfont\catcode`\^=\active\def^{\ifmmode\sp\else\^{}\fi}\catcode`\%=\active\def
\end{pgfscope}%
\begin{pgfscope}%
\pgfpathrectangle{\pgfqpoint{0.769600in}{0.565123in}}{\pgfqpoint{2.291265in}{1.928253in}}%
\pgfusepath{clip}%
\pgfsetrectcap%
\pgfsetroundjoin%
\pgfsetlinewidth{0.803000pt}%
\definecolor{currentstroke}{rgb}{0.690196,0.690196,0.690196}%
\pgfsetstrokecolor{currentstroke}%
\pgfsetdash{}{0pt}%
\pgfpathmoveto{\pgfqpoint{0.769600in}{1.360363in}}%
\pgfpathlineto{\pgfqpoint{3.060864in}{1.360363in}}%
\pgfusepath{stroke}%
\end{pgfscope}%
\begin{pgfscope}%
\pgfsetbuttcap%
\pgfsetroundjoin%
\definecolor{currentfill}{rgb}{0.000000,0.000000,0.000000}%
\pgfsetfillcolor{currentfill}%
\pgfsetlinewidth{0.803000pt}%
\definecolor{currentstroke}{rgb}{0.000000,0.000000,0.000000}%
\pgfsetstrokecolor{currentstroke}%
\pgfsetdash{}{0pt}%
\pgfsys@defobject{currentmarker}{\pgfqpoint{-0.048611in}{0.000000in}}{\pgfqpoint{-0.000000in}{0.000000in}}{%
\pgfpathmoveto{\pgfqpoint{-0.000000in}{0.000000in}}%
\pgfpathlineto{\pgfqpoint{-0.048611in}{0.000000in}}%
\pgfusepath{stroke,fill}%
}%
\begin{pgfscope}%
\pgfsys@transformshift{0.769600in}{1.360363in}%
\pgfsys@useobject{currentmarker}{}%
\end{pgfscope}%
\end{pgfscope}%
\begin{pgfscope}%
\definecolor{textcolor}{rgb}{0.000000,0.000000,0.000000}%
\pgfsetstrokecolor{textcolor}%
\pgfsetfillcolor{textcolor}%
\pgftext[x=0.384375in, y=1.312138in, left, base]{\color{textcolor}{\rmfamily\fontsize{10.000000}{12.000000}\selectfont\catcode`\^=\active\def^{\ifmmode\sp\else\^{}\fi}\catcode`\%=\active\def
\end{pgfscope}%
\begin{pgfscope}%
\pgfpathrectangle{\pgfqpoint{0.769600in}{0.565123in}}{\pgfqpoint{2.291265in}{1.928253in}}%
\pgfusepath{clip}%
\pgfsetrectcap%
\pgfsetroundjoin%
\pgfsetlinewidth{0.803000pt}%
\definecolor{currentstroke}{rgb}{0.690196,0.690196,0.690196}%
\pgfsetstrokecolor{currentstroke}%
\pgfsetdash{}{0pt}%
\pgfpathmoveto{\pgfqpoint{0.769600in}{1.757984in}}%
\pgfpathlineto{\pgfqpoint{3.060864in}{1.757984in}}%
\pgfusepath{stroke}%
\end{pgfscope}%
\begin{pgfscope}%
\pgfsetbuttcap%
\pgfsetroundjoin%
\definecolor{currentfill}{rgb}{0.000000,0.000000,0.000000}%
\pgfsetfillcolor{currentfill}%
\pgfsetlinewidth{0.803000pt}%
\definecolor{currentstroke}{rgb}{0.000000,0.000000,0.000000}%
\pgfsetstrokecolor{currentstroke}%
\pgfsetdash{}{0pt}%
\pgfsys@defobject{currentmarker}{\pgfqpoint{-0.048611in}{0.000000in}}{\pgfqpoint{-0.000000in}{0.000000in}}{%
\pgfpathmoveto{\pgfqpoint{-0.000000in}{0.000000in}}%
\pgfpathlineto{\pgfqpoint{-0.048611in}{0.000000in}}%
\pgfusepath{stroke,fill}%
}%
\begin{pgfscope}%
\pgfsys@transformshift{0.769600in}{1.757984in}%
\pgfsys@useobject{currentmarker}{}%
\end{pgfscope}%
\end{pgfscope}%
\begin{pgfscope}%
\definecolor{textcolor}{rgb}{0.000000,0.000000,0.000000}%
\pgfsetstrokecolor{textcolor}%
\pgfsetfillcolor{textcolor}%
\pgftext[x=0.384375in, y=1.709758in, left, base]{\color{textcolor}{\rmfamily\fontsize{10.000000}{12.000000}\selectfont\catcode`\^=\active\def^{\ifmmode\sp\else\^{}\fi}\catcode`\%=\active\def
\end{pgfscope}%
\begin{pgfscope}%
\pgfpathrectangle{\pgfqpoint{0.769600in}{0.565123in}}{\pgfqpoint{2.291265in}{1.928253in}}%
\pgfusepath{clip}%
\pgfsetrectcap%
\pgfsetroundjoin%
\pgfsetlinewidth{0.803000pt}%
\definecolor{currentstroke}{rgb}{0.690196,0.690196,0.690196}%
\pgfsetstrokecolor{currentstroke}%
\pgfsetdash{}{0pt}%
\pgfpathmoveto{\pgfqpoint{0.769600in}{2.155604in}}%
\pgfpathlineto{\pgfqpoint{3.060864in}{2.155604in}}%
\pgfusepath{stroke}%
\end{pgfscope}%
\begin{pgfscope}%
\pgfsetbuttcap%
\pgfsetroundjoin%
\definecolor{currentfill}{rgb}{0.000000,0.000000,0.000000}%
\pgfsetfillcolor{currentfill}%
\pgfsetlinewidth{0.803000pt}%
\definecolor{currentstroke}{rgb}{0.000000,0.000000,0.000000}%
\pgfsetstrokecolor{currentstroke}%
\pgfsetdash{}{0pt}%
\pgfsys@defobject{currentmarker}{\pgfqpoint{-0.048611in}{0.000000in}}{\pgfqpoint{-0.000000in}{0.000000in}}{%
\pgfpathmoveto{\pgfqpoint{-0.000000in}{0.000000in}}%
\pgfpathlineto{\pgfqpoint{-0.048611in}{0.000000in}}%
\pgfusepath{stroke,fill}%
}%
\begin{pgfscope}%
\pgfsys@transformshift{0.769600in}{2.155604in}%
\pgfsys@useobject{currentmarker}{}%
\end{pgfscope}%
\end{pgfscope}%
\begin{pgfscope}%
\definecolor{textcolor}{rgb}{0.000000,0.000000,0.000000}%
\pgfsetstrokecolor{textcolor}%
\pgfsetfillcolor{textcolor}%
\pgftext[x=0.384375in, y=2.107378in, left, base]{\color{textcolor}{\rmfamily\fontsize{10.000000}{12.000000}\selectfont\catcode`\^=\active\def^{\ifmmode\sp\else\^{}\fi}\catcode`\%=\active\def
\end{pgfscope}%
\begin{pgfscope}%
\definecolor{textcolor}{rgb}{0.000000,0.000000,0.000000}%
\pgfsetstrokecolor{textcolor}%
\pgfsetfillcolor{textcolor}%
\pgftext[x=0.273457in,y=1.529250in,,bottom,rotate=90.000000]{\color{textcolor}{\rmfamily\fontsize{10.000000}{12.000000}\selectfont\catcode`\^=\active\def^{\ifmmode\sp\else\^{}\fi}\catcode`\%=\active\def
\end{pgfscope}%
\begin{pgfscope}%
\pgfpathrectangle{\pgfqpoint{0.769600in}{0.565123in}}{\pgfqpoint{2.291265in}{1.928253in}}%
\pgfusepath{clip}%
\pgfsetrectcap%
\pgfsetroundjoin%
\pgfsetlinewidth{2.007500pt}%
\definecolor{currentstroke}{rgb}{0.000000,0.600000,0.533333}%
\pgfsetstrokecolor{currentstroke}%
\pgfsetdash{}{0pt}%
\pgfpathmoveto{\pgfqpoint{0.769600in}{2.392459in}}%
\pgfpathlineto{\pgfqpoint{1.096923in}{2.273105in}}%
\pgfpathlineto{\pgfqpoint{1.424247in}{1.634172in}}%
\pgfpathlineto{\pgfqpoint{1.751570in}{1.534490in}}%
\pgfpathlineto{\pgfqpoint{2.078894in}{1.497979in}}%
\pgfpathlineto{\pgfqpoint{2.406217in}{1.505525in}}%
\pgfpathlineto{\pgfqpoint{2.733541in}{1.516775in}}%
\pgfpathlineto{\pgfqpoint{3.060864in}{1.534489in}}%
\pgfpathlineto{\pgfqpoint{3.070864in}{1.535149in}}%
\pgfusepath{stroke}%
\end{pgfscope}%
\begin{pgfscope}%
\pgfpathrectangle{\pgfqpoint{0.769600in}{0.565123in}}{\pgfqpoint{2.291265in}{1.928253in}}%
\pgfusepath{clip}%
\pgfsetrectcap%
\pgfsetroundjoin%
\pgfsetlinewidth{2.007500pt}%
\definecolor{currentstroke}{rgb}{0.000000,0.466667,0.733333}%
\pgfsetstrokecolor{currentstroke}%
\pgfsetdash{}{0pt}%
\pgfpathmoveto{\pgfqpoint{0.769600in}{2.395409in}}%
\pgfpathlineto{\pgfqpoint{1.096923in}{2.264785in}}%
\pgfpathlineto{\pgfqpoint{1.424247in}{1.595559in}}%
\pgfpathlineto{\pgfqpoint{1.751570in}{1.394660in}}%
\pgfpathlineto{\pgfqpoint{2.078894in}{1.195053in}}%
\pgfpathlineto{\pgfqpoint{2.406217in}{0.995911in}}%
\pgfpathlineto{\pgfqpoint{2.733541in}{0.797025in}}%
\pgfpathlineto{\pgfqpoint{3.060864in}{0.600355in}}%
\pgfpathlineto{\pgfqpoint{3.070864in}{0.594289in}}%
\pgfusepath{stroke}%
\end{pgfscope}%
\begin{pgfscope}%
\pgfpathrectangle{\pgfqpoint{0.769600in}{0.565123in}}{\pgfqpoint{2.291265in}{1.928253in}}%
\pgfusepath{clip}%
\pgfsetbuttcap%
\pgfsetroundjoin%
\pgfsetlinewidth{2.007500pt}%
\definecolor{currentstroke}{rgb}{0.800000,0.200000,0.066667}%
\pgfsetstrokecolor{currentstroke}%
\pgfsetdash{{7.400000pt}{3.200000pt}}{0.000000pt}%
\pgfpathmoveto{\pgfqpoint{0.769600in}{2.402342in}}%
\pgfpathlineto{\pgfqpoint{1.096923in}{2.386581in}}%
\pgfpathlineto{\pgfqpoint{1.424247in}{1.593932in}}%
\pgfpathlineto{\pgfqpoint{1.751570in}{1.393072in}}%
\pgfpathlineto{\pgfqpoint{2.078894in}{1.197487in}}%
\pgfpathlineto{\pgfqpoint{2.406217in}{0.997452in}}%
\pgfpathlineto{\pgfqpoint{2.733541in}{0.797976in}}%
\pgfpathlineto{\pgfqpoint{3.060864in}{0.600713in}}%
\pgfpathlineto{\pgfqpoint{3.070864in}{0.594605in}}%
\pgfusepath{stroke}%
\end{pgfscope}%
\begin{pgfscope}%
\pgfpathrectangle{\pgfqpoint{0.769600in}{0.565123in}}{\pgfqpoint{2.291265in}{1.928253in}}%
\pgfusepath{clip}%
\pgfsetbuttcap%
\pgfsetroundjoin%
\pgfsetlinewidth{1.003750pt}%
\definecolor{currentstroke}{rgb}{0.000000,0.000000,0.000000}%
\pgfsetstrokecolor{currentstroke}%
\pgfsetdash{{1.000000pt}{1.650000pt}}{0.000000pt}%
\pgfpathmoveto{\pgfqpoint{0.769600in}{1.990982in}}%
\pgfpathlineto{\pgfqpoint{1.096923in}{1.792185in}}%
\pgfpathlineto{\pgfqpoint{1.424247in}{1.593360in}}%
\pgfpathlineto{\pgfqpoint{1.751570in}{1.394560in}}%
\pgfpathlineto{\pgfqpoint{2.078894in}{1.195752in}}%
\pgfpathlineto{\pgfqpoint{2.406217in}{0.996934in}}%
\pgfpathlineto{\pgfqpoint{2.733541in}{0.798127in}}%
\pgfpathlineto{\pgfqpoint{3.060864in}{0.599312in}}%
\pgfpathlineto{\pgfqpoint{3.070864in}{0.593239in}}%
\pgfusepath{stroke}%
\end{pgfscope}%
\begin{pgfscope}%
\pgfsetrectcap%
\pgfsetmiterjoin%
\pgfsetlinewidth{0.803000pt}%
\definecolor{currentstroke}{rgb}{0.000000,0.000000,0.000000}%
\pgfsetstrokecolor{currentstroke}%
\pgfsetdash{}{0pt}%
\pgfpathmoveto{\pgfqpoint{0.769600in}{0.565123in}}%
\pgfpathlineto{\pgfqpoint{0.769600in}{2.493376in}}%
\pgfusepath{stroke}%
\end{pgfscope}%
\begin{pgfscope}%
\pgfsetrectcap%
\pgfsetmiterjoin%
\pgfsetlinewidth{0.803000pt}%
\definecolor{currentstroke}{rgb}{0.000000,0.000000,0.000000}%
\pgfsetstrokecolor{currentstroke}%
\pgfsetdash{}{0pt}%
\pgfpathmoveto{\pgfqpoint{3.060864in}{0.565123in}}%
\pgfpathlineto{\pgfqpoint{3.060864in}{2.493376in}}%
\pgfusepath{stroke}%
\end{pgfscope}%
\begin{pgfscope}%
\pgfsetrectcap%
\pgfsetmiterjoin%
\pgfsetlinewidth{0.803000pt}%
\definecolor{currentstroke}{rgb}{0.000000,0.000000,0.000000}%
\pgfsetstrokecolor{currentstroke}%
\pgfsetdash{}{0pt}%
\pgfpathmoveto{\pgfqpoint{0.769600in}{0.565123in}}%
\pgfpathlineto{\pgfqpoint{3.060864in}{0.565123in}}%
\pgfusepath{stroke}%
\end{pgfscope}%
\begin{pgfscope}%
\pgfsetrectcap%
\pgfsetmiterjoin%
\pgfsetlinewidth{0.803000pt}%
\definecolor{currentstroke}{rgb}{0.000000,0.000000,0.000000}%
\pgfsetstrokecolor{currentstroke}%
\pgfsetdash{}{0pt}%
\pgfpathmoveto{\pgfqpoint{0.769600in}{2.493376in}}%
\pgfpathlineto{\pgfqpoint{3.060864in}{2.493376in}}%
\pgfusepath{stroke}%
\end{pgfscope}%
\begin{pgfscope}%
\pgfsetbuttcap%
\pgfsetmiterjoin%
\definecolor{currentfill}{rgb}{1.000000,1.000000,1.000000}%
\pgfsetfillcolor{currentfill}%
\pgfsetfillopacity{0.800000}%
\pgfsetlinewidth{1.003750pt}%
\definecolor{currentstroke}{rgb}{0.800000,0.800000,0.800000}%
\pgfsetstrokecolor{currentstroke}%
\pgfsetstrokeopacity{0.800000}%
\pgfsetdash{}{0pt}%
\pgfpathmoveto{\pgfqpoint{0.509752in}{2.558734in}}%
\pgfpathlineto{\pgfqpoint{3.320712in}{2.558734in}}%
\pgfpathquadraticcurveto{\pgfqpoint{3.348490in}{2.558734in}}{\pgfqpoint{3.348490in}{2.586512in}}%
\pgfpathlineto{\pgfqpoint{3.348490in}{2.974629in}}%
\pgfpathquadraticcurveto{\pgfqpoint{3.348490in}{3.002407in}}{\pgfqpoint{3.320712in}{3.002407in}}%
\pgfpathlineto{\pgfqpoint{0.509752in}{3.002407in}}%
\pgfpathquadraticcurveto{\pgfqpoint{0.481974in}{3.002407in}}{\pgfqpoint{0.481974in}{2.974629in}}%
\pgfpathlineto{\pgfqpoint{0.481974in}{2.586512in}}%
\pgfpathquadraticcurveto{\pgfqpoint{0.481974in}{2.558734in}}{\pgfqpoint{0.509752in}{2.558734in}}%
\pgfpathlineto{\pgfqpoint{0.509752in}{2.558734in}}%
\pgfpathclose%
\pgfusepath{stroke,fill}%
\end{pgfscope}%
\begin{pgfscope}%
\pgfsetrectcap%
\pgfsetroundjoin%
\pgfsetlinewidth{2.007500pt}%
\definecolor{currentstroke}{rgb}{0.000000,0.600000,0.533333}%
\pgfsetstrokecolor{currentstroke}%
\pgfsetdash{}{0pt}%
\pgfpathmoveto{\pgfqpoint{0.537530in}{2.898241in}}%
\pgfpathlineto{\pgfqpoint{0.676419in}{2.898241in}}%
\pgfpathlineto{\pgfqpoint{0.815308in}{2.898241in}}%
\pgfusepath{stroke}%
\end{pgfscope}%
\begin{pgfscope}%
\definecolor{textcolor}{rgb}{0.000000,0.000000,0.000000}%
\pgfsetstrokecolor{textcolor}%
\pgfsetfillcolor{textcolor}%
\pgftext[x=0.926419in,y=2.849629in,left,base]{\color{textcolor}{\rmfamily\fontsize{10.000000}{12.000000}\selectfont\catcode`\^=\active\def^{\ifmmode\sp\else\^{}\fi}\catcode`\%=\active\def
\end{pgfscope}%
\begin{pgfscope}%
\pgfsetrectcap%
\pgfsetroundjoin%
\pgfsetlinewidth{2.007500pt}%
\definecolor{currentstroke}{rgb}{0.000000,0.466667,0.733333}%
\pgfsetstrokecolor{currentstroke}%
\pgfsetdash{}{0pt}%
\pgfpathmoveto{\pgfqpoint{0.537530in}{2.697623in}}%
\pgfpathlineto{\pgfqpoint{0.676419in}{2.697623in}}%
\pgfpathlineto{\pgfqpoint{0.815308in}{2.697623in}}%
\pgfusepath{stroke}%
\end{pgfscope}%
\begin{pgfscope}%
\definecolor{textcolor}{rgb}{0.000000,0.000000,0.000000}%
\pgfsetstrokecolor{textcolor}%
\pgfsetfillcolor{textcolor}%
\pgftext[x=0.926419in,y=2.649012in,left,base]{\color{textcolor}{\rmfamily\fontsize{10.000000}{12.000000}\selectfont\catcode`\^=\active\def^{\ifmmode\sp\else\^{}\fi}\catcode`\%=\active\def
\end{pgfscope}%
\begin{pgfscope}%
\pgfsetbuttcap%
\pgfsetroundjoin%
\pgfsetlinewidth{2.007500pt}%
\definecolor{currentstroke}{rgb}{0.800000,0.200000,0.066667}%
\pgfsetstrokecolor{currentstroke}%
\pgfsetdash{{7.400000pt}{3.200000pt}}{0.000000pt}%
\pgfpathmoveto{\pgfqpoint{2.130124in}{2.898241in}}%
\pgfpathlineto{\pgfqpoint{2.269013in}{2.898241in}}%
\pgfpathlineto{\pgfqpoint{2.407902in}{2.898241in}}%
\pgfusepath{stroke}%
\end{pgfscope}%
\begin{pgfscope}%
\definecolor{textcolor}{rgb}{0.000000,0.000000,0.000000}%
\pgfsetstrokecolor{textcolor}%
\pgfsetfillcolor{textcolor}%
\pgftext[x=2.519013in,y=2.849629in,left,base]{\color{textcolor}{\rmfamily\fontsize{10.000000}{12.000000}\selectfont\catcode`\^=\active\def^{\ifmmode\sp\else\^{}\fi}\catcode`\%=\active\def
\end{pgfscope}%
\begin{pgfscope}%
\pgfsetbuttcap%
\pgfsetroundjoin%
\pgfsetlinewidth{1.003750pt}%
\definecolor{currentstroke}{rgb}{0.000000,0.000000,0.000000}%
\pgfsetstrokecolor{currentstroke}%
\pgfsetdash{{1.000000pt}{1.650000pt}}{0.000000pt}%
\pgfpathmoveto{\pgfqpoint{2.130124in}{2.704568in}}%
\pgfpathlineto{\pgfqpoint{2.269013in}{2.704568in}}%
\pgfpathlineto{\pgfqpoint{2.407902in}{2.704568in}}%
\pgfusepath{stroke}%
\end{pgfscope}%
\begin{pgfscope}%
\definecolor{textcolor}{rgb}{0.000000,0.000000,0.000000}%
\pgfsetstrokecolor{textcolor}%
\pgfsetfillcolor{textcolor}%
\pgftext[x=2.519013in,y=2.655957in,left,base]{\color{textcolor}{\rmfamily\fontsize{10.000000}{12.000000}\selectfont\catcode`\^=\active\def^{\ifmmode\sp\else\^{}\fi}\catcode`\%=\active\def
\end{pgfscope}%
\end{pgfpicture}%
\makeatother%
\endgroup%

%% file: figures/meas/meas.tex
\def\scalefig{1}
\centering
\subfloat[t][Delay\label{fig:analysis:delay}]{%
    \scalebox{\scalefig}{\input{figures/meas/pinn_estimates_Run7_strategy=multipath_synth_lowsnr_rx=4_delay.pgf}}
    }%
\subfloat[t][Setup illustration\label{fig:analysis:setup}]{%
    \scalebox{\scalefig}{\input{figures/meas/turntable_rx=4.pgf}}
    }%
\subfloat[t][Doppler-shift\label{fig:analysis:doppler}]{%
    \scalebox{\scalefig}{\input{figures/meas/pinn_estimates_Run7_strategy=multipath_synth_lowsnr_rx=4_doppler.pgf}}
        }%
\caption{Measurement setup and estimates. Black lines in \Cref{fig:analysis:delay} and \Cref{fig:analysis:doppler} show the sphere groundtruths, and the distinguishable motor and beam. Magnitude-colored dots are estimates, and y-grid lines are spaced to the measurement resolution of \qty{6.25}{\nano\second} for delay and \qty{156.25}{\hertz} for Doppler-shift. The sphere estimates achieve an \gls{rmse} of \qty{1.13}{\nano\second} and \qty{36.38}{\hertz}.}
\label{fig:analysis}

%% file: figures/meas/turntable_rx=4.pgf
\begingroup%
\makeatletter%
\begin{pgfpicture}%
\pgfpathrectangle{\pgfpointorigin}{\pgfqpoint{2.008269in}{3.347115in}}%
\pgfusepath{use as bounding box, clip}%
\begin{pgfscope}%
\pgfsetbuttcap%
\pgfsetmiterjoin%
\pgfsetlinewidth{0.000000pt}%
\definecolor{currentstroke}{rgb}{0.000000,0.000000,0.000000}%
\pgfsetstrokecolor{currentstroke}%
\pgfsetstrokeopacity{0.000000}%
\pgfsetdash{}{0pt}%
\pgfpathmoveto{\pgfqpoint{0.000000in}{0.000000in}}%
\pgfpathlineto{\pgfqpoint{2.008269in}{0.000000in}}%
\pgfpathlineto{\pgfqpoint{2.008269in}{3.347115in}}%
\pgfpathlineto{\pgfqpoint{0.000000in}{3.347115in}}%
\pgfpathlineto{\pgfqpoint{0.000000in}{0.000000in}}%
\pgfpathclose%
\pgfusepath{}%
\end{pgfscope}%
\begin{pgfscope}%
\pgfpathrectangle{\pgfqpoint{0.251034in}{0.368183in}}{\pgfqpoint{1.556408in}{2.577279in}}%
\pgfusepath{clip}%
\pgfsetbuttcap%
\pgfsetroundjoin%
\definecolor{currentfill}{rgb}{0.000000,0.466667,0.733333}%
\pgfsetfillcolor{currentfill}%
\pgfsetlinewidth{1.003750pt}%
\definecolor{currentstroke}{rgb}{0.000000,0.000000,0.000000}%
\pgfsetstrokecolor{currentstroke}%
\pgfsetdash{}{0pt}%
\pgfsys@defobject{currentmarker}{\pgfqpoint{-0.096225in}{-0.055556in}}{\pgfqpoint{0.096225in}{0.111111in}}{%
\pgfpathmoveto{\pgfqpoint{0.000000in}{0.111111in}}%
\pgfpathlineto{\pgfqpoint{-0.096225in}{-0.055556in}}%
\pgfpathlineto{\pgfqpoint{0.096225in}{-0.055556in}}%
\pgfpathlineto{\pgfqpoint{0.000000in}{0.111111in}}%
\pgfpathclose%
\pgfusepath{stroke,fill}%
}%
\begin{pgfscope}%
\pgfsys@transformshift{1.052978in}{2.471111in}%
\pgfsys@useobject{currentmarker}{}%
\end{pgfscope}%
\end{pgfscope}%
\begin{pgfscope}%
\pgfpathrectangle{\pgfqpoint{0.251034in}{0.368183in}}{\pgfqpoint{1.556408in}{2.577279in}}%
\pgfusepath{clip}%
\pgfsetbuttcap%
\pgfsetroundjoin%
\definecolor{currentfill}{rgb}{0.000000,0.466667,0.733333}%
\pgfsetfillcolor{currentfill}%
\pgfsetlinewidth{1.003750pt}%
\definecolor{currentstroke}{rgb}{0.000000,0.000000,0.000000}%
\pgfsetstrokecolor{currentstroke}%
\pgfsetdash{}{0pt}%
\pgfsys@defobject{currentmarker}{\pgfqpoint{-0.078567in}{-0.107325in}}{\pgfqpoint{0.107325in}{0.078567in}}{%
\pgfpathmoveto{\pgfqpoint{-0.078567in}{0.078567in}}%
\pgfpathlineto{\pgfqpoint{-0.028758in}{-0.107325in}}%
\pgfpathlineto{\pgfqpoint{0.107325in}{0.028758in}}%
\pgfpathlineto{\pgfqpoint{-0.078567in}{0.078567in}}%
\pgfpathclose%
\pgfusepath{stroke,fill}%
}%
\begin{pgfscope}%
\pgfsys@transformshift{0.417513in}{2.096437in}%
\pgfsys@useobject{currentmarker}{}%
\end{pgfscope}%
\end{pgfscope}%
\begin{pgfscope}%
\pgfpathrectangle{\pgfqpoint{0.251034in}{0.368183in}}{\pgfqpoint{1.556408in}{2.577279in}}%
\pgfusepath{clip}%
\pgfsetbuttcap%
\pgfsetroundjoin%
\definecolor{currentfill}{rgb}{0.000000,0.600000,0.533333}%
\pgfsetfillcolor{currentfill}%
\pgfsetlinewidth{1.003750pt}%
\definecolor{currentstroke}{rgb}{0.000000,0.000000,0.000000}%
\pgfsetstrokecolor{currentstroke}%
\pgfsetdash{}{0pt}%
\pgfsys@defobject{currentmarker}{\pgfqpoint{-0.041667in}{-0.041667in}}{\pgfqpoint{0.041667in}{0.041667in}}{%
\pgfpathmoveto{\pgfqpoint{0.000000in}{-0.041667in}}%
\pgfpathcurveto{\pgfqpoint{0.011050in}{-0.041667in}}{\pgfqpoint{0.021649in}{-0.037276in}}{\pgfqpoint{0.029463in}{-0.029463in}}%
\pgfpathcurveto{\pgfqpoint{0.037276in}{-0.021649in}}{\pgfqpoint{0.041667in}{-0.011050in}}{\pgfqpoint{0.041667in}{0.000000in}}%
\pgfpathcurveto{\pgfqpoint{0.041667in}{0.011050in}}{\pgfqpoint{0.037276in}{0.021649in}}{\pgfqpoint{0.029463in}{0.029463in}}%
\pgfpathcurveto{\pgfqpoint{0.021649in}{0.037276in}}{\pgfqpoint{0.011050in}{0.041667in}}{\pgfqpoint{0.000000in}{0.041667in}}%
\pgfpathcurveto{\pgfqpoint{-0.011050in}{0.041667in}}{\pgfqpoint{-0.021649in}{0.037276in}}{\pgfqpoint{-0.029463in}{0.029463in}}%
\pgfpathcurveto{\pgfqpoint{-0.037276in}{0.021649in}}{\pgfqpoint{-0.041667in}{0.011050in}}{\pgfqpoint{-0.041667in}{0.000000in}}%
\pgfpathcurveto{\pgfqpoint{-0.041667in}{-0.011050in}}{\pgfqpoint{-0.037276in}{-0.021649in}}{\pgfqpoint{-0.029463in}{-0.029463in}}%
\pgfpathcurveto{\pgfqpoint{-0.021649in}{-0.037276in}}{\pgfqpoint{-0.011050in}{-0.041667in}}{\pgfqpoint{0.000000in}{-0.041667in}}%
\pgfpathlineto{\pgfqpoint{0.000000in}{-0.041667in}}%
\pgfpathclose%
\pgfusepath{stroke,fill}%
}%
\end{pgfscope}%
\begin{pgfscope}%
\pgfpathrectangle{\pgfqpoint{0.251034in}{0.368183in}}{\pgfqpoint{1.556408in}{2.577279in}}%
\pgfusepath{clip}%
\pgfsetbuttcap%
\pgfsetroundjoin%
\definecolor{currentfill}{rgb}{0.000000,0.000000,0.000000}%
\pgfsetfillcolor{currentfill}%
\pgfsetlinewidth{1.003750pt}%
\definecolor{currentstroke}{rgb}{0.000000,0.000000,0.000000}%
\pgfsetstrokecolor{currentstroke}%
\pgfsetdash{}{0pt}%
\pgfpathmoveto{\pgfqpoint{1.495516in}{2.058395in}}%
\pgfpathcurveto{\pgfqpoint{1.395134in}{2.142625in}}{\pgfqpoint{1.272759in}{2.196471in}}{\pgfqpoint{1.142841in}{2.213575in}}%
\pgfpathcurveto{\pgfqpoint{1.012922in}{2.230679in}}{\pgfqpoint{0.880782in}{2.210341in}}{\pgfqpoint{0.762020in}{2.154961in}}%
\pgfpathcurveto{\pgfqpoint{0.643258in}{2.099581in}}{\pgfqpoint{0.542739in}{2.011429in}}{\pgfqpoint{0.472332in}{1.900911in}}%
\pgfpathcurveto{\pgfqpoint{0.401924in}{1.790394in}}{\pgfqpoint{0.364512in}{1.662038in}}{\pgfqpoint{0.364512in}{1.530999in}}%
\pgfpathlineto{\pgfqpoint{0.374008in}{1.530999in}}%
\pgfpathcurveto{\pgfqpoint{0.374008in}{1.660231in}}{\pgfqpoint{0.410904in}{1.786816in}}{\pgfqpoint{0.480340in}{1.895809in}}%
\pgfpathcurveto{\pgfqpoint{0.549777in}{2.004802in}}{\pgfqpoint{0.648909in}{2.091739in}}{\pgfqpoint{0.766033in}{2.146355in}}%
\pgfpathcurveto{\pgfqpoint{0.883157in}{2.200971in}}{\pgfqpoint{1.013475in}{2.221028in}}{\pgfqpoint{1.141601in}{2.204160in}}%
\pgfpathcurveto{\pgfqpoint{1.269728in}{2.187292in}}{\pgfqpoint{1.390414in}{2.134189in}}{\pgfqpoint{1.489412in}{2.051120in}}%
\pgfpathlineto{\pgfqpoint{1.495516in}{2.058395in}}%
\pgfpathclose%
\pgfusepath{stroke,fill}%
\end{pgfscope}%
\begin{pgfscope}%
\pgfpathrectangle{\pgfqpoint{0.251034in}{0.368183in}}{\pgfqpoint{1.556408in}{2.577279in}}%
\pgfusepath{clip}%
\pgfsetbuttcap%
\pgfsetroundjoin%
\definecolor{currentfill}{rgb}{0.000000,0.000000,0.000000}%
\pgfsetfillcolor{currentfill}%
\pgfsetlinewidth{1.003750pt}%
\definecolor{currentstroke}{rgb}{0.000000,0.000000,0.000000}%
\pgfsetstrokecolor{currentstroke}%
\pgfsetdash{}{0pt}%
\pgfpathmoveto{\pgfqpoint{0.321779in}{1.530999in}}%
\pgfpathlineto{\pgfqpoint{0.416740in}{1.530999in}}%
\pgfpathlineto{\pgfqpoint{0.369260in}{1.483518in}}%
\pgfpathlineto{\pgfqpoint{0.321779in}{1.530999in}}%
\pgfpathclose%
\pgfusepath{stroke,fill}%
\end{pgfscope}%
\begin{pgfscope}%
\pgfpathrectangle{\pgfqpoint{0.251034in}{0.368183in}}{\pgfqpoint{1.556408in}{2.577279in}}%
\pgfusepath{clip}%
\pgfsetbuttcap%
\pgfsetroundjoin%
\pgfsetlinewidth{1.505625pt}%
\definecolor{currentstroke}{rgb}{0.000000,0.000000,0.000000}%
\pgfsetstrokecolor{currentstroke}%
\pgfsetstrokeopacity{0.600000}%
\pgfsetdash{{5.550000pt}{2.400000pt}}{0.000000pt}%
\pgfpathmoveto{\pgfqpoint{0.364512in}{1.530999in}}%
\pgfpathlineto{\pgfqpoint{0.369260in}{1.611716in}}%
\pgfpathlineto{\pgfqpoint{0.374008in}{1.644952in}}%
\pgfpathlineto{\pgfqpoint{0.383504in}{1.691592in}}%
\pgfpathlineto{\pgfqpoint{0.393000in}{1.726996in}}%
\pgfpathlineto{\pgfqpoint{0.407244in}{1.769774in}}%
\pgfpathlineto{\pgfqpoint{0.421488in}{1.805237in}}%
\pgfpathlineto{\pgfqpoint{0.440481in}{1.845375in}}%
\pgfpathlineto{\pgfqpoint{0.459473in}{1.879907in}}%
\pgfpathlineto{\pgfqpoint{0.478465in}{1.910367in}}%
\pgfpathlineto{\pgfqpoint{0.502205in}{1.944079in}}%
\pgfpathlineto{\pgfqpoint{0.525945in}{1.973969in}}%
\pgfpathlineto{\pgfqpoint{0.554434in}{2.005803in}}%
\pgfpathlineto{\pgfqpoint{0.582922in}{2.034023in}}%
\pgfpathlineto{\pgfqpoint{0.611410in}{2.059207in}}%
\pgfpathlineto{\pgfqpoint{0.644646in}{2.085301in}}%
\pgfpathlineto{\pgfqpoint{0.677883in}{2.108311in}}%
\pgfpathlineto{\pgfqpoint{0.711119in}{2.128592in}}%
\pgfpathlineto{\pgfqpoint{0.744355in}{2.146415in}}%
\pgfpathlineto{\pgfqpoint{0.777592in}{2.161989in}}%
\pgfpathlineto{\pgfqpoint{0.815576in}{2.177239in}}%
\pgfpathlineto{\pgfqpoint{0.853560in}{2.189951in}}%
\pgfpathlineto{\pgfqpoint{0.891545in}{2.200271in}}%
\pgfpathlineto{\pgfqpoint{0.929529in}{2.208307in}}%
\pgfpathlineto{\pgfqpoint{0.967513in}{2.214140in}}%
\pgfpathlineto{\pgfqpoint{1.005498in}{2.217826in}}%
\pgfpathlineto{\pgfqpoint{1.043482in}{2.219400in}}%
\pgfpathlineto{\pgfqpoint{1.081466in}{2.218875in}}%
\pgfpathlineto{\pgfqpoint{1.119451in}{2.216248in}}%
\pgfpathlineto{\pgfqpoint{1.157435in}{2.211495in}}%
\pgfpathlineto{\pgfqpoint{1.195419in}{2.204569in}}%
\pgfpathlineto{\pgfqpoint{1.233404in}{2.195403in}}%
\pgfpathlineto{\pgfqpoint{1.271388in}{2.183902in}}%
\pgfpathlineto{\pgfqpoint{1.309372in}{2.169941in}}%
\pgfpathlineto{\pgfqpoint{1.347357in}{2.153355in}}%
\pgfpathlineto{\pgfqpoint{1.380593in}{2.136519in}}%
\pgfpathlineto{\pgfqpoint{1.413829in}{2.117320in}}%
\pgfpathlineto{\pgfqpoint{1.447066in}{2.095517in}}%
\pgfpathlineto{\pgfqpoint{1.480302in}{2.070795in}}%
\pgfpathlineto{\pgfqpoint{1.508790in}{2.046965in}}%
\pgfpathlineto{\pgfqpoint{1.537278in}{2.020324in}}%
\pgfpathlineto{\pgfqpoint{1.565767in}{1.990383in}}%
\pgfpathlineto{\pgfqpoint{1.589507in}{1.962418in}}%
\pgfpathlineto{\pgfqpoint{1.613247in}{1.931104in}}%
\pgfpathlineto{\pgfqpoint{1.636987in}{1.895579in}}%
\pgfpathlineto{\pgfqpoint{1.655980in}{1.863226in}}%
\pgfpathlineto{\pgfqpoint{1.674972in}{1.826142in}}%
\pgfpathlineto{\pgfqpoint{1.689216in}{1.794034in}}%
\pgfpathlineto{\pgfqpoint{1.703460in}{1.756518in}}%
\pgfpathlineto{\pgfqpoint{1.712956in}{1.726996in}}%
\pgfpathlineto{\pgfqpoint{1.722452in}{1.691592in}}%
\pgfpathlineto{\pgfqpoint{1.731948in}{1.644952in}}%
\pgfpathlineto{\pgfqpoint{1.736696in}{1.611716in}}%
\pgfpathlineto{\pgfqpoint{1.736696in}{1.530999in}}%
\pgfpathlineto{\pgfqpoint{1.731948in}{1.450282in}}%
\pgfpathlineto{\pgfqpoint{1.727200in}{1.417046in}}%
\pgfpathlineto{\pgfqpoint{1.717704in}{1.370405in}}%
\pgfpathlineto{\pgfqpoint{1.708208in}{1.335002in}}%
\pgfpathlineto{\pgfqpoint{1.693964in}{1.292224in}}%
\pgfpathlineto{\pgfqpoint{1.679720in}{1.256761in}}%
\pgfpathlineto{\pgfqpoint{1.660728in}{1.216622in}}%
\pgfpathlineto{\pgfqpoint{1.641735in}{1.182090in}}%
\pgfpathlineto{\pgfqpoint{1.622743in}{1.151630in}}%
\pgfpathlineto{\pgfqpoint{1.599003in}{1.117919in}}%
\pgfpathlineto{\pgfqpoint{1.575263in}{1.088029in}}%
\pgfpathlineto{\pgfqpoint{1.546775in}{1.056194in}}%
\pgfpathlineto{\pgfqpoint{1.518286in}{1.027975in}}%
\pgfpathlineto{\pgfqpoint{1.489798in}{1.002791in}}%
\pgfpathlineto{\pgfqpoint{1.456562in}{0.976697in}}%
\pgfpathlineto{\pgfqpoint{1.423325in}{0.953687in}}%
\pgfpathlineto{\pgfqpoint{1.390089in}{0.933405in}}%
\pgfpathlineto{\pgfqpoint{1.356853in}{0.915582in}}%
\pgfpathlineto{\pgfqpoint{1.323617in}{0.900009in}}%
\pgfpathlineto{\pgfqpoint{1.285632in}{0.884759in}}%
\pgfpathlineto{\pgfqpoint{1.247648in}{0.872046in}}%
\pgfpathlineto{\pgfqpoint{1.209663in}{0.861727in}}%
\pgfpathlineto{\pgfqpoint{1.171679in}{0.853691in}}%
\pgfpathlineto{\pgfqpoint{1.133695in}{0.847858in}}%
\pgfpathlineto{\pgfqpoint{1.095710in}{0.844172in}}%
\pgfpathlineto{\pgfqpoint{1.057726in}{0.842598in}}%
\pgfpathlineto{\pgfqpoint{1.019742in}{0.843122in}}%
\pgfpathlineto{\pgfqpoint{0.981757in}{0.845749in}}%
\pgfpathlineto{\pgfqpoint{0.943773in}{0.850503in}}%
\pgfpathlineto{\pgfqpoint{0.905789in}{0.857429in}}%
\pgfpathlineto{\pgfqpoint{0.867804in}{0.866595in}}%
\pgfpathlineto{\pgfqpoint{0.829820in}{0.878095in}}%
\pgfpathlineto{\pgfqpoint{0.791836in}{0.892056in}}%
\pgfpathlineto{\pgfqpoint{0.753851in}{0.908643in}}%
\pgfpathlineto{\pgfqpoint{0.720615in}{0.925479in}}%
\pgfpathlineto{\pgfqpoint{0.687379in}{0.944678in}}%
\pgfpathlineto{\pgfqpoint{0.654142in}{0.966481in}}%
\pgfpathlineto{\pgfqpoint{0.620906in}{0.991202in}}%
\pgfpathlineto{\pgfqpoint{0.592418in}{1.015033in}}%
\pgfpathlineto{\pgfqpoint{0.563930in}{1.041674in}}%
\pgfpathlineto{\pgfqpoint{0.535441in}{1.071615in}}%
\pgfpathlineto{\pgfqpoint{0.511701in}{1.099580in}}%
\pgfpathlineto{\pgfqpoint{0.487961in}{1.130893in}}%
\pgfpathlineto{\pgfqpoint{0.464221in}{1.166418in}}%
\pgfpathlineto{\pgfqpoint{0.445229in}{1.198771in}}%
\pgfpathlineto{\pgfqpoint{0.426236in}{1.235855in}}%
\pgfpathlineto{\pgfqpoint{0.411992in}{1.267964in}}%
\pgfpathlineto{\pgfqpoint{0.397748in}{1.305479in}}%
\pgfpathlineto{\pgfqpoint{0.388252in}{1.335002in}}%
\pgfpathlineto{\pgfqpoint{0.378756in}{1.370405in}}%
\pgfpathlineto{\pgfqpoint{0.369260in}{1.417046in}}%
\pgfpathlineto{\pgfqpoint{0.364512in}{1.450282in}}%
\pgfpathlineto{\pgfqpoint{0.364512in}{1.530999in}}%
\pgfpathlineto{\pgfqpoint{0.364512in}{1.530999in}}%
\pgfusepath{stroke}%
\end{pgfscope}%
\begin{pgfscope}%
\pgfpathrectangle{\pgfqpoint{0.251034in}{0.368183in}}{\pgfqpoint{1.556408in}{2.577279in}}%
\pgfusepath{clip}%
\pgfsetrectcap%
\pgfsetroundjoin%
\pgfsetlinewidth{1.505625pt}%
\definecolor{currentstroke}{rgb}{0.000000,0.000000,0.000000}%
\pgfsetstrokecolor{currentstroke}%
\pgfsetdash{}{0pt}%
\pgfpathmoveto{\pgfqpoint{0.654915in}{0.965561in}}%
\pgfpathlineto{\pgfqpoint{1.451041in}{2.096437in}}%
\pgfusepath{stroke}%
\end{pgfscope}%
\begin{pgfscope}%
\pgfpathrectangle{\pgfqpoint{0.251034in}{0.368183in}}{\pgfqpoint{1.556408in}{2.577279in}}%
\pgfusepath{clip}%
\pgfsetbuttcap%
\pgfsetroundjoin%
\definecolor{currentfill}{rgb}{0.000000,0.600000,0.533333}%
\pgfsetfillcolor{currentfill}%
\pgfsetlinewidth{1.003750pt}%
\definecolor{currentstroke}{rgb}{0.000000,0.000000,0.000000}%
\pgfsetstrokecolor{currentstroke}%
\pgfsetdash{}{0pt}%
\pgfsys@defobject{currentmarker}{\pgfqpoint{-0.062500in}{-0.062500in}}{\pgfqpoint{0.062500in}{0.062500in}}{%
\pgfpathmoveto{\pgfqpoint{0.000000in}{-0.062500in}}%
\pgfpathcurveto{\pgfqpoint{0.016575in}{-0.062500in}}{\pgfqpoint{0.032474in}{-0.055915in}}{\pgfqpoint{0.044194in}{-0.044194in}}%
\pgfpathcurveto{\pgfqpoint{0.055915in}{-0.032474in}}{\pgfqpoint{0.062500in}{-0.016575in}}{\pgfqpoint{0.062500in}{0.000000in}}%
\pgfpathcurveto{\pgfqpoint{0.062500in}{0.016575in}}{\pgfqpoint{0.055915in}{0.032474in}}{\pgfqpoint{0.044194in}{0.044194in}}%
\pgfpathcurveto{\pgfqpoint{0.032474in}{0.055915in}}{\pgfqpoint{0.016575in}{0.062500in}}{\pgfqpoint{0.000000in}{0.062500in}}%
\pgfpathcurveto{\pgfqpoint{-0.016575in}{0.062500in}}{\pgfqpoint{-0.032474in}{0.055915in}}{\pgfqpoint{-0.044194in}{0.044194in}}%
\pgfpathcurveto{\pgfqpoint{-0.055915in}{0.032474in}}{\pgfqpoint{-0.062500in}{0.016575in}}{\pgfqpoint{-0.062500in}{0.000000in}}%
\pgfpathcurveto{\pgfqpoint{-0.062500in}{-0.016575in}}{\pgfqpoint{-0.055915in}{-0.032474in}}{\pgfqpoint{-0.044194in}{-0.044194in}}%
\pgfpathcurveto{\pgfqpoint{-0.032474in}{-0.055915in}}{\pgfqpoint{-0.016575in}{-0.062500in}}{\pgfqpoint{0.000000in}{-0.062500in}}%
\pgfpathlineto{\pgfqpoint{0.000000in}{-0.062500in}}%
\pgfpathclose%
\pgfusepath{stroke,fill}%
}%
\begin{pgfscope}%
\pgfsys@transformshift{0.654915in}{0.965561in}%
\pgfsys@useobject{currentmarker}{}%
\end{pgfscope}%
\begin{pgfscope}%
\pgfsys@transformshift{1.451041in}{2.096437in}%
\pgfsys@useobject{currentmarker}{}%
\end{pgfscope}%
\end{pgfscope}%
\begin{pgfscope}%
\pgfpathrectangle{\pgfqpoint{0.251034in}{0.368183in}}{\pgfqpoint{1.556408in}{2.577279in}}%
\pgfusepath{clip}%
\pgfsetrectcap%
\pgfsetroundjoin%
\pgfsetlinewidth{1.505625pt}%
\definecolor{currentstroke}{rgb}{0.000000,0.000000,0.000000}%
\pgfsetstrokecolor{currentstroke}%
\pgfsetdash{}{0pt}%
\pgfpathmoveto{\pgfqpoint{1.052978in}{1.530999in}}%
\pgfusepath{stroke}%
\end{pgfscope}%
\begin{pgfscope}%
\pgfpathrectangle{\pgfqpoint{0.251034in}{0.368183in}}{\pgfqpoint{1.556408in}{2.577279in}}%
\pgfusepath{clip}%
\pgfsetbuttcap%
\pgfsetroundjoin%
\definecolor{currentfill}{rgb}{0.000000,0.000000,0.000000}%
\pgfsetfillcolor{currentfill}%
\pgfsetlinewidth{1.003750pt}%
\definecolor{currentstroke}{rgb}{0.000000,0.000000,0.000000}%
\pgfsetstrokecolor{currentstroke}%
\pgfsetdash{}{0pt}%
\pgfsys@defobject{currentmarker}{\pgfqpoint{-0.048611in}{-0.048611in}}{\pgfqpoint{0.048611in}{0.048611in}}{%
\pgfpathmoveto{\pgfqpoint{0.000000in}{-0.048611in}}%
\pgfpathcurveto{\pgfqpoint{0.012892in}{-0.048611in}}{\pgfqpoint{0.025257in}{-0.043489in}}{\pgfqpoint{0.034373in}{-0.034373in}}%
\pgfpathcurveto{\pgfqpoint{0.043489in}{-0.025257in}}{\pgfqpoint{0.048611in}{-0.012892in}}{\pgfqpoint{0.048611in}{0.000000in}}%
\pgfpathcurveto{\pgfqpoint{0.048611in}{0.012892in}}{\pgfqpoint{0.043489in}{0.025257in}}{\pgfqpoint{0.034373in}{0.034373in}}%
\pgfpathcurveto{\pgfqpoint{0.025257in}{0.043489in}}{\pgfqpoint{0.012892in}{0.048611in}}{\pgfqpoint{0.000000in}{0.048611in}}%
\pgfpathcurveto{\pgfqpoint{-0.012892in}{0.048611in}}{\pgfqpoint{-0.025257in}{0.043489in}}{\pgfqpoint{-0.034373in}{0.034373in}}%
\pgfpathcurveto{\pgfqpoint{-0.043489in}{0.025257in}}{\pgfqpoint{-0.048611in}{0.012892in}}{\pgfqpoint{-0.048611in}{0.000000in}}%
\pgfpathcurveto{\pgfqpoint{-0.048611in}{-0.012892in}}{\pgfqpoint{-0.043489in}{-0.025257in}}{\pgfqpoint{-0.034373in}{-0.034373in}}%
\pgfpathcurveto{\pgfqpoint{-0.025257in}{-0.043489in}}{\pgfqpoint{-0.012892in}{-0.048611in}}{\pgfqpoint{0.000000in}{-0.048611in}}%
\pgfpathlineto{\pgfqpoint{0.000000in}{-0.048611in}}%
\pgfpathclose%
\pgfusepath{stroke,fill}%
}%
\begin{pgfscope}%
\pgfsys@transformshift{1.052978in}{1.530999in}%
\pgfsys@useobject{currentmarker}{}%
\end{pgfscope}%
\end{pgfscope}%
\begin{pgfscope}%
\definecolor{textcolor}{rgb}{0.000000,0.000000,0.000000}%
\pgfsetstrokecolor{textcolor}%
\pgfsetfillcolor{textcolor}%
\pgftext[x=1.195419in,y=2.471111in,left,base]{\color{textcolor}{\rmfamily\fontsize{8.000000}{9.600000}\selectfont\catcode`\^=\active\def^{\ifmmode\sp\else\^{}\fi}\catcode`\%=\active\def
\end{pgfscope}%
\begin{pgfscope}%
\definecolor{textcolor}{rgb}{0.000000,0.000000,0.000000}%
\pgfsetstrokecolor{textcolor}%
\pgfsetfillcolor{textcolor}%
\pgftext[x=0.417513in,y=2.286359in,left,base]{\color{textcolor}{\rmfamily\fontsize{8.000000}{9.600000}\selectfont\catcode`\^=\active\def^{\ifmmode\sp\else\^{}\fi}\catcode`\%=\active\def
\end{pgfscope}%
\begin{pgfscope}%
\pgfsetroundcap%
\pgfsetroundjoin%
\pgfsetlinewidth{1.003750pt}%
\definecolor{currentstroke}{rgb}{0.000000,0.000000,0.000000}%
\pgfsetstrokecolor{currentstroke}%
\pgfsetdash{}{0pt}%
\pgfpathmoveto{\pgfqpoint{0.923275in}{1.710703in}}%
\pgfpathquadraticcurveto{\pgfqpoint{0.979996in}{1.632115in}}{\pgfqpoint{1.036718in}{1.553528in}}%
\pgfusepath{stroke}%
\end{pgfscope}%
\begin{pgfscope}%
\pgfsetbuttcap%
\pgfsetmiterjoin%
\definecolor{currentfill}{rgb}{1.000000,1.000000,1.000000}%
\pgfsetfillcolor{currentfill}%
\pgfsetlinewidth{1.104125pt}%
\definecolor{currentstroke}{rgb}{0.000000,0.000000,0.000000}%
\pgfsetstrokecolor{currentstroke}%
\pgfsetdash{}{0pt}%
\pgfpathmoveto{\pgfqpoint{0.617579in}{1.738721in}}%
\pgfpathlineto{\pgfqpoint{1.037047in}{1.738721in}}%
\pgfpathlineto{\pgfqpoint{1.037047in}{1.948597in}}%
\pgfpathlineto{\pgfqpoint{0.617579in}{1.948597in}}%
\pgfpathlineto{\pgfqpoint{0.617579in}{1.738721in}}%
\pgfpathclose%
\pgfusepath{stroke,fill}%
\end{pgfscope}%
\begin{pgfscope}%
\definecolor{textcolor}{rgb}{0.000000,0.000000,0.000000}%
\pgfsetstrokecolor{textcolor}%
\pgfsetfillcolor{textcolor}%
\pgftext[x=0.673135in,y=1.815881in,left,base]{\color{textcolor}{\rmfamily\fontsize{8.000000}{9.600000}\selectfont\catcode`\^=\active\def^{\ifmmode\sp\else\^{}\fi}\catcode`\%=\active\def
\end{pgfscope}%
\begin{pgfscope}%
\pgfsetroundcap%
\pgfsetroundjoin%
\pgfsetlinewidth{1.003750pt}%
\definecolor{currentstroke}{rgb}{0.000000,0.000000,0.000000}%
\pgfsetstrokecolor{currentstroke}%
\pgfsetdash{}{0pt}%
\pgfpathmoveto{\pgfqpoint{1.107122in}{1.284467in}}%
\pgfpathquadraticcurveto{\pgfqpoint{1.022724in}{1.288387in}}{\pgfqpoint{0.953837in}{1.291586in}}%
\pgfusepath{stroke}%
\end{pgfscope}%
\begin{pgfscope}%
\pgfsetroundcap%
\pgfsetroundjoin%
\pgfsetlinewidth{1.003750pt}%
\definecolor{currentstroke}{rgb}{0.000000,0.000000,0.000000}%
\pgfsetstrokecolor{currentstroke}%
\pgfsetdash{}{0pt}%
\pgfpathmoveto{\pgfqpoint{0.997203in}{1.267326in}}%
\pgfpathlineto{\pgfqpoint{0.953837in}{1.291586in}}%
\pgfpathlineto{\pgfqpoint{0.999265in}{1.311722in}}%
\pgfusepath{stroke}%
\end{pgfscope}%
\begin{pgfscope}%
\pgfsetbuttcap%
\pgfsetmiterjoin%
\definecolor{currentfill}{rgb}{1.000000,1.000000,1.000000}%
\pgfsetfillcolor{currentfill}%
\pgfsetlinewidth{1.003750pt}%
\definecolor{currentstroke}{rgb}{0.000000,0.000000,0.000000}%
\pgfsetstrokecolor{currentstroke}%
\pgfsetdash{}{0pt}%
\pgfpathmoveto{\pgfqpoint{1.139864in}{1.168956in}}%
\pgfpathlineto{\pgfqpoint{1.529721in}{1.168956in}}%
\pgfpathlineto{\pgfqpoint{1.529721in}{1.378832in}}%
\pgfpathlineto{\pgfqpoint{1.139864in}{1.378832in}}%
\pgfpathlineto{\pgfqpoint{1.139864in}{1.168956in}}%
\pgfpathclose%
\pgfusepath{stroke,fill}%
\end{pgfscope}%
\begin{pgfscope}%
\definecolor{textcolor}{rgb}{0.000000,0.000000,0.000000}%
\pgfsetstrokecolor{textcolor}%
\pgfsetfillcolor{textcolor}%
\pgftext[x=1.195419in,y=1.246116in,left,base]{\color{textcolor}{\rmfamily\fontsize{8.000000}{9.600000}\selectfont\catcode`\^=\active\def^{\ifmmode\sp\else\^{}\fi}\catcode`\%=\active\def
\end{pgfscope}%
\begin{pgfscope}%
\pgfsetbuttcap%
\pgfsetmiterjoin%
\definecolor{currentfill}{rgb}{1.000000,1.000000,1.000000}%
\pgfsetfillcolor{currentfill}%
\pgfsetfillopacity{0.800000}%
\pgfsetlinewidth{1.003750pt}%
\definecolor{currentstroke}{rgb}{0.800000,0.800000,0.800000}%
\pgfsetstrokecolor{currentstroke}%
\pgfsetstrokeopacity{0.800000}%
\pgfsetdash{}{0pt}%
\pgfpathmoveto{\pgfqpoint{0.654322in}{0.423738in}}%
\pgfpathlineto{\pgfqpoint{1.404154in}{0.423738in}}%
\pgfpathquadraticcurveto{\pgfqpoint{1.426376in}{0.423738in}}{\pgfqpoint{1.426376in}{0.445960in}}%
\pgfpathlineto{\pgfqpoint{1.426376in}{0.589788in}}%
\pgfpathquadraticcurveto{\pgfqpoint{1.426376in}{0.612010in}}{\pgfqpoint{1.404154in}{0.612010in}}%
\pgfpathlineto{\pgfqpoint{0.654322in}{0.612010in}}%
\pgfpathquadraticcurveto{\pgfqpoint{0.632100in}{0.612010in}}{\pgfqpoint{0.632100in}{0.589788in}}%
\pgfpathlineto{\pgfqpoint{0.632100in}{0.445960in}}%
\pgfpathquadraticcurveto{\pgfqpoint{0.632100in}{0.423738in}}{\pgfqpoint{0.654322in}{0.423738in}}%
\pgfpathlineto{\pgfqpoint{0.654322in}{0.423738in}}%
\pgfpathclose%
\pgfusepath{stroke,fill}%
\end{pgfscope}%
\begin{pgfscope}%
\pgfsetbuttcap%
\pgfsetroundjoin%
\definecolor{currentfill}{rgb}{0.000000,0.600000,0.533333}%
\pgfsetfillcolor{currentfill}%
\pgfsetlinewidth{1.003750pt}%
\definecolor{currentstroke}{rgb}{0.000000,0.000000,0.000000}%
\pgfsetstrokecolor{currentstroke}%
\pgfsetdash{}{0pt}%
\pgfsys@defobject{currentmarker}{\pgfqpoint{-0.041667in}{-0.041667in}}{\pgfqpoint{0.041667in}{0.041667in}}{%
\pgfpathmoveto{\pgfqpoint{0.000000in}{-0.041667in}}%
\pgfpathcurveto{\pgfqpoint{0.011050in}{-0.041667in}}{\pgfqpoint{0.021649in}{-0.037276in}}{\pgfqpoint{0.029463in}{-0.029463in}}%
\pgfpathcurveto{\pgfqpoint{0.037276in}{-0.021649in}}{\pgfqpoint{0.041667in}{-0.011050in}}{\pgfqpoint{0.041667in}{0.000000in}}%
\pgfpathcurveto{\pgfqpoint{0.041667in}{0.011050in}}{\pgfqpoint{0.037276in}{0.021649in}}{\pgfqpoint{0.029463in}{0.029463in}}%
\pgfpathcurveto{\pgfqpoint{0.021649in}{0.037276in}}{\pgfqpoint{0.011050in}{0.041667in}}{\pgfqpoint{0.000000in}{0.041667in}}%
\pgfpathcurveto{\pgfqpoint{-0.011050in}{0.041667in}}{\pgfqpoint{-0.021649in}{0.037276in}}{\pgfqpoint{-0.029463in}{0.029463in}}%
\pgfpathcurveto{\pgfqpoint{-0.037276in}{0.021649in}}{\pgfqpoint{-0.041667in}{0.011050in}}{\pgfqpoint{-0.041667in}{0.000000in}}%
\pgfpathcurveto{\pgfqpoint{-0.041667in}{-0.011050in}}{\pgfqpoint{-0.037276in}{-0.021649in}}{\pgfqpoint{-0.029463in}{-0.029463in}}%
\pgfpathcurveto{\pgfqpoint{-0.021649in}{-0.037276in}}{\pgfqpoint{-0.011050in}{-0.041667in}}{\pgfqpoint{0.000000in}{-0.041667in}}%
\pgfpathlineto{\pgfqpoint{0.000000in}{-0.041667in}}%
\pgfpathclose%
\pgfusepath{stroke,fill}%
}%
\begin{pgfscope}%
\pgfsys@transformshift{0.787655in}{0.518954in}%
\pgfsys@useobject{currentmarker}{}%
\end{pgfscope}%
\end{pgfscope}%
\begin{pgfscope}%
\definecolor{textcolor}{rgb}{0.000000,0.000000,0.000000}%
\pgfsetstrokecolor{textcolor}%
\pgfsetfillcolor{textcolor}%
\pgftext[x=0.987655in,y=0.489788in,left,base]{\color{textcolor}{\rmfamily\fontsize{8.000000}{9.600000}\selectfont\catcode`\^=\active\def^{\ifmmode\sp\else\^{}\fi}\catcode`\%=\active\def
\end{pgfscope}%
\end{pgfpicture}%
\makeatother%
\endgroup%

%% file: tex/conclusion.tex
Our work introduces a new approach for joined, two-dimensional wireless propagation parameter and model order selection using a \gls{cnn}.
Using a cell-based encoding of the parameters, it utilizes supervised learning for the training on a synthetic dataset.
The approach performs on par with a model-based \gls{ml} estimator in the low \gls{snr} domain, but suffers from a bias at higher \gls{snr}.
Yet, the estimates are sufficient to warm start a second-order Gauss Newton scheme, combining the \gls{cnn} with \gls{ml} techniques, effectively mitigating the bias. 
The combined algorithmic architecture also delivers all the advantageous properties, such as statistical consistency and efficiency, while demonstrating lower computational complexity. 
Using a \gls{cnn} with its reduced number of trainable parameters, it represents a significant improvement in terms of parameter dimension, quantity, and data complexity, e.g. compared to~\cite{barthelme_machine_2021}.
Our approach exhibits the same superior model order estimation, especially in the low-\gls{snr} regime, already shown in previous work by other authors.
However, the most significant feature is the applicability of synthetically trained \gls{cnn} to measurement data without further modification. 
This is a very promising feature, as it indicates that extensive, site-specific measurement data collection and labeling are not necessary for the successful application of Machine Learning algorithms in the field, if trained carefully.

Due to the \gls{cnn} structure of our approach, it also scales well to more than two dimensions, as demonstrated in~\cite{schieler4d2023}. 
Ultimately, this could allow processing of full \gls{mimo} measurements, and hence, also spatially resolved and polarimetric channel sounding data, similar to~\cite{richter_estimation_2005}.
For this, the preprocessing must be extended with realistic antenna beampatterns by a suitable beamspace transformation as facilitated by the \gls{eadf}~\cite{landmann2004EADF}.
Moreover, an architecture ablation study is required to quantify the performance impacts of the individual architecture blocks, possibly alleviating the aforementioned bias, whose source currently remains unknown.
A further opportunity is in the area of wideband channel estimation, where additional dispersion in delay and Doppler domains, as well as coupling of the two domains and violation of the narrowband assumption, present an even more challenging task.